\documentclass[letterpaper,12pt]{article}%
\usepackage{amsmath,bm}
\usepackage{amssymb}
\usepackage{lscape}
\usepackage{float}
\usepackage{graphicx}
\usepackage{rotating}
\usepackage{pdflscape}
\usepackage{subcaption}
\usepackage{amsthm}
\usepackage{comment}
\usepackage[letterpaper,margin=1.25in]{geometry}
\usepackage{setspace}
\usepackage[colorlinks=true,citecolor=blue, linkcolor=black]{hyperref}
\usepackage[authoryear]{natbib}
\usepackage{color}
\usepackage{colortbl}
\usepackage{appendix}
\usepackage{paralist}
\usepackage{threeparttable}
\usepackage{booktabs}
\usepackage{xcolor}
\usepackage{multirow}
\usepackage{tabularx}
\usepackage{pifont}%
\usepackage{array}
\newcolumntype{P}[1]{>{\centering\arraybackslash}p{#1}}
\usepackage{titlecaps}
\usepackage{titling}
\Addlcwords{for a an and with in of to the for}
\usepackage{silence}
\providecommand{\U}[1]{\protect\rule{.1in}{.1in}}

\RequirePackage{hypernat}
\newtheorem{theorem}{Theorem}

\newtheorem{assumption}{Assumption}

\newtheorem{corollary}[theorem]{Corollary}

\newtheorem{lemma}[theorem]{Lemma}

\newtheorem{proposition}[theorem]{Proposition}

\newtheorem{definition}{Definition}
\newtheorem{example}{Example}
\newtheorem{remark}{Remark}

\catcode`~=11
\newcommand{\urltilde}{\kern -.15em\lower .7ex\hbox{~}\kern .04em}
\catcode`~=13
\makeatletter
\renewcommand{\@seccntformat}[1]{\csname the#1\endcsname.\quad}
\makeatother
\numberwithin{equation}{section}
\hypersetup{pdftitle={When Is Neyman-Orthogonal Inference Feasible? Existence and Relevance}, pdfsubject={Existence and relevance of Neyman-orthogonal moments}, pdfauthor={Facundo Arganaraz and Juan Carlos Escanciano}, pdfkeywords={Neyman orthogonality; Debiased machine learning; Relevance; Weak identification; Sample selection; RKHS} }

\DeclareCaptionLabelFormat{table}{TABLE~#2}
\begin{document}
	\title{When Is Neyman-Orthogonal Inference Feasible?\\Existence and Relevance\thanks{This paper substantially revises and extends our earlier paper,
``On the Existence and Information of Orthogonal Moments''
(2023, arXiv:2303.11418). Research funded by Ministerio de Ciencia e Innovaci\'{o}n MICIN/AEI/10.13039/501100011033, grant CEX2021-001181-M; Comunidad de Madrid, grants EPUC3M11 (V PRICIT) and H2019/HUM-5891; and grant PID2021-127794NB-I00 (MCI/AEI/FEDER, UE)}}
	\author{Facundo Arga\~{n}araz\thanks{Department of Economics. E-mail:
			\href{mailto:facundo.arganaraz@sciencespo.fr}{facundo.arganaraz@sciencespo.fr}.}\\\textit{Sciences Po}
		\and Juan Carlos Escanciano\thanks{Corresponding author: Department of Economics.
			E-mail: \href{mailto:jescanci@eco.uc3m.es}{jescanci@eco.uc3m.es}. }\\\textit{Universidad Carlos III de Madrid}}
\date{September 25, 2026}
\maketitle
\vspace{-2em}
\begin{abstract}
Neyman-orthogonal moments underpin debiased and double machine learning,
yet their existence and informativeness are typically presumed. We ask
three questions: when do nontrivial orthogonal moments exist, which
target directions contain first-order information, and does a proposed
moment detect all such directions? Restricted local overidentification
characterizes existence under attainability. Effective Fisher
information determines relevance; its rank counts informative target
directions. A rank sandwich combines moment Jacobians with target
changes offset by nuisance perturbations, distinguishing information
missed by a moment from information absent from the experiment. In
sample selection without covariate exclusions, this yields
identification-robust inference under rank deficiency. In
latent-variable models, the informative target space is an RKHS. For
Gaussian repeated measurements, additional measurements enlarge the
class of smoothed distributional targets that are regularly learnable,
while CDFs, quantiles, and threshold-policy functionals remain
uninformative at the parametric rate for every finite number of
measurements.
\end{abstract}

\vspace{-0.2em}

\vspace{0.2em}

\newpage
\section{Introduction}\label{sec:intro}
Modern empirical work increasingly relies on Neyman-orthogonal moments
to conduct inference with high-dimensional nuisance parameters and
machine-learning first steps. Orthogonality makes a moment's expectation
locally insensitive to nuisance perturbations that leave the target
unchanged. Debiased and double machine learning (DML) typically begin
with an identifying moment and construct an orthogonal adjustment.
Before feasible estimation, however, three logically distinct oracle
questions arise. First, can a given target--model pair admit any
nontrivial orthogonal moment? Second, after nuisance variation is
allowed, which directions of the target contain first-order
information? Third, does a proposed orthogonal moment detect all of
those informative directions, or does it leave some of them unused?
We refer to these three properties as existence, relevance, and strong
relevance, respectively. Strong relevance is distinct from efficiency:
it asks whether informative target directions are missed, not whether
the information in detected directions is used optimally.

These distinctions matter because failure can occur at each stage.
Orthogonal moments may exist yet have no local power against departures
from the target-restricted model. A relevant moment may detect some but
not all informative target directions. Conversely, a rank-deficient
moment may already detect every informative direction because those it
misses can be offset by nuisance changes and are therefore
uninformative in the statistical experiment. Orthogonal informative
inference on particular coefficient combinations may consequently be
possible even when the full coefficient vector is not point identified.
Our analysis separates these possibilities and provides ways to verify
them. Its organizing principle is simple: orthogonalization isolates
information but cannot create it. The paper makes four contributions.

First, we characterize existence and relevance for admissible moment
functions. For existence, the key object is the restricted model obtained
by holding the target fixed while allowing everything else to vary.
Under our attainability conditions, a nontrivial orthogonal moment exists
if and only if this restricted model imposes a first-order restriction
on the observed-data law---a property we call \emph{restricted local
overidentification}. Unlike local overidentification of the full model
\citep{chen2018overidentification}, this is a property of the
target--model pair. The orthocomplement geometry is classical; attainability is not.
We establish the pathwise moment identity---which relates a
moment's first-order sensitivity to parameter perturbations
to its covariance with the corresponding score---under primitive
conditions permitting unbounded moments and unbounded likelihood
ratios, and give richness conditions ensuring that admissible
orthogonal moments are dense in the orthocomplement. We then characterize which moments are
\emph{relevant}, meaning that their oracle tests have nontrivial local
power against some parametric-rate departure from the restricted model.
Under suitable target--nuisance coordinates and integrability conditions,
a relevant moment exists if and only if effective Fisher information
is nonzero. Its rank---the \emph{relevance rank}---counts independent
informative target directions and may be smaller than the target dimension.
These are oracle properties; feasible inference additionally requires
control of the effect of estimating nuisance components.

Second, we provide a practical way to determine whether a proposed
orthogonal moment captures every informative direction. Its target
Jacobian gives a lower bound on the relevance rank. Target directions
whose effects can be offset by nuisance changes give an upper bound.
When these bounds coincide, our \emph{rank sandwich} identifies the
relevance rank and certifies that the moment captures every informative
direction, without computing the efficient score or full tangent spaces.
We call this property \emph{strong relevance}: it requires
neither full-rank information nor efficiency. For affine
conditional restrictions, we provide a finite-matrix
sufficient check based on selected nuisance perturbations.

Third, we characterize orthogonal moments and develop inference in
semiparametric sample selection without credible covariate exclusion restrictions
\citep{blundell2007changes,jones2015selection}, where a standard response
is partial identification \citep{lee2009training,honore2020selection}.
Residualizing covariates on the propensity score yields a single
covariance matrix whose range identifies coefficient combinations from
the orthogonal moments and whose rank equals the relevance rank.
A canonical residualized-instrument vector captures every informative
direction, even when the full coefficient vector is not point identified.
We construct Anderson--Rubin--Neyman-Orthogonal (AR--NO) confidence sets
\citep{anderson1949estimation,neyman1959} by centering the expansion at
a minimum-norm observationally equivalent coefficient vector, so
unidentified nuisance coefficients need not be consistently estimated.
A distinctive difficulty is that target sensitivity and score variance
can vanish together. We establish uniform coverage as positive relevance
weakens, under explicit first-step and spectral-separation conditions
that permit recovery of the population range. Monte Carlo experiments
illustrate the informative subspace, inference under rank deficiency,
and the limits of range recovery. In Current Population Survey (CPS)
data, the AR--NO intervals for Mexican American wage differentials
exclude zero for both men and women and overlap the Honor\'e--Hu
estimated identified sets.

Fourth, we characterize the informative target space in latent-variable
models for functionals of the latent distribution, with teacher value-added
as a leading application \citep{rockoff2004impact,chetty2014measuring}.
In the repeated-measurement model, orthogonal moments exist for every
target, so relevance is the binding issue. We show that the exact range
of informative linear targets is an RKHS whose kernel is the
$\chi^2$-affinity of the measurement experiment. This yields both
possibility and impossibility results. In the Gaussian repeated-measurement model, a distributional target smoothed at scale $h$ admits a relevant orthogonal moment when $h>\theta_0/\sqrt J$, but not when $h<\theta_0/\sqrt J$, where $\theta_0$ is the measurement-error standard deviation and $J$ the number of measurements. As $J$ increases, this threshold falls, expanding the class of smoothed distributional targets amenable to regular inference. Repeated measurements therefore affect which targets can be learned, not only the precision with which they can be estimated. By contrast,
unsmoothed CDFs, quantiles, and threshold-policy functionals remain
outside the informative space for every finite $J$. Existence and
relevance are thus fundamentally distinct properties, and understanding
the limits of orthogonalization can matter as much as understanding its
possibilities.
\subsection{Relation to the literature}

Orthogonalization originates in \citet{neyman1959} and classical semiparametric
geometry \citep{bickel1993efficient,van1991differentiable,newey1994asymptotic};
modern work develops inference with high-dimensional or machine-learning first steps
\citep{belloni2015uniform,chernozhukov2018double}. Adjustment-based approaches correct an identifying moment
for nuisance estimation \citep{chernozhukov2022locally}. Our existence
criterion requires neither a starting identifying moment, point
identification, nor a regular target--nuisance parameterization.
It concerns the entire admissible orthogonal-moment class associated
with the target-restricted observed-law model.
\citet{chenxie2026} study efficiency gains from model-level local
overidentification; \citet{chen2026equivalence} link orthogonality and
pathwise differentiability under a full-rank Jacobian and bounded
estimating equations, whereas we allow singular Jacobians and
unbounded moments under explicit integrability conditions.

Our relevance analysis is closest to
\citet{lee2024locallyregularefficienttests}, who constructs locally
regular nuisance-orthogonal \(C(\alpha)\) tests for nonregular models
and derives power bounds, including under singular efficient
information. Our complementary contribution concerns admissible
moment functions and verification: the rank sandwich can establish
that a proposed moment detects every informative target direction
using its Jacobian and target changes offset by nuisance perturbations.
When the bounds coincide, this verification requires neither the
efficient score nor full tangent-space calculations.

The AR--NO procedure connects orthogonal inference to identification-robust
testing \citep{anderson1949estimation}. \citet{belloni2015uniform} proposed related inferences, but require a
full-rank identifying derivative, while \citet{ma2025} allows a weak Jacobian with
nondegenerate score covariance. Here, the Jacobian and score variance weaken
together through the same residual covariate variation.

In sample selection, rather than imposing full-vector point identification, we
characterize all orthogonal moments, their informative coefficient directions,
and inference under singular information. This complements identification from
functional form \citep{ahn1993semiparametric,escanciano2016identification},
partial identification \citep{lee2009training,honore2020selection}, and locally
robust estimation under point-identifying conditions
\citep{pan2024locally,kim2025point}.

For latent-variable models, we connect teacher value-added and
latent-distribution recovery
\citep{rockoff2004impact,chetty2014measuring,carrasco2011spectral,gilraine2020new,gaillac2025predicting}
to semiparametric range conditions and RKHS theory
\citep{aronszajn1950,carrasco2007linear,bonhomme2012functional}. We characterize
the entire informative space and a smoothness frontier, complementing
\citet{escanciano2023irregular} and deconvolution lower bounds
\citep{fan1991optimal}.

Two companion papers develop constructive methods based on the general theory of this paper:
\citet{arganaraz2025debiased} derive feasible DML for functionals of
nonparametric unobserved heterogeneity, while
\citet{arganarazescancianoCMR2024} treat conditional moment restrictions.

\subsection{Roadmap}\label{subsec:roadmap}

Section~\ref{SecExistence} develops the general theory;
Sections~\ref{sectionExampleSelection} and~\ref{TVAcexample}
apply it to sample selection and latent-variable models, respectively.
Section~\ref{sec:conclusions} concludes.
Appendices~\ref{app:prelim}--\ref{app:section3} give general
preliminaries and proofs; the Supplementary Appendix contains
application proofs and primitive conditions, mixture-model
tangent theory, the one-sided cone extension, and additional
Monte Carlo results.	%%%%%%%%%%%%%%%%%%%%%%%%%%%%%%%%%%%%%%%%%%%%%%%%%%%%%%
	\section{Existence and Relevance}
	\label{SecExistence}

\subsection{Setting}
\label{Setting}

Let $Z,Z_1,\ldots,Z_n$ be iid observations on a measurable space
$(\mathcal Z,\mathcal F)$ with distribution
$\mathbb P_0=\mathbb P_{\lambda_0}$ in the semiparametric model $\mathcal P=\{\mathbb P_\lambda:\lambda\in\Lambda\}.$ The structural parameter $\lambda$ may be infinite-dimensional and
need not be identified: distinct values may generate $\mathbb P_0$.
The target is $\psi_0:=\psi(\lambda_0)\in\mathbb R^{d_\psi}$,
$d_\psi<\infty$.\footnote{Finite dimensionality is used for the rank
results; the existence theory extends to Banach-valued targets.}
Throughout, $\mathbb E$ denotes expectation under $\mathbb P_0$.

\medskip\noindent\textbf{Notation.}
Write $f_\lambda=d\mathbb P_\lambda/d\mu$ for a $\sigma$-finite
dominating measure $\mu$, and $f_0=f_{\lambda_0}$.
Let $L_p=L_p(\mathbb P_0)$, $1\le p\le\infty$, with norm $\|g\|_{L_p}=\left(\mathbb E\!\left[|g(Z)|^p\right]\right)^{1/p}$, $1\le p<\infty$; write $L_p(P)$ for another measure $P$.
On $L_2$, set $\langle g,h\rangle=\mathbb E[gh]$ and
$L_2^0=\{g\in L_2:\mathbb E[g]=0\}$.
For $A\subseteq L_2^0$, $A^\perp$ is its orthocomplement in $L_2^0$
and $\Pi_A$ projects onto its closed linear span.
Abbreviate
$\overline{\mathcal T}=\overline{\operatorname{span}}\,\mathcal T$
and
$\overline{\mathcal T_0}=\overline{\operatorname{span}}\,\mathcal T_0$.
For an operator $L$, $\ker(L)$ and $\operatorname{Range}(L)$ denote
its kernel and range. For a matrix $A$, $A^\dagger$ is its
Moore--Penrose inverse, $A'$ its transpose, and
$\|A\|=\{\operatorname{tr}(A'A)\}^{1/2}$.
The vector $e_j$ is the $j$th standard basis vector.

\subsection{Tangent spaces and overview of the results}
\label{Regularity}

Fixing the target defines the restricted model
\[
\mathcal P_0
:=\{\mathbb P_\lambda:\lambda\in\Lambda,\ \psi(\lambda)=\psi_0\}.
\]
A \emph{restricted path} is
$\{\mathbb P_\tau=\mathbb P_{\lambda_\tau}:
\tau\in[0,\varepsilon)\}\subseteq\mathcal P_0$,
with $\mathbb P_{\tau=0}=\mathbb P_0$ and
$\psi(\lambda_\tau)=\psi_0$ throughout. Writing $f_\tau=f_{\lambda_\tau}$, the path is
\emph{differentiable in quadratic mean} (DQM) with score
$s_0\in L_2^0$ if
\begin{equation}
\int
\left[
\frac{f_\tau^{1/2}-f_0^{1/2}}{\tau}
-\frac12 s_0f_0^{1/2}
\right]^2d\mu\longrightarrow0,
\qquad \tau\downarrow0.
\label{msd}
\end{equation}
The scores of all restricted DQM paths form the
\emph{restricted tangent set} $\mathcal T_0$.
Allowing all DQM paths in $\mathcal P$ through $\mathbb P_0$
gives the \emph{full tangent set} $\mathcal T$, so
$\mathcal T_0\subseteq\mathcal T$.
Orthogonality to target-preserving perturbations is described by
$\overline{\mathcal T_0}^{\perp}$.

When $\psi(\lambda_\tau)=\psi_0+\tau a+o(\tau)$, we call
$a\in\mathbb R^{d_\psi}$ the path's \emph{target derivative}.
A DQM path is \emph{first-order observationally flat} if its score
is zero, equivalently if
$\int(\sqrt{f_\tau}-\sqrt{f_0})^2\,d\mu=o(\tau^2)$.
Such a path may have a nonzero target derivative, which the
tangent set alone does not record.

\begin{assumption}[Regularity]
\label{Regularity_model}
(i) The data $\{Z_i\}_{i=1}^{n}$ are iid from
$\mathbb P_0\in\mathcal P$;
(ii) the tangent sets $\mathcal T$ and $\mathcal T_0$ are linear
subspaces of $L_2^0$.
\end{assumption}

Part~(i) supplies the iid local experiment.
Part~(ii) requires linearity although paths are indexed by
$\tau\ge0$. Supplementary Appendix~\ref{app:cone} treats one-sided
orthogonality using closed convex tangent cones and their polars.

\begin{remark}[Generating classes]
\label{rem:generating}
A \emph{generating class} $(\mathfrak G,\mathfrak G_0)$ consists
of full-model and restricted DQM paths whose scores have closed
linear spans $\overline{\mathcal T}$ and
$\overline{\mathcal T_0}$, respectively.
The existence and intrinsic relevance results depend only on
these spans, so their score-continuous conditions may be verified
on a generating class.
\end{remark}

Table~\ref{tab:main_results} separates properties of the
target--model pair from those of a proposed moment.
The following subsections establish these characterizations.

\begin{table}[H]
		\centering
		\caption{Existence and relevance: a summary}
		\label{tab:main_results}
		\footnotesize
		\renewcommand{\arraystretch}{1.2}
		\begin{minipage}{0.96\textwidth}
			\centering
			\resizebox{\linewidth}{!}{%
			\begin{tabular}{@{}llll@{}}
				\toprule
				Property & Intrinsic criterion & In coordinates & Meaning \\
				\midrule
				\multicolumn{4}{@{}l}{\emph{Panel A. Properties of the target--model pair $(\psi,\mathcal P)$}}\\[2pt]
				Existence
				& $\overline{\mathcal T_0}\neq L_2^{0}$
				& ---
				& some nontrivial $g\in\mathcal G_{\mathrm{LR}}$ exists \\
				Relevance
				& $\widetilde S_\psi\neq\{0\}$
				& $I_\psi\neq0$
				& some direction of $\psi_0$ is detectable \\
				Full information
				& $r_\psi=d_\psi$
				& $I_\psi\succ0$
				& every direction of $\psi_0$ is detectable \\[4pt]
				\multicolumn{4}{@{}l}{
\emph{Panel B. Properties of a given orthogonal moment vector
$g=(g_1,\ldots,g_m)'$}}\\[2pt]
                Relevant
				& $\Pi_{\widetilde S_\psi}g\neq0$
				& $J_{\psi,g}\neq0$
				& $g$ detects at least one such direction \\
				Strongly relevant
				& $L_g$ injective
				& $\operatorname{rank}(J_{\psi,g})=r_\psi$
				& $g$ detects all $r_\psi$ of them \\
				\bottomrule
			\end{tabular}}
			\par\vspace{2pt}\raggedright\footnotesize
\textit{Notes:}
Entries hold under the assumptions of the corresponding results below.
$\mathcal G_{\mathrm{LR}}$ denotes admissible orthogonal moments;
$\widetilde S_\psi
=\overline{\mathcal T}\cap\overline{\mathcal T_0}^{\perp}$
is the informative score space, with relevance rank
$r_\psi=\dim(\widetilde S_\psi)$.
$I_\psi$ is effective Fisher information,
$J_{\psi,g}$ is the target Jacobian of the population moment
with nuisance held fixed, and $L_g(s)=\mathbb E[gs]$.
Detection concerns oracle local power at the $n^{-1/2}$ scale.
Strong relevance requires neither full-rank information nor efficiency.
For any verified subspace $\mathcal C$ of target directions
offset by nuisance perturbations to first order,
$\operatorname{rank}(J_{\psi,g})\le r_\psi
\le d_\psi-\dim(\mathcal C)$;
matching bounds certify strong relevance.
\end{minipage}
	\end{table}

\subsection{Admissible moments and local robustness}
\label{AdmissibleMoments}

We first specify valid moment maps and the regularity needed to
differentiate their expectations, then define local robustness.
Let $\mathbb E_\tau$ denote expectation under $\mathbb P_\tau$.

\begin{definition}[Admissible moments]
\label{def:admissible}
An admissible moment is a parameter-indexed map
$g:\mathcal Z\times\Lambda\to\mathbb R$ whose true evaluation
$g_0:=g(\cdot,\lambda_0)$ is a nonzero element of $L_2^0$
and such that, for every path in
$\mathfrak G\cup\mathfrak G_0$ with score $s$:
\begin{enumerate}
\item[(i)]
$\mathbb E_\tau[g(Z,\lambda_\tau)]=0$ for all sufficiently
small $\tau\ge0$;
\label{eq:admissible_validity}
\item[(ii)]
the derivatives
$\frac{d}{d\tau}\mathbb E[g(Z,\lambda_\tau)]\big|_{\tau=0^+}$
and
$\frac{d}{d\tau}\mathbb E_\tau[g(Z,\lambda_0)]\big|_{\tau=0^+}$
exist, and
\begin{equation}
\frac{d}{d\tau}\mathbb E_\tau[g(Z,\lambda_0)]
\Big|_{\tau=0^+}
=\mathbb E[g(Z,\lambda_0)\,s(Z)];
\label{GI}
\end{equation}
\item[(iii)]
$\|g(\cdot,\lambda_\tau)-g(\cdot,\lambda_0)\|_{L_2(\mathbb P_\tau)}
+\|g(\cdot,\lambda_\tau)-g(\cdot,\lambda_0)\|_{L_2(\mathbb P_0)}
\to0$.
\label{cont}
\end{enumerate}
We write $\mathcal G_{\mathrm{adm}}$ for the class of admissible
maps and $\mathcal G_{\mathrm{adm}}^0$ for their true evaluations.
When no confusion can arise, $g$ denotes both the map and its
true evaluation.
\end{definition}

For the existence statements in this subsection,
admissibility requires conditions~(i)--(iii) of
Definition~\ref{def:admissible} only along a restricted
generating class $\mathfrak G_0$.

\begin{remark}
\label{remarkdif}
Condition (i) requires correct specification of the moment along
the path; the conditions in (ii) are standard
(see, e.g., \citet{newey1994asymptotic}); and (iii) is a mild
continuity requirement. These conditions hold, for example, whenever
\begin{equation}
g(Z,\lambda)=m(Z)-\int m(z)f_\lambda(z)\,d\mu(z),
\label{const_adm}
\end{equation}
provided that, for all paths in $\mathfrak G\cup\mathfrak G_0$,
\begin{equation}
\limsup_{\tau\downarrow0}\int m^2(z)f_\tau(z)\,d\mu(z)<\infty;
\label{bounded_mom}
\end{equation}
see Lemma~I.7.2 of \citet{ibragimovkhasminskii1981}.
Admissibility therefore does not require boundedness of $g$,
and \eqref{const_adm} provides a general construction of admissible
moments; see Lemma~\ref{lem:autoadm} in Appendix~\ref{app:richness}.
\end{remark}

Condition \eqref{bounded_mom} constrains the moment function.
The next definition instead constrains the \emph{paths}, making
\eqref{bounded_mom} automatic for $m\in L_{2q}$ by H\"older's
inequality, where $1/p+1/q=1$. Because the geometric results
depend only on closed score spans, it is enough that one such
generating class exists.

\begin{definition}[Locally dominated generating classes]
\label{def:regularclass}
A generating class $(\mathfrak G,\mathfrak G_0)$ is
\emph{locally dominated} at $\mathbb P_0$ if there exist
$p\in(1,\infty]$, $\varepsilon>0$, and a nonnegative
$M_p\in L_p$ such that
\[
\frac{d\mathbb P_\tau}{d\mathbb P_0}\le M_p(Z)
\quad\mathbb P_0\text{-a.s. for all }\tau\in[0,\varepsilon)
\]
and every path in $\mathfrak G\cup\mathfrak G_0$.
It is \emph{locally bounded} if $M_\infty$ can be chosen
to be constant.
\end{definition}

\begin{assumption}[Regular generating class]
\label{ass:regclass}
There is a locally dominated generating class
$(\mathfrak G,\mathfrak G_0)$ for
$(\overline{\mathcal T},\overline{\mathcal T_0})$ with exponent $p\in(1,\infty]$.
\end{assumption}

\begin{assumption}[Richness for existence]
\label{ass:dense0}
$\overline{\mathcal T_0}^{\perp}\cap L_{2q}$ is dense in
$\overline{\mathcal T_0}^{\perp}$, where $q$ satisfies
$1/p+1/q=1$ and $p$ is the exponent supplied by
Assumption~\ref{ass:regclass}.
\end{assumption}

Local boundedness permits $q=1$, making
Assumption~\ref{ass:dense0} automatic. The weaker domination
condition is needed for Gaussian noise-scale perturbations in
Section~\ref{TVAcexample}, whose likelihood ratios are unbounded.
For $q>1$, Lemma~\ref{lem:density} gives sufficient conditions
for richness.

Together, Assumptions~\ref{ass:regclass}--\ref{ass:dense0}
ensure that every nonzero element of
$\overline{\mathcal T_0}^{\perp}\cap L_{2q}$ is realized by
an admissible moment, with
$\overline{\mathcal T_0}^{\perp}\cap L_{2q}$ dense in
$\overline{\mathcal T_0}^{\perp}$.
Under local boundedness, every nonzero element of the
orthocomplement is realized; see Lemma~\ref{lem:autoadm} and
Corollary~\ref{cor:richness} in Appendix~\ref{app:richness}.

\begin{definition}[Locally robust moments]
\label{def:LR}
An admissible moment map $g\in\mathcal G_{\mathrm{adm}}$
is \emph{locally robust} for $\psi(\lambda_0)$ if
\begin{equation}
\frac{d}{d\tau}\mathbb E[g(Z,\lambda_\tau)]
\Big|_{\tau=0^+}=0
\label{DLR}
\end{equation}
for every path in $\mathfrak G_0$.
We denote the set of true evaluations of such maps by
$\mathcal G_{\mathrm{LR}}\subset L_2^0$.
\end{definition}

This definition uses target-preserving paths without requiring
a parameterization $\lambda=\Xi(\psi,\eta)$ that separates
the target $\psi$ from a nuisance parameter $\eta$.
When such regular coordinates exist, local robustness is
equivalent to vanishing nuisance G\^ateaux derivatives;
see Proposition~\ref{prop:LR_derivative} in
Appendix~\ref{app:der}. Existence depends on the following property of the restricted model.

\begin{definition}[Restricted local overidentification]
\label{def:surjectivity}
The pair $(\psi,\mathcal P)$ satisfies
\emph{restricted local overidentification} at $\psi(\lambda_0)$
if $\overline{\mathcal T_0}\neq L_2^0$.
\end{definition}

The condition leaves a nonzero direction orthogonal to all
target-preserving scores. It depends on both the target and
$\lambda_0$: changing either may change whether it holds.

\subsection{Existence of Locally Robust Moments}
\label{sec:existence}

The next theorem characterizes the admissible orthogonal-moment
class and its existence; its proof is in Appendix~\ref{app:existence}.

\begin{theorem}[Existence of locally robust moments]
\label{mainExistence}
Let Assumptions~\ref{Regularity_model}--\ref{ass:dense0} hold. Then
\[
\mathcal G_{\mathrm{LR}}
=\overline{\mathcal T_0}^{\perp}\cap\mathcal G_{\mathrm{adm}}^0
\ \supseteq\
\bigl(\overline{\mathcal T_0}^{\perp}\cap L_{2q}\bigr)
\setminus\{0\}.
\]
If the generating class is locally bounded, then
$\mathcal G_{\mathrm{LR}}
=\overline{\mathcal T_0}^{\perp}\setminus\{0\}$.
A nontrivial locally robust moment for $\psi(\lambda_0)$
exists iff the pair $(\psi,\mathcal P)$ satisfies restricted
local overidentification.
\end{theorem}

The orthocomplement characterization follows from the key pathwise
identity in Lemma~\ref{lem:pathwise}, established there under
Definition~\ref{def:admissible} rather than under boundedness of
the moment or of the likelihood ratios. The existence equivalence additionally requires attainability:
Assumptions~\ref{ass:regclass}--\ref{ass:dense0} ensure that
a nontrivial orthocomplement contains an admissible moment.
This is an oracle statement; feasible construction and estimation
require further, model-specific conditions.

\begin{remark}[Comparison with adjustment-based approaches]
\label{rem:riesz}
Influence-function adjustments, including constructions based on
Riesz representers, correct a starting identifying moment for nuisance estimation
\citep{chernozhukov2022locally}.
Theorem~\ref{mainExistence} characterizes the entire admissible
orthogonal-moment class for a target--model pair, without requiring
a starting identifying moment or point identification.
A particular adjusted moment may vanish even when other nonzero
orthogonal moments exist.
\end{remark}
In conditional moment models
\citep{chamberlain1987,aichen2003,aichen2012,chenpouzo2015},
linearizing the restrictions often characterizes
$\overline{\mathcal T_0}^{\perp}$ directly.
Its admissible elements give all locally robust moments;
the relevance criteria below determine which are informative.

Since $\mathcal T_0\subseteq\mathcal T$, model-level local
overidentification---$\overline{\mathcal T}\neq L_2^0$ in
\citet{chen2018overidentification}---gives an immediate
sufficient condition.

\begin{corollary}
\label{corchenandsantos}
Under the assumptions of Theorem~\ref{mainExistence}, if
$\mathcal P$ is locally overidentified in the sense of
\citet{chen2018overidentification}, then nontrivial locally
robust moments exist for every target $\psi(\lambda_0)$.
\end{corollary}

The corollary supplies existence in Section~\ref{TVAcexample}.
Full-model overidentification is unnecessary: fixing the
target may itself impose first-order restrictions.%%%%%%%%%%%%%%%%%%%%%%%%%%%%%%%%%%%%%%%%%%%%%%%%
\subsection{Relevance of Locally Robust Moments}
\label{Relevance}

\subsubsection{Intrinsic Geometry}

Let $g\in\mathcal G_{\mathrm{LR}}$ be scalar and
$\{\mathbb P_\tau\}\in\mathfrak G$ a full-model DQM path
through $\mathbb P_0$ with score $s$.
Write $\rightsquigarrow_{\mathbb P_{n,s}}$ for weak convergence
under $\mathbb P_{n,s}:=\mathbb P_{1/\sqrt n}^{\otimes n}$.
Le Cam's third lemma gives
\begin{equation}
n^{-1/2}\sum_{i=1}^{n}g(Z_i,\lambda_0)
\rightsquigarrow_{\mathbb P_{n,s}}
N\bigl(\langle g,s\rangle,\ \|g\|_{L_2}^{2}\bigr).
\label{oracle}
\end{equation}
By \eqref{eq:appendix_pathwise_identity},
$\frac{d}{d\tau}\mathbb E[g(Z,\lambda_\tau)]
\big|_{\tau=0^+}=-\langle g,s\rangle$.
Thus $\langle g,s\rangle$ determines both the oracle local
mean shift and the moment's first-order sensitivity. This motivates the following definition:

\begin{definition}[Relevance]
\label{def:relevance}
$g\in\mathcal G_{\mathrm{LR}}$ is \emph{relevant} if
$\langle g,s\rangle\neq0$ for some
$s\in\overline{\mathcal T}\setminus\overline{\mathcal T_0}$.
\end{definition}

\begin{proposition}[Orthogonal decomposition]
\label{prop:decomposition}
Under Assumption~\ref{Regularity_model},
\[
\overline{\mathcal T_0}^{\perp}
=\widetilde S_\psi\oplus\overline{\mathcal T}^{\perp},
\qquad
\widetilde S_\psi
:=\overline{\mathcal T}\cap\overline{\mathcal T_0}^{\perp}.
\]
\end{proposition}

The component $\overline{\mathcal T}^{\perp}$ is orthogonal
to every full-model score and contributes no local power.
All target information therefore lies in $\widetilde S_\psi$.
Its dimension $r_\psi:=\dim(\widetilde S_\psi)$ is the
\emph{relevance rank}. Attaining this geometry requires
the following richness condition.

\begin{assumption}[Richness for relevance]
\label{ass:denseS}
$\widetilde S_\psi\cap L_{2q}$ is dense in $\widetilde S_\psi$.
\end{assumption}

This condition is automatic under local boundedness but is
not implied by Assumption~\ref{ass:dense0}.
Under the target coordinates below, $\widetilde S_\psi$
is finite-dimensional, so richness reduces to a moment
condition by Lemma~\ref{lem:density}(b).

\begin{definition}[Independently relevant moments]
\label{def:indep_relevant}
Scalar moments $g_1,\ldots,g_k\in\mathcal G_{\mathrm{LR}}$
are \emph{independently relevant} if
$\Pi_{\widetilde S_\psi}g_1,\ldots,
\Pi_{\widetilde S_\psi}g_k$ are linearly independent in $L_2$.
\end{definition}

\begin{theorem}[Geometric characterization of relevance]
\label{thm:relevance}
Under Assumptions~\ref{Regularity_model},
\ref{ass:regclass}, and~\ref{ass:denseS}, a relevant locally
robust moment exists iff $\widetilde S_\psi\neq\{0\}$,
and the maximal number of independently relevant locally
robust moments is $r_\psi$.
\end{theorem}

Under the stated attainability conditions, existence without relevance occurs exactly when the full model is locally overidentified and $\widetilde S_\psi=\{0\}$, as in Section~\ref{TVAcexample}. When $\overline{\mathcal T}=L_2^0$,
existence and relevance coincide: fixing the target may
restrict the observed law even in a saturated model. Target coordinates give this intrinsic criterion an
information-theoretic interpretation and yield practical
rank bounds.
\subsubsection{Information Geometry}
\label{sec:info_geometry}

\begin{assumption}[Target coordinates]
\label{ass:coordinate_paths}
Let $(\mathfrak G,\mathfrak G_0)$ be a generating class in
the sense of Remark~\ref{rem:generating}, not necessarily
locally dominated. Suppose there exist
$\Psi\subset\mathbb R^{d_\psi}$, a linear space $\mathcal H$,
a point $(\psi_0,\eta_0)\in\Psi\times\mathcal H$, and
$\Xi:\Psi\times\mathcal H\to\Lambda$ such that there are
local coordinates $\lambda=\Xi(\psi,\eta)$ with
$\Xi(\psi_0,\eta_0)=\lambda_0$ and
$\psi(\Xi(\psi,\eta))=\psi$. Also,
\begin{enumerate}
\item[(i)]
for every $a\in\mathbb R^{d_\psi}$, the path
$\tau\mapsto\Xi(\psi_0+\tau a,\eta_0)$ belongs to
$\mathfrak G$ with score $a'\ell_\psi$;
\item[(ii)]
every path in $\mathfrak G$ has score
$s=a'\ell_\psi+s_\eta$ with $a\in\mathbb R^{d_\psi}$
and $s_\eta\in\overline{\mathcal T_0}$, where $a$ is the
coordinate derivative of the path, so that
$\psi(\lambda_\tau)=\psi_0+\tau a+o(\tau)$.
\item[(iii)]
for every $c\in\mathbb R^{d_\psi}\setminus\{0\}$, if
$\overline{\mathcal T_0}^{\,c}$ denotes the closed tangent
space generated by all DQM paths satisfying
$c'\psi(\lambda_\tau)=c'\psi_0$, then
$\overline{\mathcal T_0}^{\,c}
=\overline{\mathcal T_0}
+\operatorname{span}\{a'\ell_\psi:c'a=0\}$.
\end{enumerate}
\end{assumption}

Define
\[
\ell_\psi^{\rm eff}
:=\ell_\psi-\Pi_{\overline{\mathcal T_0}}\ell_\psi,
\qquad
I_\psi
:=\mathbb E[\ell_\psi^{\rm eff}\ell_\psi^{\rm eff\prime}],
\qquad
Aa:=a'\ell_\psi^{\rm eff}.
\]
Here $I_\psi$ is the effective Fisher information,
$A:\mathbb R^{d_\psi}\to L_2^0$ is linear, and
$I_\psi\succ0$ denotes positive definiteness.

\begin{proposition}[Information geometry]
\label{prop:information_geometry}
Under Assumptions~\ref{Regularity_model}
and~\ref{ass:coordinate_paths},
\[
\overline{\mathcal T}
=\overline{\mathcal T_0}\oplus\widetilde S_\psi,
\qquad
\widetilde S_\psi
=\operatorname{span}\{\ell_\psi^{\rm eff}\},
\qquad
r_\psi=\operatorname{rank}(I_\psi)\le d_\psi.
\]
If Assumptions~\ref{ass:regclass} and~\ref{ass:denseS}
also hold, a relevant locally robust moment exists iff
$I_\psi\neq0$.
\end{proposition}

Thus relevance requires nonzero, possibly singular information.
Since $\widetilde S_\psi$ is finite-dimensional,
Assumption~\ref{ass:denseS} reduces to
$\mathbb E[\|\ell_\psi^{\rm eff}\|^{2q}]<\infty$
by Lemma~\ref{lem:density}(b).

To discuss regular estimation, $c'\psi$ must define an
observed-law functional. We therefore require
\emph{local observational invariance}:
$c'\psi(\lambda_1)=c'\psi(\lambda_2)$ whenever
$\mathbb P_{\lambda_1}=\mathbb P_{\lambda_2}$
in the local experiment, and write
$\vartheta_c(\mathbb P_\lambda)=c'\psi(\lambda)$.

\begin{definition}[Pathwise differentiability]
\label{def:pathwise_diff}
The observed-law functional $\vartheta_c$ is
\emph{pathwise differentiable} at $\mathbb P_0$ relative to
$\overline{\mathcal T}\subseteq L_2^0$ if, along every DQM
path through $\mathbb P_0$, its derivative depends only on
the score $s\in\mathcal T$ and the map
\[
\dot\vartheta_c:\mathcal T\to\mathbb R,
\qquad
\dot\vartheta_c(s)
:=\frac{d}{d\tau}\vartheta_c(\mathbb P_\tau)
\Big|_{\tau=0^+},
\]
is linear and continuous under the $L_2$ norm.
\end{definition}

The next result connects the classical range criterion
\citep{van1991differentiable} to relevance for individual
linear combinations when information may be singular.

\begin{theorem}[Relevance and pathwise differentiability]
\label{thm:relev_estim}
Let Assumptions~\ref{Regularity_model}, \ref{ass:regclass},
\ref{ass:denseS}, and~\ref{ass:coordinate_paths} hold, and let
$c'\psi$ satisfy local observational invariance for
$c\in\mathbb R^{d_\psi}\setminus\{0\}$. Suppose additionally
that Assumption~\ref{ass:coordinate_paths}(ii) holds with
$\mathfrak G$ replaced by the class of all DQM paths through
$\mathbb P_0$.
Then the following are equivalent:
\begin{enumerate}
\item[(i)]
a relevant locally robust moment exists for $c'\psi_0$;
\item[(ii)]
$c\in\operatorname{Range}(I_\psi)$;
\item[(iii)]
$\vartheta_c$ is pathwise differentiable at $\mathbb P_0$
relative to $\overline{\mathcal T}$.
\end{enumerate}
If these conditions hold, $\widetilde S_{c'\psi}$ is spanned
by $\tilde\psi_c=b'\ell_\psi^{\rm eff}$, where $c=I_\psi b$,
and $\tilde\psi_c$ is the efficient influence function
of $c'\psi_0$.
\end{theorem}

Relative to \citet{chen2026equivalence}, the theorem allows
singular information and characterizes relevance direction
by direction. Failure of (ii) rules out regular estimation
of $\vartheta_c$ on this local experiment
\citep[Theorem~2(ii)]{chamberlain1986asymptotic}; failure of observational
invariance means $c'\psi$ is not an observed-law functional.
The scalar informative space is at most one-dimensional,
so Lemma~\ref{lem:density}(b) applies to
Assumption~\ref{ass:denseS}.

\subsubsection{Strong Relevance and Rank Verification}

We now ask whether a moment vector detects every informative
target direction. For $g=(g_1, \cdots, g_m)^{\prime}$ and $h \in L^0_2,$ write $\left<g,h \right> = \left(\left<g_1,h \right>, \cdots, \left<g_m,h \right>\right)^{\prime}$ and define $L_g: \tilde{S}_\psi \mapsto \mathbb{R}^m$ by $L_g(s) := \left<g,s \right>. $

\begin{definition}[Strong relevance]
\label{def:strong_relevance}
A vector-valued locally robust moment $g$ is
\emph{strongly relevant} if $L_g$ is injective, equivalently,
\[
\langle g,s\rangle\neq0
\qquad
\text{for every nonzero }s\in\widetilde S_\psi.
\]
\end{definition}

By Proposition~\ref{prop:decomposition}, this is equivalent to
$\langle g,s\rangle\neq0$ for every
$s\in\overline{\mathcal T}\setminus\overline{\mathcal T_0}$.
If $r_\psi<\infty$, strong relevance necessarily requires
$m\ge r_\psi$. It means that no informative direction is missed;
it does not require efficiency.

The target Jacobian measures how many informative directions
a moment captures. Let
$M_g(\psi,\eta):=\mathbb E[g(Z,\Xi(\psi,\eta))]$ and define
\[
J_{\psi,g}
:=\left.
\frac{\partial M_g(\psi,\eta_0)}{\partial\psi'}
\right|_{\psi=\psi_0}.
\]

\begin{proposition}[Jacobian verification]
\label{prop:jacobian_verification}
Under Assumptions~\ref{Regularity_model}
and~\ref{ass:coordinate_paths}, let the vector
$g=(g_1,\ldots,g_m)'$ with $g_j\in\mathcal G_{\mathrm{LR}}$,
and suppose $J_{\psi,g}$ exists.
Then $J_{\psi,g}=-L_g\circ A$,
$\ker(I_\psi)\subseteq\ker(J_{\psi,g})$, and
$\operatorname{rank}(J_{\psi,g})\le r_\psi$.
Some $g_j$ is relevant iff $J_{\psi,g}\neq0$, while
$g$ is strongly relevant iff
$\operatorname{rank}(J_{\psi,g})=r_\psi$.
If $m\ge d_\psi$, then
$\operatorname{rank}(J_{\psi,g})=d_\psi$ iff
$r_\psi=d_\psi$ and $g$ is strongly relevant.
\end{proposition}

A rank-deficient Jacobian is ambiguous. If
$a\in\ker(J_{\psi,g})$, the proposed moment does not detect the
target perturbation $a$, but this may have two different explanations:
$g$ may miss an informative direction, or the direction may itself be
uninformative because its effect on the observed law can be offset by
a nuisance perturbation. To distinguish these cases, we complement the
Jacobian lower bound on $r_\psi$ with an upper bound obtained from such
nuisance-compensable target directions. Each independent
nuisance-compensable direction removes one possible informative
direction. If the two bounds coincide, every direction missed by the
moment is intrinsically uninformative, so the moment is strongly
relevant.
\begin{definition}[Nuisance-compensable directions]
\label{def:nuisance_compensable}
Under Assumption~\ref{ass:coordinate_paths}, a target direction
$a\in\mathbb R^{d_\psi}$ is \emph{nuisance-compensable} if
$a'\ell_\psi\in\overline{\mathcal T_0}$. Let
\[
\mathcal N_\psi
:=\{a\in\mathbb R^{d_\psi}:
a'\ell_\psi\in\overline{\mathcal T_0}\}.
\]
Equivalently, target-preserving scores approximate
$-a'\ell_\psi$ in $L_2$, offsetting the target perturbation
to first order.
\end{definition}

\begin{theorem}[Relevance-rank sandwich]
\label{thm:rank_sandwich}
Under Assumptions~\ref{Regularity_model}
and~\ref{ass:coordinate_paths}, let the vector
$g=(g_1,\ldots,g_m)'\in\mathcal G_{\mathrm{LR}}^m$
and suppose $J_{\psi,g}$ exists.
Then, for every linear subspace
$\mathcal C\subseteq\mathcal N_\psi$,
\[
\mathcal N_\psi=\ker(I_\psi),\qquad
\operatorname{rank}(J_{\psi,g})
\le r_\psi
=d_\psi-\dim(\mathcal N_\psi)
\le d_\psi-\dim(\mathcal C).
\]
If the outer bounds coincide, then
$\mathcal C=\mathcal N_\psi$,
$r_\psi=\operatorname{rank}(J_{\psi,g})$,
and $g$ is strongly relevant. Equivalently,
\[
\mathcal C\subseteq\mathcal N_\psi=\ker(I_\psi)
\subseteq\ker(J_{\psi,g}),
\]
and $g$ is strongly relevant iff
$\ker(J_{\psi,g})=\ker(I_\psi)$.
Thus, to certify strong relevance, it suffices to show that
every direction in $\ker(J_{\psi,g})$ is nuisance-compensable.
\end{theorem}

To certify strong relevance, it suffices to find, for each direction
$a$ in a basis of $\ker(J_{\psi,g})$, an element
$s_{\eta,a}\in\overline{\mathcal T_0}$ satisfying
$a'\ell_\psi+s_{\eta,a}=0$.
A path covered by
Assumption~\ref{ass:coordinate_paths}\textup{(ii)}
with target derivative $a$ and zero score supplies such
a certificate; observationally equivalent paths suffice.
If only a subspace $\mathcal C$ is certified, the number
of informative directions missed by $g$ is at most
$\dim\ker(J_{\psi,g})-\dim(\mathcal C)$.

\begin{remark}[Finite-matrix verification]
Consider conditional restrictions
\[
\mathbb E[\rho_j(Z,\psi,\eta)\mid W_j]=0,
\qquad j=1,\ldots,J,
\]
and choose nuisance directions $v_1,\ldots,v_K$. Suppose that,
under $\mathbb P_0$,
\[
\mathbb E\!\left[
\rho_j\!\left(Z,\psi_0+\tau a,
\eta_0+\tau\sum_{\ell=1}^K b_\ell v_\ell\right)
\mid W_j\right]
=
\tau\{D_j(W_j)a+B_j(W_j)b\},
\]
and assume the following observational equivalence condition:
if $D_j(W_j)a+B_j(W_j)b=0$ almost surely for every $j$,
then, for all sufficiently small $\tau$, the displayed
changes in $\psi$ and $\eta$ remain within the model and
generate exactly the same distribution of $Z$ as
$(\psi_0,\eta_0)$. Assume these paths satisfy
Assumption~\ref{ass:coordinate_paths}\textup{(ii)}.
Define
\[
G:=\sum_{j=1}^J\mathbb E[(D_j,B_j)'(D_j,B_j)],
\qquad
G_{\eta\eta}:=\sum_{j=1}^J\mathbb E[B_j'B_j].
\]
Then target directions generated by $\ker(G)$ are
nuisance-compensable, and Theorem~\ref{thm:rank_sandwich} gives
\[
\operatorname{rank}(J_{\psi,g})
\le r_\psi
\le
\operatorname{rank}(G)-\operatorname{rank}(G_{\eta\eta}).
\]
Thus equality of the outer bounds certifies strong relevance.
Under this fixed-law condition, the rank calculation requires only
conditional derivatives and finite matrices; the sample-selection
verification below is a special case.\end{remark}
For feasible inference, the oracle limit \eqref{oracle} must be complemented by
oracle equivalence of the cross-fitted moment. Partition
$\{1,\ldots,n\}$ into folds $I_1,\ldots,I_L$, let $\ell(i)$ satisfy
$i\in I_{\ell(i)}$, and let $\widehat g_{-\ell}$ estimate
$g(\cdot,\lambda_0)$ using observations in $I_\ell^c$. Setting
$\widehat g_i=\widehat g_{-\ell(i)}(Z_i)$, we require
\begin{equation}
	\label{eq:oracle_equivalence}
	n^{-1/2}\sum_{i=1}^{n}
	\{\widehat g_i-g(Z_i,\lambda_0)\}
	=o_{\mathbb P_0}(1).
\end{equation}
Only the features needed to construct the moment, not the full $\lambda_0$,
must be estimated. Unlike existence and relevance,
\eqref{eq:oracle_equivalence} requires first-step rates and remainder control, see \citet{chernozhukov2022locally};
cross-fitting controls empirical-process error, while orthogonality makes the
leading bias second order. Section~\ref{sectionExampleSelection} gives
sufficient conditions.
%%%%%%%%%%%%%%%%%%%%%%%%%%%%%%%%%%%%%%%%%%%%%%%%%%%%%%
\section[Sample Selection without Exclusion Restrictions]{Sample Selection without Exclusion\\Restrictions}
\label{sectionExampleSelection}

\subsection{Model, targets, and maintained assumptions}
\label{sec:ss_model}

Let $Z=(Y,D,X)\sim\mathbb P_0$, where $D\in\{0,1\}$
indicates selection, $Y=Y^*D$, and
$X\in\mathcal X\subset\mathbb R^{d_\theta}$. Consider
\[
Y^*=X'\theta_0+U,
\qquad
D=\mathbf 1\{h_0(X)-V\ge0\},
\]
where $\mathbf 1\{\cdot\}$ is the indicator function,
$\theta_0$ and $h_0$ are unknown, and the latent disturbances
$(U,V)$ may be dependent \citep{heckman1979sample}.
Without exclusions, we maintain
\begin{align}
\mathbb E[D-p_0(X)\mid X]&=0,
\label{eq:ss_cmr1}\\
\mathbb E[Y-p_0(X)X'\theta_0-q_0(p_0(X))\mid X]&=0.
\label{eq:ss_cmr2}
\end{align}
Here $p_0(x)=\mathbb P_0(D=1\mid X=x)$ and $q_0$ is
an unknown selection correction. Dependence of this correction
on $X$ only through $p_0(X)$ supplies the identifying restriction,
as under familiar index-sufficiency conditions
\citep{powell1987,ahn1993semiparametric,newey2009two}.

To parameterize the observed law, let
$\lambda=(\theta,\eta)$, with
$\theta\in\Theta\subset\mathbb R^{d_\theta}$ and
$\eta=(p,q,f_1,Q_X)$, where
$p:\mathcal X\to I^+$, $q:I^+\to\mathbb R$,
$Q_X$ is otherwise unrestricted and dominated, and
$f_1(\cdot\mid x)$ is a positive conditional density with
$\int u f_1(u\mid x)du=0$. Set $\mu_\lambda(x):=x'\theta+q(p(x))/p(x)$.
Under $\mathbb P_\lambda$, $X\sim Q_X$,
$D\mid X=x\sim\operatorname{Bernoulli}\{p(x)\}$,
$Y=0$ if $D=0$, and $Y\mid(D=1,X=x)$ has density
$f_1(y-\mu_\lambda(x)\mid x)$.
The target is $\theta$ or one of its components;
$\lambda$ need not be identified.

Write $\eta_0=(p_0,q_0,f_{1,0},P_X)$ and $T:=p_0(X)$.
Let $\sigma(X)$ and $\sigma(T)$ denote the generated
$\sigma$-fields, and $\dot a,\ddot a$ the first two derivatives
of a function $a$, componentwise for vector-valued functions.
Set
\begin{equation}
\varepsilon\equiv
\begin{pmatrix}
\varepsilon_1\\
\varepsilon_2
\end{pmatrix}
:=
\begin{pmatrix}
D-T\\
Y-TX'\theta_0-q_0(T)
\end{pmatrix},
\qquad
A_0(X):=X'\theta_0+\dot q_0(T),
\label{eq:A0_SS}
\end{equation}
where $q_0(t)=\mathbb E[Y-TX'\theta_0\mid T=t]$.

\begin{assumption}[Regularity]
\label{ass:ss_moments}
\begin{compactenum}[(i)]
\item
$\Theta$ is open, $\{Z_i\}_{i=1}^n$ is iid from
$\mathbb P_{\lambda_0}$, and
$\mathbb E[Y^4+\|X\|^4]<\infty$.

\item
For some $0<c\le C<\infty$,
$cI_2\le\Omega(X):=\mathbb E[\varepsilon\varepsilon'\mid X]
\le CI_2$ and
$\mathbb E[\|\varepsilon\|^3\mid X]\le C$ a.s.

\item
For some $0<c_p<\underline p<\bar p<C_p<1$,
$I^+=(c_p,C_p)$ and
$\operatorname{supp}(P_T)=I:=[\underline p,\bar p]$,
where $P_T$ is the law of $T$.
Uniformly over $\lambda\in\Lambda$, $q\in C^2(I^+)$,
$q$ and its first two derivatives are bounded on $I^+$,
and the conditional second moments of $f_1(\cdot\mid x)$
are bounded uniformly in $x$.
The corresponding bounds are strict at $\lambda_0$.

\item
The function $h_{0X}(t):=\mathbb E[X\mid T=t]$ is twice
continuously differentiable, with its first two derivatives
uniformly bounded on $I^+$. Moreover,
$f_{1,0}^{1/2}(\cdot\mid x)$ is absolutely continuous
for $P_X$-a.e.\ $x$ and
\[
\mathcal I_{1,0}(x)
:=4\int\{\partial_u f_{1,0}^{1/2}(u\mid x)\}^2du,
\qquad
\mathbb E[(1+\|X\|^2)\mathcal I_{1,0}(X)]<\infty.
\]
\end{compactenum}
\end{assumption}

The information condition and strict interior bounds permit
the regular paths constructed in Supplementary
Appendix~\ref{app:selection}.

\subsection{Existence}
\label{sec:existencesampleselection}

The unrestricted conditional density and covariate distribution
force orthogonal moments to use the conditional-moment residuals.
Orthogonality to $q_0$ requires residualizing instruments on $T$;
orthogonality to $p_0$ requires a first-step correction. Define
\[
\mathcal V_0
:=\{\varphi\in L_2(P_X):
\mathbb E[\varphi(X)\mid T]=0,\ A_0(X)\varphi(X)\in L_2\},
\]
\[
\mathcal V
:=\{\varphi\in\mathcal V_0:
Y\varphi(X)\in L_2,\ \|X\|\varphi(X)\in L_2\}.
\]
For $\varphi\in\mathcal V_0$, set
\begin{equation}
g_\varphi(Z,\lambda_0)
:=\varphi(X)\{\varepsilon_2-A_0(X)\varepsilon_1\}.
\label{eq:SS_moment_class}
\end{equation}
The correction $A_0(X)\varepsilon_1$ accounts for estimating
the generated regressor $p_0$
\citep{hahn2013asymptotic,escanciano2023automatic}.
Taking $\varphi(X)=T\{X-\mathbb E[X\mid T]\}$ gives the
score studied by \citet{pan2024locally} under full-vector
point identification; here it belongs to a complete moment class that we characterize below.

\begin{theorem}[Existence]
\label{thm:SS_existence}
Under Assumption~\ref{ass:ss_moments}, for the full-vector
target $\psi_0=\theta_0$,
\[
\overline{\mathcal T_0}^{\perp}
=\{g_\varphi:\varphi\in\mathcal V_0\}.
\]
Every nonzero $g_\varphi$ with $\varphi\in\mathcal V$
is the true evaluation of an admissible locally robust
structural moment. Such a moment exists if and only if
$\sigma(T)\subsetneq\sigma(X)$, equivalently, if and only if
$X$ is not almost surely a measurable function of $T$.
\end{theorem}

A convenient instrument is
$\varphi(X)=\zeta(X)-\mathbb E[\zeta(X)\mid T]$,
yielding an admissible nonzero moment whenever
$\varphi\in\mathcal V\setminus\{0\}$.
Bounded $\zeta$ ensures $\varphi\in\mathcal V$.
For $v=(v_1,\ldots,v_d)'$, let $v_{-j}$ delete component $j$.

\begin{corollary}[Orthogonality for a component]
\label{cor:SS_theta1}
Under Assumption~\ref{ass:ss_moments}, a moment $g_\varphi$
characterized in Theorem~\ref{thm:SS_existence} is also
orthogonal to the regular scores generated by perturbations
of $\theta_{0,-1}$ if and only if
\begin{equation}
\mathbb E[\varphi(X)TX_{-1}]=0.
\label{eq:SS_theta1_condition}
\end{equation}
Consequently, this class contains a nonzero admissible moment
orthogonal both to $\overline{\mathcal T_0}$ and to every
regular $\theta_{0,-1}$ direction if and only if
\eqref{eq:SS_theta1_condition} holds for some
$\varphi\in\mathcal V\setminus\{0\}$.
\end{corollary}

Condition~\eqref{eq:SS_theta1_condition} treats
$\theta_{0,-1}$ as additional regular nuisance.
If no nonzero $\varphi\in\mathcal V$ satisfies it,
the target may be augmented to include additional coefficients.

\begin{assumption}[Generating paths]
\label{ass:ss_local_rep}
For the full-vector target $\psi_0=\theta_0$, the restricted
DQM paths, together with the coefficient paths
$\lambda_\tau=(\theta_0+\tau a,\eta_0)$,
$a\in\mathbb R^{d_\theta}$, generate
$\overline{\mathcal T}$ in the sense of
Remark~\ref{rem:generating}.
\end{assumption}

\subsection{Relevance}
\label{sec:ss_relevance}

For the full-vector target, the Jacobian is
\begin{equation}
J_{\theta,\varphi}
:=\left.
\frac{\partial}{\partial\theta'}
\mathbb E[g_\varphi(Z,\theta,\eta_0)]
\right|_{\theta=\theta_0}
=-\mathbb E[\varphi(X)TX']
\equiv-\Sigma_\varphi,
\label{eq:ss_jacobian_compact}
\end{equation}
where $J_{\theta,\varphi}$ abbreviates $J_{\theta,g_\varphi}$.
Let $\tilde X:=X-\mathbb E[X\mid T]$ and
$\Sigma_{\tilde X}:=\mathbb E[T\tilde X\tilde X']$.
If $a\in\ker(\Sigma_{\tilde X})$, then $a'\tilde X=0$ a.s.
The path
\[
\theta_\tau=\theta_0+\tau a,
\qquad
q_\tau(t)=q_0(t)-\tau t\,a'h_{0X}(t),
\]
with other components fixed, remains in $\Lambda$ for small
$\tau$ by Assumption~\ref{ass:ss_moments}(iii)--(iv)
and leaves $\mu_\lambda(X)$ and the observed law unchanged.
Thus $\ker(\Sigma_{\tilde X})$ is nuisance-compensable.
Since $g_{\tilde X}$ has Jacobian $-\Sigma_{\tilde X}$,
the rank-sandwich bounds coincide. Thus every coefficient direction missed by the canonical
orthogonal moment can be offset by a change in the selection
nuisance that leaves the observed law unchanged. Its rank deficiency
therefore reflects information absent from the statistical experiment,
rather than information lost through the choice of orthogonal moment.

\begin{theorem}[Relevance rank]
\label{thm:SS_relevance_compact}
Let Assumptions~\ref{ass:ss_moments}
and~\ref{ass:ss_local_rep} hold and $\varphi\in\mathcal V^m$.
Then
\[
\mathcal N_{\theta_0}=\ker(\Sigma_{\tilde X}),\qquad
\operatorname{rank}(\Sigma_\varphi)
\le r_{\theta_0}=\operatorname{rank}(\Sigma_{\tilde X}),\qquad
\ker(\Sigma_{\tilde X})\subseteq\ker(\Sigma_\varphi).
\]
Hence $g_\varphi$ is relevant iff $\Sigma_\varphi\neq0$;
if $r_{\theta_0}>0$, it is strongly relevant iff
$\ker(\Sigma_\varphi)=\ker(\Sigma_{\tilde X})$.
In particular, if $r_{\theta_0}>0$, $g_{\tilde X}$ is
strongly relevant, and all directions are informative iff
$\Sigma_{\tilde X}$ is nonsingular.
\end{theorem}

For $\theta_{0,1}$, instruments satisfying
$\mathbb E[\varphi_j(X)TX_{-1}]=0$ for every $j$
are relevant iff $\mathbb E[\varphi(X)TX_1]\neq0$.
For the full vector, $T\ge\underline p>0$ gives
$\Sigma_{\tilde X}\neq0$ iff $\tilde X\not\equiv0$
iff $\sigma(T)\subsetneq\sigma(X)$:
existence and relevance coincide.

\begin{remark}[Single-index benchmark]
\label{rem:ss_single_index}
If $p_0(X)=G(X'\gamma_0)$ with $G$ strictly monotone, $\gamma_0 \neq 0$,
then $\gamma_0'\tilde X=0$ a.s., so
$\gamma_0\in\ker(\Sigma_{\tilde X})$.
If $X_1=1$, also $e_1\in\ker(\Sigma_{\tilde X})$,
recovering \citet{chamberlain1986asymptotic}'s
zero-information result for the intercept.
If $T$ is nonconstant these directions are independent,
so $r_{\theta_0}\le d_\theta-2$, with equality if they
are the only linear combinations of $X$ measurable
with respect to $T$.
\end{remark}

\subsection{Linear representation and identification}
\label{sec:ss_linear}

Let $h_{0Y}(t):=\mathbb E[Y\mid T=t]$ and
$\tilde Y:=Y-h_{0Y}(T)$.
The identity $h_{0Y}(t)=t h_{0X}(t)'\theta_0+q_0(t)$
makes $h_{0Y}$ twice continuously differentiable under
Assumption~\ref{ass:ss_moments}.
We define $B_0(X):=X-h_{0X}(T)-T\dot h_{0X}(T)$ and
differentiating gives
$A_0(X)=\dot h_{0Y}(T)+B_0(X)'\theta_0$.
Define
\begin{equation}
m_0(Z):=\tilde Y-\dot h_{0Y}(T)(D-T),
\qquad
r_0(Z):=T\tilde X+B_0(X)(D-T).
\label{eq:m0r0}
\end{equation}
Then \eqref{eq:SS_moment_class} has the linear representation
\begin{equation}
g_\varphi(Z,\theta,\eta_0)
:=\varphi(X)\{m_0(Z)-r_0(Z)'\theta\},
\qquad
\mathbb E[g_\varphi(Z,\theta_0,\eta_0)]=0.
\label{eq:ss_g_linear_theta}
\end{equation}
Thus \eqref{eq:ss_g_linear_theta} is an IV equation with generated
outcome $m_0$, generated regressors $r_0$, and residualized instrument
$\varphi(X)$, all constructed from regressions of $Y$ and $X$ on $T$
and their derivatives. Linearity will permit inference without first consistently
estimating the full coefficient vector.
Let $\Theta_{\mathrm{OM}}\subseteq\Theta$ be the set
satisfying all moments in \eqref{eq:ss_g_linear_theta}.

\begin{proposition}[Identification from orthogonal moments]
\label{prop:SS_identification}
Under Assumption~\ref{ass:ss_moments},
\[
\Theta_{\mathrm{OM}}
=\{\theta\in\Theta:
\mathbb E[g_{\tilde X}(Z,\theta,\eta_0)]=0\}
=(\theta_0+\ker(\Sigma_{\tilde X}))\cap\Theta.
\]
Thus $a'\theta_0$ is identified by these moments iff
$a\in\operatorname{Range}(\Sigma_{\tilde X})$.
\end{proposition}

The diagnostic is therefore a residual covariance matrix:
its kernel gives coefficient directions left unrestricted
by the orthogonal moments, its range gives identified
linear combinations, and its rank gives the relevance rank.
Small eigenvalues signal weak relevance and motivate the
inference procedures below.
Identification is relative to the orthogonal moments;
additional model restrictions may further shrink the
identified set.

\subsection{Anderson--Rubin--Neyman-Orthogonal inference}
\label{subsec:ss_inference}

For inference on $\theta_{0,1}$, define
\[
\Sigma_{-1,-1}
=\mathbb E[T\tilde X_{-1}\tilde X_{-1}'],
\qquad
b^*=\Sigma_{-1,-1}^{\dagger}
\mathbb E[T\tilde X_{-1}\tilde X_1],
\qquad
\varphi^{(1)}=\tilde X_1-b^{*\prime}\tilde X_{-1}.
\]
Thus, $\varphi^{(1)}$ is the population OLS residual from projecting $\tilde X_1$ on $\tilde X_{-1}$ among observations with $D=1$. If $a\in\ker(\Sigma_{-1,-1})$, positivity of $T$ implies
$a'\tilde X_{-1}=0$ a.s., hence
$a'\mathbb E[T\tilde X_{-1}\tilde X_1]=0$.
The Moore--Penrose coefficient therefore satisfies the
population normal equations, giving
\begin{equation}
\mathbb E[\varphi^{(1)}r_{0,-1}']=0,\qquad
\nu^2:=\mathbb E[\{\varphi^{(1)}\}^2],\qquad
\Sigma^{(1)}:=\mathbb E[T\{\varphi^{(1)}\}^2].
\label{eq:ss_population_profile_orthogonality}
\end{equation}
For $z=(1,-b^{*\prime})'$,
$\Sigma_{\tilde X}z=\Sigma^{(1)}e_1$.
Since $T\ge\underline p>0$, $\nu>0$ iff
$e_1\in\operatorname{Range}(\Sigma_{\tilde X})$:
positive relevance identifies $\theta_{0,1}$ even when
$\theta_{0,-1}$ remains unidentified.

Profiling coefficients need not converge to the particular
representative $\theta_{0,-1}$. We instead center at the
minimum-norm representative determined by the observed law:
\begin{equation}
\Gamma:=\mathbb E[\tilde Xm_0],\qquad
\bar\theta:=\Sigma_{\tilde X}^{\dagger}\Gamma,\qquad
\bar\Delta:=\bar\theta-\theta_0.
\label{eq:ss_canonical_representative}
\end{equation}
Assume $\bar\theta\in\Theta$.
Lemma~\ref{lem:ss_canonical_representative} proves
$\Gamma=\Sigma_{\tilde X}\theta_0$,
$\bar\theta=\Pi_{\operatorname{Range}(\Sigma_{\tilde X})}\theta_0$,
and $\tilde X'\bar\Delta=0$ a.s.
If $\nu>0$, then $\bar\theta_1=\theta_{0,1}$, and
$\bar q(t)=q_0(t)-t h_{0X}(t)'\bar\Delta$ makes
$(\bar\theta,\bar q)$ observationally equivalent to
$(\theta_0,q_0)$. Define the corrected oracle score and variance:
\begin{equation}
\begin{split}
g^*(Z)&=\varphi^{(1)}(X)
\{m_0(Z)-r_{0,1}(Z)\theta_{0,1}
-r_{0,-1}(Z)'\bar\theta_{-1}\},\\
\Omega^*&=\mathbb E[(g^*)^2].
\end{split}
\label{eq:ss_corrected_oracle}
\end{equation}
Then $\mathbb E[g^*]=0$, and under the moment conditions below,
\begin{equation}
c\nu^2\le\Omega^*\le C\nu^2,\qquad
\mathbb E[|g^*|^4]\le C\nu^4.
\label{eq:ss_corrected_variance_order}
\end{equation}
Thus $\hat\theta_{-1,-\ell}$ may converge to
$\bar\theta_{-1}\neq\theta_{0,-1}$; this nonvanishing difference is
included in the oracle score and variance.

\paragraph{Implementation.}
For each evaluation fold $I_\ell$, nested splitting partitions
$I_\ell^c$ into disjoint samples $H_\ell$ and $B_\ell$:
estimate nuisance functions on $H_\ell$ and calibrate
projection and minimum-norm coefficients on $B_\ell$.
Simple cross-fitting uses $I_\ell^c$ for both tasks.
Let $J_\ell=B_\ell$ or $J_\ell=I_\ell^c$, respectively.
For $i\in J_\ell\cup I_\ell$, form
\[
\begin{aligned}
\hat T_{i\ell}&=\hat p_\ell(X_i),\quad
\hat X_{i\ell}=X_i-\hat h_{X,\ell}(\hat T_{i\ell}),\quad
\hat Y_{i\ell}=Y_i-\hat h_{Y,\ell}(\hat T_{i\ell}),\quad
\hat B_{i\ell}=\hat X_{i\ell}
-\hat T_{i\ell}\widehat{\dot h}_{X,\ell}(\hat T_{i\ell}),\\
\hat r_{i\ell}&=\hat T_{i\ell}\hat X_{i\ell}
+\hat B_{i\ell}(D_i-\hat T_{i\ell}),\quad
\hat m_{i\ell}=\hat Y_{i\ell}
-\widehat{\dot h}_{Y,\ell}(\hat T_{i\ell})
(D_i-\hat T_{i\ell}).
\end{aligned}
\]
These are the generated regressors and outcome in the IV representation
\eqref{eq:ss_g_linear_theta}. To isolate $\theta_{0,1}$, we construct an
instrument orthogonal to the nuisance regressors $\hat r_{-1}$.
Define
\[
\hat M_\ell
:=|J_\ell|^{-1}\sum_{i\in J_\ell}
\hat X_{-1,i\ell}\hat r_{-1,i\ell}',
\qquad
\hat d_\ell
:=|J_\ell|^{-1}\sum_{i\in J_\ell}
\hat r_{-1,i\ell}\hat X_{1,i\ell},
\]
\[
\hat c_\ell:=(\hat M_\ell')_{\rho_n}^\dagger\hat d_\ell,
\qquad
\hat\varphi_{i\ell}^{(1)}
:=\hat X_{1,i\ell}-\hat c_\ell'\hat X_{-1,i\ell}.
\]
The calibration inverse retains only singular values above
the same threshold $\rho_n$ used below for profiling. Next define
\[
\hat\Sigma_{-\ell}
:=|J_\ell|^{-1}\sum_{i\in J_\ell}
\hat X_{i\ell}\hat r_{i\ell}',
\qquad
\hat\Gamma_{-\ell}
:=|J_\ell|^{-1}\sum_{i\in J_\ell}
\hat X_{i\ell}\hat m_{i\ell},
\qquad
\hat\theta_{-\ell}
:=\hat\Sigma_{-\ell,\rho_n}^{\dagger}\hat\Gamma_{-\ell}.
\]
Here $\hat\Sigma_{-\ell,\rho_n}^{\dagger}$ inverts only
singular values above the prespecified threshold $\rho_n$:
the raw sample matrix is generically full rank even when
its population counterpart is singular.
Let $\hat\theta_{-1,-\ell}$ exclude the first coordinate
of $\hat\theta_{-\ell}$. For $i\in I_\ell$, define the profiled generated outcome
\[
\hat y_{i\ell}^{(1)}
:=
\hat m_{i\ell}
-\hat r_{-1,i\ell}'\hat\theta_{-1,-\ell}.
\]
The fold-specific IV-type score is then
\[
\hat g_{i\ell}^{(1)}(\theta_1)
=
\hat\varphi_{i\ell}^{(1)}
\{\hat y_{i\ell}^{(1)}-\hat r_{1,i\ell}\theta_1\}.
\]
and
\[
\hat g_n^{(1)}(\theta_1)
=n^{-1}\sum_\ell\sum_{i\in I_\ell}
\hat g_{i\ell}^{(1)}(\theta_1),\qquad
\hat\Omega_n^{(1)}(\theta_1)
=n^{-1}\sum_\ell\sum_{i\in I_\ell}
\{\hat g_{i\ell}^{(1)}(\theta_1)
-\hat g_n^{(1)}(\theta_1)\}^2.
\]
The AR--NO statistic and confidence set are
\begin{equation}
\mathrm{AR}_n(\theta_1)
=n\frac{\{\hat g_n^{(1)}(\theta_1)\}^2}
{\hat\Omega_n^{(1)}(\theta_1)},\qquad
CR_{1-\zeta}^{\mathrm{AR}}
=\{\theta_1:\mathrm{AR}_n(\theta_1)
\le\chi^2_{1,1-\zeta}\},
\label{eq:ss_AR_stat_theta1}
\end{equation}
where $\chi^2_{1,1-\zeta}$ is the $(1-\zeta)$-quantile
of the chi-squared distribution with one degree of freedom.
The score is affine and its variance quadratic in $\theta_1$,
so inversion solves a quadratic inequality analytically,
including degenerate cases. The set may be bounded,
unbounded, empty, or the whole real line.

Target relevance relative to score noise is summarized by
the population concentration parameter
$\mathrm{CP}_n:=n(\Sigma^{(1)})^2/\Omega^*$.
Using $\Sigma^{(1)}=\mathbb E[T\{\varphi^{(1)}\}^2]$, define
\[
\hat\Sigma_n^{(1)}
:=n^{-1}\sum_\ell\sum_{i\in I_\ell}
\hat T_{i\ell}\{\hat\varphi_{i\ell}^{(1)}\}^2,
\qquad
\widehat{\mathrm{CP}}_n(\theta_1)
:=n\frac{\{\hat\Sigma_n^{(1)}\}^2}
{\hat\Omega_n^{(1)}(\theta_1)}.
\]
For descriptive reporting,
$\widehat{\mathrm{CP}}_{n,\mathrm{plug}}$ uses the same
variance formula after evaluating each fold-$I_\ell$ score
at $\theta_1=\hat\theta_{1,-\ell}$, the first coordinate
of the corresponding thresholded minimum-norm vector.

Let $\delta_n$ be a common $L_4$ rate for
$\hat\varphi^{(1)}-\varphi^{(1)}$, $\hat m-m_0$, and $\hat r-r_0$; see Assumption~\ref{ass:app_score_rates} in the Supplementary Appendix.
Let $q_n\to0$ bound estimation of
$(\Sigma_{\tilde X},\Gamma)$; let $\nu_n,\lambda_n>0$
lower-bound $\|\varphi^{(1)}\|_{L_2}$ and the smallest
nonzero singular value of $\Sigma_{\tilde X}$; and let
$\kappa_n\ge0$ control the normalized cross-moment between
$\hat\varphi^{(1)}$ and $\hat r_{-1}$.
The threshold is $\rho_n$ as above.

\begin{proposition}[Pointwise AR--NO inference]
\label{prop:ss_AR_pointwise}
Fix a distribution with $\nu>0$ and with its smallest
nonzero singular value of $\Sigma_{\tilde X}$ bounded
away from zero. Suppose the pointwise versions of
(U1)--(U4) stated below hold with
$\delta_n=o(n^{-1/4})$,
$q_n=\delta_n+n^{-1/2}$, and
$\kappa_n=n^{-1/2}$ under nested splitting or
$\kappa_n=\delta_n/\nu$ under simple cross-fitting.
If $q_n=o(\rho_n)$ and $\rho_n=o(1)$, then
\[
\mathrm{AR}_n(\theta_{0,1})\rightsquigarrow\chi_1^2,
\qquad
\mathbb P_0\{\theta_{0,1}\in CR_{1-\zeta}^{\mathrm{AR}}\}
\to1-\zeta.
\]
Under the local alternatives
$\mathbb P_{\lambda_{1/\sqrt n}}$ along a regular location DQM
path with score $\bar\delta'\ell_\theta$ and target derivative
$\bar\delta_1$, and along which the preceding first-step and
threshold conditions continue to hold,
\[
\mathrm{AR}_n(\theta_{0,1})\rightsquigarrow
\chi_1^2\!\left(
\frac{\bar\delta_1^2(\Sigma^{(1)})^2}{\Omega^*}
\right),
\]
a noncentral chi-squared distribution with one degree
of freedom and the displayed noncentrality parameter.
\end{proposition}

No identification of $\theta_{0,-1}$ or preliminary
consistent estimator of $\theta_{0,1}$ is required.
Here \emph{identification-robust} means validity without
identification or consistent estimation of nuisance coefficient
directions. It does not imply uniformity across zero
information or arbitrary rank transitions. The next result
allows shrinking positive relevance under first-step and
spectral-separation conditions.

\subsection{Uniform validity and the relevance boundary}
\label{sec:uniformity}

Index population quantities by $\mathbb P$ and let
$\mathcal F_\ell$ be the sigma-field generated by the training
data for fold $\ell$.
For a distribution class $\mathcal Q_n$, write
$A_n=O_p(a_n)$ \emph{uniformly over} $\mathcal Q_n$ if
\begin{equation}
\lim_{C\to\infty}\ \limsup_{n\to\infty}\
\sup_{\mathbb P\in\mathcal Q_n}
\mathbb P(|A_n|>Ca_n)=0,
\label{eq:ss_uniform_Op}
\end{equation}
with the same convention for vector and matrix norms.

Define $\mathcal P_n^+$ by population restrictions alone:
\begin{compactenum}[(P1)]
\item
Assumption~\ref{ass:ss_moments} holds with common constants,
$\|\theta_0(\mathbb P)\|\le C$,
$\nu_{\mathbb P}\ge\nu_n$, and
$\|g_{\mathbb P}^*\|_{\mathbb P,4}\le C\nu_{\mathbb P}$;
\item
every nonzero singular value of
$\Sigma_{\tilde X,\mathbb P}$ is at least $\lambda_n$,
and all remaining singular values are zero.
\end{compactenum}
Membership is therefore independent of the estimators.
Require (U2)--(U4) below uniformly over $\mathcal P_n^+$
and folds. Their pointwise versions in
Proposition~\ref{prop:ss_AR_pointwise} replace uniform
$O_p$ by ordinary $O_p$ at a fixed distribution.

\begin{compactenum}[(U1)]
\item
$\mathbb P\in\mathcal P_n^+$; that is, (P1) and (P2) hold.

\item
With
\[
\tilde g_{\ell,\mathbb P}
=\hat\varphi_\ell^{(1)}
\{\hat m_\ell-\hat r_{1,\ell}\theta_{0,1}(\mathbb P)
-\hat r_{-1,\ell}'\bar\theta_{-1,\mathbb P}\},
\]
\[
\|\tilde g_{\ell,\mathbb P}-g_{\mathbb P}^*\|_{\mathbb P,2}
=O_p(\delta_n),\qquad
\left|\mathbb E_{\mathbb P}
[\tilde g_{\ell,\mathbb P}-g_{\mathbb P}^*
\mid\mathcal F_\ell]\right|=O_p(\delta_n^2).
\]

\item
$\|\hat\Sigma_{-\ell}-\Sigma_{\tilde X,\mathbb P}\|
+\|\hat\Gamma_{-\ell}-\Gamma_{\mathbb P}\|=O_p(q_n)$,
and the threshold of Section~\ref{subsec:ss_inference}
satisfies $q_n\ll\rho_n\ll\lambda_n$.

\item
The fresh-draw projection bias and second moment satisfy
\[
\nu_{\mathbb P}^{-1}
\left\|\mathbb E_{\mathbb P}
[\hat\varphi_\ell^{(1)}\hat r_{-1,\ell}'
\mid\mathcal F_\ell]\right\|
=O_p(\kappa_n),\qquad
\|\hat\varphi_\ell^{(1)}\hat r_{-1,\ell}'\|_{\mathbb P,2}
=O_p(\nu_{\mathbb P}).
\]
\end{compactenum}

The separation $q_n\ll\lambda_n$ permits recovery of the
population range; it is not an identification requirement.
Rank may vary across distributions, but the theorem excludes
rank transitions where a nonzero singular value is of the
same or smaller order than matrix-estimation error.

Corollary~\ref{cor:ss_primitive_firststep} gives sufficient
conditions for (U2)--(U4) with sparse logistic lasso and cubic
B-splines on a stable-nuisance-block subclass.
Under its maintained generated-index, local spline-design,
score, and calibration conditions, four derivatives and spline
dimension $J_n\asymp n^{1/9}$ yield
$\delta_n=q_n=n^{-1/3+o(1)}$.
\begin{theorem}[Uniform AR--NO inference]
\label{thm:ss_AR}
Suppose (U1)--(U4) hold and
\begin{equation}
\frac{\delta_n}{\nu_n}\to0,\qquad
\frac{\sqrt n\,\delta_n^2}{\nu_n}\to0,\qquad
a_n:=\frac{q_n}{\lambda_n}\to0,\qquad
(\sqrt n\,\kappa_n+1)a_n\to0.
\label{eq:ss_uniform_rates_corrected}
\end{equation}
Then, uniformly over $\mathbb P\in\mathcal P_n^+$,
\[
\sqrt n\,\hat g_n^{(1)}\{\theta_{0,1}(\mathbb P)\}
=n^{-1/2}\sum_{i=1}^n g_{\mathbb P}^*(Z_i)
+o_p(\nu_{\mathbb P}),
\qquad
\frac{\hat\Omega_n^{(1)}\{\theta_{0,1}(\mathbb P)\}}
{\Omega_{\mathbb P}^*}\to_p1,
\]
and
\[
\sup_{\mathbb P\in\mathcal P_n^+}
\left|\mathbb P\{\theta_{0,1}(\mathbb P)
\in CR_{1-\zeta}^{\mathrm{AR}}\}-(1-\zeta)\right|\to0.
\]
\end{theorem}

The theorem permits rank deficiency and shrinking positive
relevance without requiring
$\hat\theta_{-1,-\ell}\to\theta_{0,-1}$.
It establishes uniform coverage at the true target,
not uniform approximations to confidence-set length or
boundedness. The final rate condition controls estimation
of $\bar\theta_{-1}$. Under nested splitting it reduces to
$q_n/\lambda_n\to0$; under simple cross-fitting it becomes
\begin{equation}
\left(\frac{\sqrt n\,\delta_n}{\nu_n}+1\right)
\frac{q_n}{\lambda_n}\to0.
\label{eq:ss_simple_crossfit_rate}
\end{equation}

The rates $\nu_n^2$ and $\lambda_n$ measure different
features: target-specific residual variation and separation
of all nonzero eigenvalues, respectively. They need not
have the same order, since another coefficient direction
may approach singularity faster. Indeed, when $\nu>0$,
$
\Sigma^{(1)}
=\{e_1'\Sigma_{\tilde X}^{\dagger}e_1\}^{-1}
\ge\lambda_{\min}^+(\Sigma_{\tilde X}),
$
where $\lambda_{\min}^+$ is the smallest positive eigenvalue.
The separate spectral condition accounts for estimating
the minimum-norm representative.

At exact irrelevance, $\nu_{\mathbb P}=0$, the oracle score
and variance vanish, so studentization degenerates.
The zero-information bound of
\citet{lee2024locallyregularefficienttests} rules out
nontrivial parametric-rate local power for correctly sized
tests when effective information is zero.
Inference valid across this boundary must allow uninformative
confidence sets there. Theorem~\ref{thm:ss_AR} covers only
positive relevance satisfying
\eqref{eq:ss_uniform_rates_corrected}; distinguishing it
from exact irrelevance after nuisance estimation requires
additional rate information and is beyond our scope.

\begin{remark}[Primitive rate frontier]
\label{rem:rate_configurations}
Under Corollary~\ref{cor:ss_primitive_firststep},
$\delta_n=q_n=n^{-1/3+o(1)}$.
For $\nu_n=n^{-v}$ and $\lambda_n=n^{-\xi}$, nested splitting
requires $v<1/6$ and $\xi<1/3$, whereas simple cross-fitting
requires $v+\xi<1/6$.
Along identified sequences with sharp population rates
$\nu_{\mathbb P_n}\asymp n^{-v}$ and
$\lambda_{\min}^+(\Sigma_{\tilde X,\mathbb P_n})\asymp n^{-\xi}$,
the population geometry also requires $2v\le\xi$.
Thus the corresponding concentration frontiers are
$\mathrm{CP}_n\gg n^{2/3}$ and
$\mathrm{CP}_n\gg n^{8/9}$, respectively, up to logarithmic factors.

For example, $\nu_n=n^{-1/25}$,
$\lambda_n=n^{-1/10}$, and $\rho_n=n^{-1/4}$ satisfy
$q_n\ll\rho_n\ll\lambda_n$ and the simple-cross-fitting
restriction $1/25+1/10<1/6$.
If $\nu_{\mathbb P_n}\asymp n^{-1/25}$, then
$\mathrm{CP}_n\asymp n^{23/25}$.
Strict polynomial slack absorbs logarithmic factors.
\end{remark}
%%%%%%%%%%%%%%%%%%%%%%%%%%%%%%%%%%%%%%%%%%%%%%
\subsection{Monte Carlo and Empirical Application}
\label{sec:MCApp}

\subsubsection{Monte Carlo}
\label{sec:montecarlo}

Three experiments examine recovery of informative directions
and inference under full rank, rank deficiency, and unresolved
rank transitions. Theorem~\ref{thm:SS_relevance_compact} and
Proposition~\ref{prop:SS_identification} link these configurations
to the rank, range, and kernel of $\Sigma_{\tilde X}$.
We report AR--NO results; Supplementary
Appendix~\ref{app:mc_additional} compares ordinary-inverse DML.

\paragraph{Designs and implementation.}
We use $n\in\{5{,}000,10{,}000\}$, 1{,}000 replications
per configuration, and five-fold simple cross-fitting.
Within each training sample, logit lasso estimates the
propensity from a polynomial dictionary and cross-fits it
before use as a generated regressor. Lasso estimates
$h_{0X}$ and $h_{0Y}$ from cubic B-splines in that propensity.
Penalties are selected by five-fold cross-validation within
the relevant training sample. AR--NO uses $\rho_n=n^{-1/4}$
and exact analytic inversion.

All designs start from
$X_1,X_2\stackrel{\mathrm{iid}}{\sim}N(0,1)$,
independent of standard bivariate normal $(U,V)$
with correlation $0.75$, and
\[
Y^*=0.5+0.5X_1+0.25X_2+2U,\qquad
h_c(X)=1.5+0.5X_1+cX_1^2+X_2.
\]
Except in Design~A, $D=\mathbf 1\{V\le h_c(X)\}$.
The intercept is absorbed into $q_0$; Designs~A and C
target the coefficient $\theta_{0,1}=0.5$ on $X_1$.

\emph{Design A (triangular sequence)} replaces $h_c$ by
$\Phi^{-1}[\Lambda\{h_{c_n}(X)\}]$, where $\Phi$ and
$\Lambda$ are the standard normal and logistic distribution
functions, $c_n=0.60n^{-1/20}$, and the spline dimension is
$J_n=\lceil4n^{1/9}\rceil$. The population relevance rank
is two, with $\nu_n\asymp n^{-1/20}$ and
$\lambda_n\asymp n^{-1/10}$.
If $\delta_n=q_n=n^{-1/3+o(1)}$, these exponents satisfy
the numerical rate restrictions of
Theorem~\ref{thm:ss_AR} with $\rho_n=n^{-1/4}$.

\emph{Design B (rank one)} sets $c=0$.
Selection depends only on $0.5X_1+X_2$, so the orthogonal
moments identify $\theta_{0,1}-0.5\theta_{0,2}$,
but not $\theta_{0,1}$ separately. Rotate the covariates to
\[
W_1=0.8(X_1-0.5X_2),\qquad
W_2=0.8(0.5X_1+X_2).
\]
Then $Y^*=0.5+\beta_1W_1+\beta_2W_2+2U$, with
$(\beta_1,\beta_2)=(0.375,0.5)$.
For $\widetilde W:=W-\mathbb E[W\mid T]$,
\[
\Sigma_{\widetilde W}
:=\mathbb E[T\widetilde W\widetilde W']
=\operatorname{diag}(0.6731,0).
\]
Thus $\beta_1$ is identified from the orthogonal moments,
while the $\beta_2$ direction lies in their kernel.
The unrotated specification serves as an irrelevance benchmark.

\emph{Design C (unresolved rank transition)} uses the original
coordinates and chooses $c>0$ separately for each $n$ to set
the population concentration parameter
$\mathrm{CP}_n\in\{10,50,200,500\}$.
Population rank is two, but each fixed-concentration sequence
has $\nu_n\asymp n^{-1/2}$ and $\lambda_n\asymp n^{-1}$.
Even with $q_n\asymp n^{-1/2}$, no threshold satisfies
$q_n\ll\rho_n\ll\lambda_n$.
Thus (U3) fails and this design serves as a stress test for our results.

\begin{figure}[t]
\centering
\includegraphics[width=.88\linewidth]
{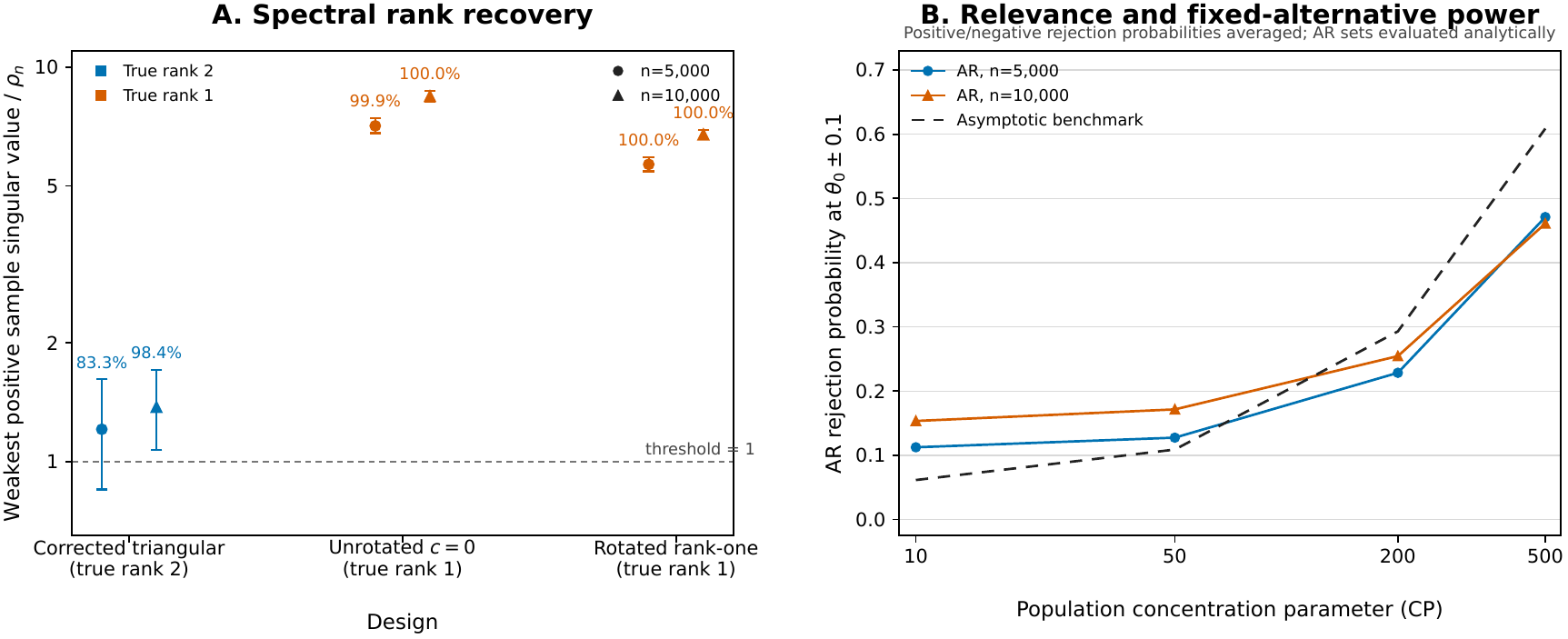}
\caption{Rank recovery and power. Panel~A reports the weakest
positive sample singular value relative to $\rho_n$,
5th--95th percentile bars, and correct-rank frequencies;
the horizontal line is the threshold. Panel~B reports average
AR--NO rejection in Design~C at $\theta_{0,1}\pm0.1$;
the dashed noncentral $\chi_1^2$ curve is an asymptotic benchmark.}
\label{fig:mc_relevance}
\end{figure}

\paragraph{Results.}
Panel~A of Figure~\ref{fig:mc_relevance} shows correct-rank
recovery increasing from $83.3\%$ to $98.4\%$ in Design~A
and reaching $99.9$--$100\%$ in both versions of Design~B. In the rotated design, the procedure recovered the unidentified $\beta_2$ direction almost exactly.\footnote{The median absolute cosine between the estimated null direction and its population counterpart, $e_2=(0,1)^{\prime}$, was 0.9996 for $n=5{,}000$ and 0.9997 for $n=10{,}000$, where one indicates exact agreement.}
In Design~C, at least $79.1\%$
of replications retain rank one despite population rank two.
Panel~B shows rejection increasing with concentration,
consistent with greater target sensitivity relative to noise.

Table~\ref{tab:mc_identified} reports coverage of
$94.4$--$97.4\%$. Lengths contract in Design~A;
Design~B permits inference on $\beta_1$ despite full-vector
nonidentification; and weaker relevance in Design~C produces
longer, more frequently unbounded sets.
Coverage in Design~C remains close to nominal despite failure
of (U3). This finite-sample finding does not establish uniform
validity without (U3), which would require controlling profiling
across unresolved rank transitions.
The unrotated Design~B is omitted because $\theta_{0,1}$
has no relevant orthogonal moment; inclusion of a reference
value would not measure coverage.

\begin{table}[t]
\centering
\footnotesize
\setlength{\tabcolsep}{3pt}
\renewcommand{\arraystretch}{0.95}
\caption{AR--NO inference}
\label{tab:mc_identified}
\begin{tabular*}{\linewidth}
{@{\extracolsep{\fill}}lrrrrrr@{}}
\toprule
& \multicolumn{3}{c}{$n=5{,}000$}
& \multicolumn{3}{c}{$n=10{,}000$} \\
\cmidrule(lr){2-4}\cmidrule(lr){5-7}
Design
& Cov. & $L_{50}$ & Unbd.
& Cov. & $L_{50}$ & Unbd. \\
\midrule
A: Triangular
& 95.1 & .19 & 0.1
& 94.4 & .13 & 0.0 \\
B: Rank one
& 96.7 & .20 & 1.1
& 96.3 & .14 & 0.6 \\
C: $\mathrm{CP}=10$--$500$
& 95.4--97.2 & $.61\to.20$ & $10.6\to0.8$
& 94.9--97.4 & $.45\to.20$ & $7.2\to1.0$ \\
\bottomrule
\end{tabular*}

\parbox{.98\linewidth}{\scriptsize
\textit{Notes:} Cov.\ and Unbd.\ denote coverage and
unboundedness percentages. $L_{50}$ is median 95\%
confidence-set length conditional on boundedness.
Design~B reports $\beta_1$ in the rotated specification.
For Design~C, $\mathrm{CP}$ denotes $\mathrm{CP}_n$;
coverage is the range across its four values, and arrows
give results at $\mathrm{CP}=10$ and $500$.
The Monte Carlo standard error at 95\% coverage is
0.69 percentage points.}
\end{table}

\subsubsection{Empirical application}
\label{sec:empirical_application}

We revisit \citet{honore2020selection}'s application to
third-generation Mexican American wage differentials without
covariate exclusions. We do not impose their linear-index restriction on participation;
Proposition~\ref{prop:SS_identification} characterizes which
coefficient combinations our orthogonal moments identify.
We use their 2003--2016 Current Population Survey (CPS)
sample of 118{,}250 men and 127{,}738 women aged 25--62
in four Southwestern states, with the same covariates
entering participation and wages.

The wage equation includes the Mexican American indicator,
quadratic terms in age and potential experience, and education,
veteran, marital-status, state, and year indicators:
27 regressors, with the constant absorbed by $q_0$. We use five-fold outer cross-fitting. Within each training sample,
logit lasso estimates the propensity from a 43-term dictionary
of discrete regressors, cubic B-splines in age and experience,
and their interactions with the Mexican American indicator.
The propensity is itself cross-fitted within the training sample
to obtain out-of-sample $\hat T$ values, on which Gaussian lasso
regresses the outcome and each regressor using ten-degree-of-freedom
cubic B-splines. All penalties minimize five-fold cross-validation error.
Before thresholding, residualized regressors are scaled to unit
$\hat T$-weighted second moments. The full-system rank diagnostic
and leave-one-fold-out AR--NO profiling inverses use
$\rho_n=n^{-1/4}$. The minimum-norm representative is selected
in these standardized coordinates; results are reported in
original units. DML uses the ordinary inverse and conventional sandwich
variance. AR--NO analytically inverts
\eqref{eq:ss_AR_stat_theta1}.

In both full samples
and every training sample, $24$ of $27$ normalized Jacobian
singular values exceed the cutoff. Excluding three directions
from inversion does not imply target irrelevance.

\begin{table}[t]
\centering
\footnotesize
\setlength{\tabcolsep}{2pt}
\caption{Third-generation Mexican American wage coefficient}
\label{tab:wage_mexican}
\begin{threeparttable}
\begin{tabular*}{\linewidth}
{@{\extracolsep{\fill}}lccccc@{}}
\toprule
& \shortstack{Honor\'e--Hu\\identified set}
& \shortstack{DML\\(s.e.)}
& 95\% Wald CI
& 95\% AR--NO CI
& $\widehat{\mathrm{CP}}_{n,\mathrm{plug}}$ \\
\midrule
Men
& $[-0.109,-0.097]$
& \shortstack{$-0.042$\\$(0.020)$}
& $[-0.081,\,-0.002]$
& $[-0.104,\,-0.050]$
& 6{,}169 \\
Women
& $[-0.086,-0.080]$
& \shortstack{$-0.068$\\$(0.008)$}
& $[-0.083,\,-0.052]$
& $[-0.083,\,-0.053]$
& 16{,}864 \\
\bottomrule
\end{tabular*}
\begin{tablenotes}[flushleft]
\footnotesize
\item \textit{Notes:} The base category is non-Hispanic White.
CI denotes confidence interval and s.e.\ standard error.
The first column gives \citet{honore2020selection}'s
estimated identified sets, not confidence intervals.
DML uses the ordinary inverse and full-sample-score sandwich;
AR--NO uses $\rho_n=n^{-1/4}$.
The concentration diagnostic
$\widehat{\mathrm{CP}}_{n,\mathrm{plug}}$ evaluates each fold
score at its leave-one-fold-out thresholded minimum-norm
coefficient vector, rather than the DML estimate.
\end{tablenotes}
\end{threeparttable}
\end{table}

Both procedures exclude zero for men and women
(Table~\ref{tab:wage_mexican}).
For women, their intervals nearly coincide and overlap
the upper part of the Honor\'e--Hu set.
For men, the intervals overlap each other, but only AR--NO
overlaps that set. These comparisons are descriptive because
the Honor\'e--Hu entries are estimated identified sets.

Leave-one-fold-out DML estimates range from $-0.075$ to
$-0.063$ for women and $-0.078$ to $-0.056$ for men;
only the women's range contains the full-sample estimate.
Evaluating variance at fold-specific coefficients changes
standard errors by less than $0.2\%$ and by $10.4\%$,
respectively, suggesting greater numerical stability for women.

The concentration diagnostic, $n$ times squared target
sensitivity divided by score variance, is larger for women,
consistent with their narrower AR--NO interval.
It measures information relative to noise on a common
target scale, with no universal cutoff.
Multiplying $\rho_n$ by values from $0.5$ to $2$ changes
retained rank from $25$ to $22$ for men and from $25$
to $23$ for women, while every AR--NO interval remains
bounded and below zero.
The male interval changes from $[-0.094,-0.050]$ to
$[-0.109,-0.053]$ at these multipliers; the female interval
is $[-0.083,-0.053]$ at both endpoints.
Weak directions in the full system therefore need not
prevent informative inference on a particular coefficient,
although magnitudes can be more method-sensitive than signs.%%%%%%%%%%%%%%%%%%%%%%%%%%%%%%%%%%%%%%%%%%%%%%%%%%%%%
	
\section{Latent-Variable Models: An RKHS Characterization}
\label{TVAcexample}

In mixture models, orthogonal moments may exist without
informing the target. We characterize the representer range
exactly as a reproducing kernel Hilbert space (RKHS), linking
informativeness to the measurement process. Its kernel is the
$\chi^2$-affinity between latent types: diagonal values measure
distinguishability from the population and off-diagonal values
their affinity. The analysis permits general mixing kernels;
Gaussian repeated measurements provide the leading example.

\subsection{Setup}
\label{subsec:tva_setup}

Let the latent type $\alpha$ take values in a compact
$\mathcal A\subset\mathbb R$, let $\mathcal M$ be the
Borel probability measures on $\mathcal A$, and let
$\Theta\subseteq\mathbb R^{d_\theta}$ be open.
The model
$\mathcal P=\{\mathbb P_{\theta,\eta}:
\theta\in\Theta,\ \eta\in\mathcal M\}$
has parameter $\lambda=(\theta,\eta)$ and density
\begin{equation}
p_{\theta,\eta}(z)
=\int p(z\mid\alpha,\theta)\,d\eta(\alpha)
\label{eq:mixture}
\end{equation}
relative to a $\sigma$-finite measure $\mu$, where $p(z\mid\alpha,\theta_0)$ is known up to $\theta_0$.
Write $p_0=p_{\theta_0,\eta_0}$.
We study scalar targets linear in the latent distribution:
\[
\psi(\lambda)=\int r(\alpha,\theta)\,d\eta(\alpha),
\qquad
\psi_0=\psi(\lambda_0),
\qquad
r_0:=r(\cdot,\theta_0)-\psi_0.
\]
The necessary range condition also extends to nonlinear
functionals $\Psi(\theta,\eta)$ under continuous
differentiability on finite-dimensional submodels formed
by $\theta$-perturbations and bounded latent-law tilts
from an $L_2(\eta_0)$-dense linear class.
For a nonzero centered derivative representer
$r_{\Psi,0}\in L_2^0(\eta_0)$, write
\begin{equation}
\left.\frac{d}{d\tau}\Psi(\theta_0,\eta_\tau)\right|_{\tau=0}
=\int r_{\Psi,0}(\alpha)b(\alpha)\,d\eta_0(\alpha),
\qquad
d\eta_\tau=(1+\tau b)d\eta_0.
\label{eq:nonlinear_latent_representer}
\end{equation}
Under the model conditions of
Lemma~\ref{lem:nonlinear_range}, relevance requires
$r_{\Psi,0}\in\operatorname{Range}(S^*)$ in $L_2(\eta_0),$ where $S^{*}$ is the operator defined below.
The lemma states the smooth-tilt condition precisely
and verifies it for the quantile and endogenous-threshold
policy examples below.

Define the feature map
$k_\alpha:=p(\cdot\mid\alpha,\theta_0)/p_0$,
the score operator $S:L_2^0(\eta_0)\to L_2^0$,
and its adjoint $S^*:L_2^0\to L_2^0(\eta_0)$ by
\[
\begin{aligned}
s_b:=Sb(z)
&=\mathbb E[b(\alpha)\mid Z=z]
=\int b(\alpha)k_\alpha(z)\,d\eta_0(\alpha),\\
S^*g(\alpha)
&=\mathbb E[g(Z)\mid\alpha]
=\langle g,k_\alpha\rangle_{L_2}.
\end{aligned}
\]
The observed-data score for $\theta_0$ is
$s_\theta(z)
:=\mathbb E[\partial_\theta\log p(Z\mid\alpha,\theta_0)
\mid Z=z]$.
Write
$\mathcal H_\eta:=\overline{\operatorname{Range}(S)}$,
$\mathcal B:=L_\infty(\eta_0)\cap L_2^0(\eta_0)$, and
$\bar r_\theta
:=\int\partial_\theta r(\alpha,\theta_0)\,d\eta_0(\alpha)$.

\begin{example}[Gaussian repeated measurements]
\label{ex:tva_gaussian}
Consider
\begin{equation}
Y_j=\alpha+\varepsilon_j,\qquad
\varepsilon_j=\theta_0u_j,\qquad
j=1,\ldots,J,\qquad
Z\equiv Y:=(Y_1,\ldots,Y_J),
\label{eq:tva_model}
\end{equation}
where the $u_j$ are independent standard normal variables,
independent of $\alpha$, and $\theta_0>0$ is the noise scale.
Then
\[
p(y\mid\alpha,\theta)
=\prod_{j=1}^J\phi_\theta(y_j-\alpha),
\qquad
\phi_\theta(v)
:=\frac{1}{\sqrt{2\pi}\theta}
\exp\left(-\frac{v^2}{2\theta^2}\right),
\]
and the observed-data scale score is
\[
s_\theta(y)
=\mathbb E\left[
\left.\frac{\sum_{j=1}^J(u_j^2-1)}{\theta_0}
\,\right|\,Y=y\right].
\]
This stylizes repeated noisy measurements of a latent
teacher effect \citep{gilraine2020new}.
For the general verification below we take $J=2$; the
measurement-resolution result allows arbitrary $J\ge2$.
\end{example}

The following conditions control arbitrary DQM paths.
Write
$H(\mathbb P,\mathbb Q)
=\{ \frac12\int(\sqrt p-\sqrt q)^2\,d\mu\}^{1/2}$
for the Hellinger distance between distributions with
densities $p$ and $q$.

\begin{assumption}[Regular mixture model]
\label{ass:mixture}
\begin{compactenum}[(i)]
\item
$\mathcal A\subset\mathbb R$ is compact, $\eta_0$ has support
$\mathcal A$, $p(z\mid\alpha,\theta)>0$, and
$(\alpha,\theta)\mapsto p(z\mid\alpha,\theta)$ is continuous
and continuously differentiable in $\theta$ for $\mu$-a.e.\ $z$.
Moreover, $\theta\mapsto\mathbb P_{\theta,\eta_0}$ is DQM
at $\theta_0$ with score $s_\theta$.

\item
$\alpha\mapsto k_\alpha$ is continuous from $\mathcal A$ into $L_2$.

\item
For every $p<\infty$, there exist a neighborhood $U$ of
$\theta_0$ and a nonnegative $M_p\in L_p$ such that
\[
\max\left\{
p_{\theta,\eta}(z),
\sup_{\alpha\in\mathcal A}
\|\partial_\theta p(z\mid\alpha,\theta)\|
\right\}
\le M_p(z)p_0(z)
\]
$\mu$-a.e., for every $\eta\in\mathcal M$ and $\theta\in U$.

\item
$\mathbb P_{\theta,\eta}=\mathbb P_0$ implies
$(\theta,\eta)=\lambda_0$. For some $C<\infty$,
$\|\theta-\theta_0\|\le C H(\mathbb P_{\theta,\eta},\mathbb P_0)$
for every $\eta\in\mathcal M$ and $\theta$ near $\theta_0$.
Moreover, for every sequence $(\theta_n,\eta_n)$ in the model,
$H(\mathbb P_{\theta_n,\eta_n},\mathbb P_0)\to0$
implies $\theta_n\to\theta_0$.

\item
$r_0\in L_2^0(\eta_0)\setminus\{0\}$.
The map $\theta\mapsto r(\alpha,\theta)$ is continuously differentiable at $\theta_0$ uniformly in $\alpha\in\mathcal A$.
Moreover, $\alpha\mapsto\partial_\theta r(\alpha,\theta_0)$
is bounded and is continuous at $\eta_0$-almost every
point of $\mathcal A$.

\item
For some $q>1$, $\ker(S^*)\cap L_{2q}$ is dense in
$\ker(S^*)$ in $L_2$.
\end{compactenum}
\end{assumption}

Full identification of $\lambda_0$ in part~\textup{(iv)} is stronger than necessary, but relaxing it while retaining the tangent characterization would require additional analysis; we maintain it for simplicity. Lemma~\ref{lem:tva_gauss} verifies parts
\textup{(i)--(iv) and (vi)} in Example~\ref{ex:tva_gaussian}
when $\eta_0$ has a density bounded away from zero on the
compact interval $\mathcal A$. Part~\textup{(v)} is a separate
condition on the target.
\begin{proposition}[Available tangent directions]
\label{prop:tva_tangent}
Let Assumption~\ref{ass:mixture}\textup{(i)--(iv) and (vi)}
hold and define
\[
V:=\overline{\operatorname{span}}
\{a's_\theta+s_b:
a\in\mathbb R^{d_\theta},\ b\in\mathcal B\}.
\]
For every $t\in\mathbb R^{d_\theta}$ and $b\in\mathcal B$,
the paths $\theta_\tau=\theta_0+\tau t$ and
$\eta_\tau=(1+\tau b)\eta_0$ belong to the model for
all sufficiently small $\tau$ and are DQM with score
$t's_\theta+s_b$.
After normalization, these paths form a locally dominated
class with an $L_p$ envelope for every $p<\infty$.
Their scores span $V$, and $V=\overline{\mathcal T}$.
\end{proposition}

\subsection{Existence}
\label{subsec:tva_existence}

Let $Z=(Y_1,\ldots,Y_J)$, $J\ge2$, with conditionally
iid measurements given $\alpha$ under every $(\alpha,\theta)$.
For bounded measurable $a$, set
$g_a(Y):=a(Y_1,Y_2)-a(Y_2,Y_1)$.

\begin{proposition}[Local overidentification]
\label{prop:tva_overid}
For every bounded measurable $a$,
$g_a\in\overline{\mathcal T}^{\perp}$.
Hence, if $\mathbb P_0(Y_1\neq Y_2)>0$, the model is
locally overidentified and, under
Assumptions~\ref{Regularity_model}--\ref{ass:dense0},
locally robust moments exist for every target
$\psi(\lambda_0)$.
\end{proposition}

Conditional exchangeability gives
$\mathbb E_{\theta,\eta}[g_a]=0$ throughout the model;
differentiation yields orthogonality to every DQM score.
Existence therefore requires neither Gaussianity nor
a tangent-space calculation. Relevance remains to be established.

\subsection{Relevance: an Exact Characterization}
\label{sec:rkhs}

\begin{proposition}[Range condition for relevance]
\label{prop:tva_key_equation}
Let Assumptions~\ref{Regularity_model}
and~\ref{ass:mixture} hold.
\begin{compactenum}[(i)]
\item
If a relevant locally robust moment exists for $\psi_0$,
or if $\psi_0$ is pathwise differentiable along every
DQM path in the mixture model, then there is
$g_0\in L_2^0$ with
\begin{equation}
\mathbb E[g_0(Z)\mid\alpha]
=r(\alpha,\theta_0)-\psi_0,
\qquad\text{for every }\alpha\in\mathcal A,
\label{eq:tva_key_equation}
\end{equation}
that is, $r_0\in\operatorname{Range}(S^*)$.
\item
Conversely, if \eqref{eq:tva_key_equation} has a solution
$g_0\in L_2^0\cap L_{2q}$ for some $q>1$ and
$\Pi_{\mathcal H_\eta}s_\theta\in L_{2q}$, then
\[
g^\star:=g_0+\kappa'\Omega^{-1}v_\theta,
\]
where
$v_\theta:=s_\theta-\Pi_{\mathcal H_\eta}s_\theta$,
$\Omega:=\mathbb E[v_\theta v_\theta']$, and
$\kappa:=\bar r_\theta-\langle g_0,s_\theta\rangle$,
is a relevant locally robust moment and an influence function
for $\psi_0$ along every DQM path in the mixture model.
\end{compactenum}
\end{proposition}

The explicit paths of Proposition~\ref{prop:tva_tangent}
give the necessary range equation; its tangent-space equality
rules out relevance through additional DQM directions.
Assumption~\ref{ass:mixture}(iv) ensures $\Omega\succ0$,
making the correction for $\theta$ well defined.
The remaining range restriction has an exact RKHS representation.

Assume
\begin{equation}
\|k_\alpha\|_{L_2}^2
=1+\chi^2(\mathbb P_\alpha\|\mathbb P_0)<\infty
\qquad\text{for }\eta_0\text{-a.e. }\alpha,
\label{eq:chi2}
\end{equation}
where $\mathbb P_\alpha$ is the conditional law of $Z$
given $\alpha$ at $\theta_0$, and
$\chi^2(\mathbb P\|\mathbb Q)
:=\int(d\mathbb P/d\mathbb Q-1)^2\,d\mathbb Q$.
Condition~\eqref{eq:chi2} holds in \eqref{eq:tva_model}
when $\eta_0$ has bounded support.

\begin{theorem}[The informative target space is an RKHS]
\label{thm:rkhs_range}
Let Assumption~\ref{ass:mixture}(i)--(ii) hold and define
\begin{equation}
K(\alpha,\alpha')
:=\langle k_\alpha,k_{\alpha'}\rangle_{L_2}
=\int
\frac{p(z\mid\alpha,\theta_0)p(z\mid\alpha',\theta_0)}
{p_0(z)}\,d\mu(z).
\label{eq:kernel}
\end{equation}
Then $K$ is a positive semidefinite kernel on $\mathcal A$, and
\[
\operatorname{Range}(S^*)
=\left\{f\in\mathcal H_K:
\textstyle\int f\,d\eta_0=0\right\},
\qquad
\|f\|_{\mathcal H_K}
=\min\{\|g\|_{L_2}:S^*g=f\},
\]
where $\mathcal H_K$ is the RKHS generated by $K$.
If Assumptions~\ref{Regularity_model}
and~\ref{ass:mixture}(iii)--(vi) also hold, then
$r(\cdot,\theta_0)\in\mathcal H_K$ is necessary for
a relevant locally robust moment for $\psi_0$ and,
under the integrability conditions of
Proposition~\ref{prop:tva_key_equation}\textup{(ii)},
sufficient and equivalent to pathwise differentiability
of the observed-law target.
Failure therefore rules out any regular estimator sequence
on the stated local experiment.
\end{theorem}

The diagonal
$K(\alpha,\alpha)=1+\chi^2(\mathbb P_\alpha\|\mathbb P_0)$
measures distinguishability of type $\alpha$ from the
population, while $K(\alpha,\alpha')$ measures affinity across
types. Hence informative targets cannot vary more sharply across
latent types than the measurement process can distinguish them.
Smoothness of the feature map therefore transfers to
$\mathcal H_K$, ruling out sufficiently irregular target
representers.
\begin{corollary}[Smoothness ladder]
\label{cor:ladder}
Let $\mathcal A^\circ$ be the interior of $\mathcal A$.
For $f\in\mathcal H_K$, let $g_f\in L_2$ be the
minimum-norm representer satisfying
$f(\alpha)=\langle g_f,k_\alpha\rangle_{L_2}$.
\begin{compactenum}[(i)]
\item
If $\alpha\mapsto k_\alpha$ is continuous from
$\mathcal A^\circ$ into $L_2$, then every
$f\in\mathcal H_K$ is continuous on $\mathcal A^\circ$, with
\[
|f(\alpha)-f(\alpha')|
\le\|f\|_{\mathcal H_K}
\|k_\alpha-k_{\alpha'}\|_{L_2}.
\]
\item
If $\alpha\mapsto k_\alpha$ is $m$ times continuously
differentiable into $L_2$, then
$\mathcal H_K\subseteq C^m(\mathcal A^\circ)$ and
\[
f^{(j)}(\alpha)
=\langle g_f,\partial_\alpha^j k_\alpha\rangle,
\qquad j\le m.
\]
\item
If $\alpha\mapsto k_\alpha$ extends holomorphically to
a complex neighborhood of each point of $\mathcal A^\circ$,
with locally bounded $L_2$-norm, then every
$f\in\mathcal H_K$ is real-analytic on $\mathcal A^\circ$.
\end{compactenum}
\end{corollary}

Continuity excludes discontinuous target representers;
differentiability and analyticity impose stronger necessary
restrictions. Whereas \citet{escanciano2023irregular}
derives individual smoothness restrictions by differentiating
the representer equation, the RKHS characterization identifies
the exact range and yields the full ladder, including
the quantitative continuity bound.

\subsection{Gaussian Measurement Error and Value-Added Targets}
\label{subsec:tva_gauss}

Gaussian measurement error yields analyticity and explicit
moments for polynomial targets and the noise scale.

\begin{corollary}[Gaussian repeated measurements]
\label{propAnal}
Let Assumptions~\ref{Regularity_model}
and~\ref{ass:mixture} hold, and let $\mathcal A$ be
a nondegenerate compact interval.
In the Gaussian model \eqref{eq:tva_model},
with $\theta_0>0$, every function in $\mathcal H_K$
is real-analytic on $\mathcal A^\circ$.
Analyticity is necessary, but not sufficient, for
membership in $\mathcal H_K$.
Hence, if $r(\cdot,\theta_0)$ does not agree
$\eta_0$-a.s.\ with a real-analytic function on
$\mathcal A^\circ$, no relevant locally robust moment
exists and $\psi_0$ is not regularly estimable.
\end{corollary}

\subsubsection{Distributional and policy targets}

Maintain Assumptions~\ref{Regularity_model} and
\ref{ass:mixture}\textup{(i)--(iv), (vi)}.
Let $F_{\alpha,0}$ and $f_{\alpha,0}$ denote the latent
type's cumulative distribution function (CDF) and density.
For $\varphi\in(0,1)$, let
$q_\varphi:=F_{\alpha,0}^{-1}(\varphi)$ be a unique
interior quantile, and assume that $f_{\alpha,0}$ is
continuous and strictly positive on a neighborhood
of $q_\varphi$.
Up to additive constants, the derivative representers
for the CDF at $\bar\alpha$, the $\varphi$-quantile,
the gain below a fixed cutoff $c$, and the gain from
replacing the bottom $\varphi$ share are, respectively,
\[
\begin{aligned}
r_F(\alpha)&=\mathbf 1\{\alpha\le\bar\alpha\},
&
r_q(\alpha)&=
\frac{\varphi-\mathbf 1\{\alpha\le q_\varphi\}}
{f_{\alpha,0}(q_\varphi)},\\
r_{G,c}(\alpha)&=(c-\alpha)\mathbf 1\{\alpha\le c\},
&
r_{G,\varphi}(\alpha)&=
\left(q_\varphi-\alpha
-\frac{\varphi}{f_{\alpha,0}(q_\varphi)}\right)
\mathbf 1\{\alpha\le q_\varphi\}.
\end{aligned}
\]
The last expression accounts for the endogenous quantile
cutoff. The first, second, and fourth representers are
discontinuous; the third has a kink.
For interior thresholds, the corresponding continuity
or differentiability condition in Corollary~\ref{cor:ladder}
excludes them from $\mathcal H_K$.
Under Gaussian noise, all four are excluded by
Corollary~\ref{propAnal} because they are nonanalytic.

These targets therefore admit neither relevant orthogonal
moments nor regular parametric-rate estimators, applying
Lemma~\ref{lem:nonlinear_range} to the nonlinear targets.
Differentiability with respect to the latent distribution
does not imply differentiability with respect to the observed
law: the latter requires the corresponding centered
derivative representer to belong to
$\operatorname{Range}(S^*)$ in $L_2(\eta_0)$
\citep{van1991differentiable}.
Repeated measurements supply orthogonal moments, but for
these targets they carry no first-order target information
and test only the model's overidentifying restrictions.

\subsubsection{A measurement-resolution frontier}

The preceding results apply for every finite number of measurements: adding measurements does not make the unsmoothed distributional and threshold-policy targets regularly estimable. Increasing the number of measurements can nevertheless expand the class of smoothed distributional targets that admit relevant orthogonal moments. To make this explicit, fix $a\in\mathcal A^\circ$ and
$h>0$ and consider the smoothed distributional target
\[
\psi_{a,h}
:=
\mathbb E\left[
\Phi\left(\frac{a-\alpha}{h}\right)
\right],
\]
where $\Phi$ is the standard normal CDF. As $h\downarrow0$,
$\psi_{a,h}$ approaches the latent CDF evaluated at $a$.
Let
\[
\bar Y_J:=J^{-1}\sum_{j=1}^JY_j,
\qquad
\sigma_J:=\frac{\theta_0}{\sqrt J}.
\]

\begin{proposition}[Measurement-resolution frontier]
\label{prop:tva_analytic}
Let Assumptions~\ref{Regularity_model} and
\ref{ass:mixture} hold in the Gaussian model
\eqref{eq:tva_model}, with $J\ge2$.
Then:
\begin{compactenum}[(i)]
\item
If $0<h<\sigma_J$, no relevant locally robust moment exists
for $\psi_{a,h}$, and $\psi_{a,h}$ is not regularly estimable.
\item
If $h>\sigma_J$, define
\[
d_J(\theta):=\left(h^2-\frac{\theta^2}{J}\right)^{1/2},
\qquad
q_{J,a,h}(Y,\theta)
:=
\Phi\left(\frac{a-\bar Y_J}{d_J(\theta)}\right).
\]
Then
\[
\mathbb E_{\theta,\eta}
[q_{J,a,h}(Y,\theta)\mid\alpha]
=
\Phi\left(\frac{a-\alpha}{h}\right)
\]
for $\theta$ in a neighborhood of $\theta_0$. Hence, with
\[
\varrho(Y,\theta):=(Y_1-Y_2)^2-2\theta^2,
\qquad
\Gamma_{J,a,h}
:=
-\frac{1}{4\theta_0}
\mathbb E[\partial_\theta q_{J,a,h}(Y,\theta_0)],
\]
the moment
\[
g_{J,a,h}(Y,\psi,\theta)
=
q_{J,a,h}(Y,\theta)-\psi
-\Gamma_{J,a,h}\varrho(Y,\theta)
\]
is an admissible, relevant locally robust moment for
$\psi_{a,h}$.
\end{compactenum}
\end{proposition}

Thus the smallest distributional smoothing scale accessible to
regular orthogonal inference is of order $\theta_0/\sqrt J$.
In particular, if
\[
\frac{\theta_0}{\sqrt{J+1}}
<h<
\frac{\theta_0}{\sqrt J},
\]
the same target admits no relevant orthogonal moment with $J$
measurements but does with $J+1$. Additional measurements therefore
enlarge the informative target space, rather than merely improving
precision for a fixed regular target. The unsmoothed CDF corresponds
to $h=0$ and remains outside this space for every finite $J$.

\section{Concluding Remarks}
\label{sec:conclusions}

This paper asks questions logically prior to constructing
Neyman-orthogonal moments: when can nontrivial orthogonal moments exist,
which target directions contain first-order information, and does a
proposed moment detect all such directions? Restricted local
overidentification, together with attainability by admissible moments,
determines existence. The relevance rank counts the number of linearly independent informative target directions, while the rank sandwich distinguishes directions
missed by a particular moment from directions that are intrinsically
uninformative because their effects can be absorbed by nuisance
variation.

The applications illustrate both sides of this distinction. In sample
selection, a rank-deficient canonical moment nevertheless captures every
informative coefficient direction, yielding identification-robust
inference even when some coefficients are not identified. In
latent-variable models, the informative target space is an RKHS
determined by the measurement experiment. Additional repeated
measurements enlarge the class of smoothed distributional targets that
are regularly learnable, while important nonsmooth distributional and
policy targets remain outside the informative space for every finite
number of measurements. Orthogonalization can isolate information
already present in the experiment, but it cannot create it; understanding
its limits can matter as much as understanding its possibilities.
%\clearpage
\begin{appendix}
\section*{Appendix: Proofs of General Results}

\section{Preliminaries}
\label{app:prelim}

\subsection{The pathwise moment identity}

\begin{lemma}[Pathwise moment identity]
\label{lem:pathwise}
Let $g\in\mathcal G_{\mathrm{adm}}$ and let
$\{\mathbb P_\tau\}\in\mathfrak G\cup\mathfrak G_0$ have score $s$. Then
\begin{equation}
\left.\frac{d}{d\tau}\mathbb E[g(Z,\lambda_\tau)]\right|_{\tau=0^+}
=-\langle g(\cdot,\lambda_0),s\rangle.
\label{eq:appendix_pathwise_identity}
\end{equation}
\end{lemma}

\noindent\textbf{Proof.}
Write $g_0=g(\cdot,\lambda_0)$ and
$h_\tau=g(\cdot,\lambda_\tau)-g_0$. Validity gives
\begin{equation}
0=(\mathbb E_\tau-\mathbb E)g_0+\mathbb E h_\tau+R_\tau,
\qquad R_\tau=(\mathbb E_\tau-\mathbb E)h_\tau.
\label{eq:proof_product_decomposition}
\end{equation}
Factoring $f_\tau-f_0$ and applying Cauchy--Schwarz yields
\[
|R_\tau|
\le \{2\mathbb E_\tau h_\tau^2+2\mathbb E h_\tau^2\}^{1/2}
\left\{\int(\sqrt{f_\tau}-\sqrt{f_0})^2\,d\mu\right\}^{1/2}
=o(\tau).
\]
The first factor is $o(1)$ by
Definition~\ref{def:admissible}(iii), and the second is $O(\tau)$
by DQM. Divide \eqref{eq:proof_product_decomposition} by $\tau$
and apply \eqref{GI}.
\hfill$\blacksquare$

\subsection{Richness of the admissible class}
\label{app:richness}

\begin{lemma}[Automatic admissibility]
\label{lem:autoadm}
Let Assumption~\ref{ass:regclass} hold. For every nonconstant
$m\in L_{2q}$, the moment \eqref{const_adm} belongs to
$\mathcal G_{\mathrm{adm}}$, with nonzero true evaluation
$g(\cdot,\lambda_0)=m-\mathbb E[m(Z)]$.
\end{lemma}

\noindent\textbf{Proof.}
For every generating path, H\"older's inequality and
Definition~\ref{def:regularclass} give, uniformly for small $\tau$,
\[
\int m^2f_\tau\,d\mu
\le \|m\|_{L_{2q}}^2\|M_p\|_{L_p}<\infty,
\qquad 1/p+1/q=1.
\]
This is \eqref{bounded_mom}. Lemma~I.7.2 of
\citet{ibragimovkhasminskii1981} gives
$\left.\frac{d}{d\tau}\mathbb E_\tau m\right|_{0^+}=\mathbb E[ms]$.
Thus $g(Z,\lambda)=m(Z)-\mathbb E_\lambda m$ is valid and satisfies
the derivative requirements. Its change along a path is a constant
tending to zero, proving both continuity requirements. Its true
evaluation is nonzero because $m$ is nonconstant. If $p=\infty$,
take $q=1$.
\hfill$\blacksquare$

\begin{corollary}[Richness holds]
\label{cor:richness}
Let Assumption~\ref{ass:regclass} hold. Every nonzero
$m\in L_2^0\cap L_{2q}$ is the value at $\lambda_0$ of an admissible
moment. Consequently, for every closed subspace $V\subseteq L_2^0$
such that $V\cap L_{2q}$ is dense in $V$,
\[
\overline{\operatorname{span}}\{g(\cdot,\lambda_0):
g\in\mathcal G_{\mathrm{adm}},\ g(\cdot,\lambda_0)\in V\}=V.
\]
\end{corollary}

\noindent\textbf{Proof.}
A nonzero centered $m$ is nonconstant, so
Lemma~\ref{lem:autoadm} realizes it as an admissible true evaluation.
The second claim follows by density.
\hfill$\blacksquare$

\begin{lemma}[Primitive conditions for richness]
\label{lem:density}
Let $V\subseteq L_2^0$ be closed and $q\in[1,\infty)$.
Then $V\cap L_{2q}$ is dense in $V$ if:
(a) $V$ is the closed linear span of functions in $L_{2q}$;
(b) $V$ is finite dimensional with a basis in $L_{2q}$; or
(c) $\Pi_V$ maps bounded mean-zero functions into $L_{2q}$,
as when $\Pi_V$ is a conditional expectation.
\end{lemma}

\noindent\textbf{Proof.}
In (a), finite linear combinations lie in $V\cap L_{2q}$;
(b) is a special case. For (c), truncate $v\in V$ at $\pm M$
and recenter to obtain $v_M$. Then $\Pi_Vv_M\in V\cap L_{2q}$ and
$\|\Pi_Vv_M-v\|_{L_2}\le\|v_M-v\|_{L_2}\to0$.
Conditional expectations preserve boundedness.

\hfill$\blacksquare$

\subsection{Derivative characterization of local robustness}
\label{app:der}

\begin{proposition}[Derivative characterization of local robustness]
\label{prop:LR_derivative}
Let $g\in\mathcal G_{\mathrm{adm}}$. Suppose there exist
$\Psi\subset\mathbb R^{d_\psi}$, a linear space $\mathcal H$,
$(\psi_0,\eta_0)\in\Psi\times\mathcal H$, and
$\Xi:\Psi\times\mathcal H\to\Lambda$ with
$\Xi(\psi_0,\eta_0)=\lambda_0$, such that:
(i) each path $\tau\mapsto\Xi(\psi_0,\eta_0+\tau h)$,
$h\in\mathcal H$, belongs to $\mathfrak G_0$ and has score $s_h$;
(ii) $\operatorname{span}\{s_h:h\in\mathcal H\}$ is dense in
$\overline{\mathcal T_0}$; and
(iii) $M_g(\psi,\eta)=\mathbb E[g(Z,\Xi(\psi,\eta))]$
has G\^ateaux derivative $D_\eta M_g(\psi_0,\eta_0)[h]$
for every $h\in\mathcal H$.
Then $g$ is locally robust iff all these derivatives vanish.
\end{proposition}

\noindent\textbf{Proof.}
Lemma~\ref{lem:pathwise} gives
\begin{equation}
D_\eta M_g(\psi_0,\eta_0)[h]
=-\langle g(\cdot,\lambda_0),s_h\rangle.
\label{eq:proof_gateaux_identity}
\end{equation}
Local robustness implies vanishing derivatives by (i).
Conversely, vanishing derivatives imply orthogonality to
$\overline{\mathcal T_0}$ by (ii) and continuity of the inner product.
Lemma~\ref{lem:pathwise} then gives local robustness along every
path in $\mathfrak G_0$.
\hfill$\blacksquare$

\section{Proofs for Section~\ref{SecExistence}}
\label{app:section3}

\subsection{Proofs for Existence}
\label{app:existence}

\noindent\textbf{Proof of Theorem~\ref{mainExistence}.}
By Lemma~\ref{lem:pathwise}, an admissible moment is locally robust
iff its true evaluation is orthogonal to every restricted generating
score, hence to their closed linear span. Therefore
\[
\mathcal G_{\mathrm{LR}}
=\overline{\mathcal T_0}^{\perp}\cap\mathcal G_{\mathrm{adm}}^0.
\]
Corollary~\ref{cor:richness} realizes every nonzero
$m\in\overline{\mathcal T_0}^{\perp}\cap L_{2q}$ as an admissible
true evaluation, proving the inclusion. Under local boundedness,
$q=1$, so every nonzero element of the orthocomplement is realized.
Finally, restricted local overidentification is equivalent to
$\overline{\mathcal T_0}^{\perp}\neq\{0\}$.
Assumption~\ref{ass:dense0} ensures a nonzero element in its
intersection with $L_{2q}$, proving sufficiency; the displayed
identity proves necessity.
\hfill$\blacksquare$

\subsection{Proofs for Relevance}

\noindent\textbf{Proof of Proposition~\ref{prop:decomposition}.}
For $h\in\overline{\mathcal T_0}^{\perp}$, write
$h=\Pi_{\overline{\mathcal T}}h+(I-\Pi_{\overline{\mathcal T}})h$.
The second term belongs to $\overline{\mathcal T}^{\perp}$.
Since $\overline{\mathcal T_0}\subseteq\overline{\mathcal T}$,
$\langle\Pi_{\overline{\mathcal T}}h,t_0\rangle
=\langle h,t_0\rangle=0$ for every $t_0\in\overline{\mathcal T_0}$,
so the first belongs to $\widetilde S_\psi$.
The terms are orthogonal, and both spaces lie in
$\overline{\mathcal T_0}^{\perp}$, proving the decomposition.
\hfill$\blacksquare$

\medskip\noindent
\textbf{Proof of Theorem~\ref{thm:relevance}.}
By Proposition~\ref{prop:decomposition}, only the
$\widetilde S_\psi$ component of a locally robust true evaluation
can have nonzero inner product with a full-model score.
Relevance therefore requires $\widetilde S_\psi\neq\{0\}$.
Conversely, Assumption~\ref{ass:denseS} provides
$0\neq m\in\widetilde S_\psi\cap L_{2q}$.
Corollary~\ref{cor:richness} realizes $m$ by an admissible moment,
which is locally robust by Lemma~\ref{lem:pathwise}.
Since $m\in\overline{\mathcal T}$ is nonzero and the full-model
scores have dense linear span, some such score $s$ satisfies
$\langle m,s\rangle\neq0$. Necessarily
$s\notin\overline{\mathcal T_0}$, proving relevance.

Independent projections into $\widetilde S_\psi$ number at most
$r_\psi$. Conversely, its dense linear subspace
$\widetilde S_\psi\cap L_{2q}$ contains a basis when $r_\psi<\infty$,
and arbitrarily large finite independent collections otherwise.
Realizing these by Corollary~\ref{cor:richness} gives the counting claim.

\hfill$\blacksquare$

\medskip\noindent
\textbf{Proof of Proposition~\ref{prop:information_geometry}.}
Let $\mathcal S=\operatorname{Range}(A)$, a finite-dimensional
closed subspace orthogonal to $\overline{\mathcal T_0}$.
Assumption~\ref{ass:coordinate_paths} represents every generating score as

\[
s=a'\ell_\psi+s_\eta
=Aa+\{\Pi_{\overline{\mathcal T_0}}(a'\ell_\psi)+s_\eta\}
\in\mathcal S\oplus\overline{\mathcal T_0}.
\]
This sum is closed, so it contains $\overline{\mathcal T}$.
The reverse inclusion follows because both
$\overline{\mathcal T_0}$ and each
$Aa=a'\ell_\psi-\Pi_{\overline{\mathcal T_0}}(a'\ell_\psi)$
lie in $\overline{\mathcal T}$. Consequently,
\[
\overline{\mathcal T}=\overline{\mathcal T_0}\oplus\mathcal S,
\qquad \widetilde S_\psi=\mathcal S,
\qquad \|Aa\|_{L_2}^2=a'I_\psi a.
\]
Thus $\ker(A)=\ker(I_\psi)$, and rank--nullity gives
$r_\psi=\operatorname{rank}(I_\psi)\le d_\psi$.
The range of $A$ is the span of the components of
$\ell_\psi^{\rm eff}$; the relevance claim follows from
Theorem~\ref{thm:relevance}.
\hfill$\blacksquare$

\begin{lemma}[Restricted tangent space for a linear target]
\label{lem:restricted_linear}
Let Assumptions~\ref{Regularity_model}
and~\ref{ass:coordinate_paths} hold. If
$\overline{\mathcal T_0}^{\,c}$ is the closed linear span of the
restricted tangent set for $c'\psi(\lambda_0)$, then
$\overline{\mathcal T_0}^{\,c}
=\overline{\mathcal T_0}\oplus A(c^\perp)$,
where $c^\perp=\{a\in\mathbb R^{d_\psi}:c'a=0\}$.
\end{lemma}

\noindent\textbf{Proof.}
Every restricted path for $\psi$ preserves $c'\psi$, so
$\overline{\mathcal T_0}\subseteq\overline{\mathcal T_0}^{\,c}$.
For $a\perp c$, Assumption~\ref{ass:coordinate_paths}(i)
provides a path with $\psi(\lambda_\tau)=\psi_0+\tau a$
and score $a'\ell_\psi$. It preserves $c'\psi$ exactly;
subtracting its projection onto $\overline{\mathcal T_0}$ gives
$Aa\in\overline{\mathcal T_0}^{\,c}$.
Conversely, part (iii) places every scalar-target restricted score
in $\overline{\mathcal T_0}+A(c^\perp)$.
This sum is orthogonal and closed because $A(c^\perp)$ is
finite dimensional, proving equality.
\hfill$\blacksquare$

\medskip\noindent
\textbf{Proof of Theorem~\ref{thm:relev_estim}.}
Proposition~\ref{prop:information_geometry} and
Lemma~\ref{lem:restricted_linear} give
\[
\widetilde S_{c'\psi}
=\operatorname{Range}(A)\cap A(c^\perp)^\perp
=\{Ab:I_\psi b\in\operatorname{span}\{c\}\},
\]
where the second equality follows from
$\langle Ab,Aa\rangle=a'I_\psi b$.
Since $c\neq0$ and $\ker(A)=\ker(I_\psi)$, this space is nonzero
iff $c\in\operatorname{Range}(I_\psi)$; it is then spanned by
$\tilde\psi_c=Ab$ with $I_\psi b=c$.
Moreover, finite dimensionality and Assumption~\ref{ass:denseS}
imply $\widetilde S_\psi\subseteq L_{2q}$, so the same holds for
this scalar informative space. Theorem~\ref{thm:relevance}
therefore proves (i)$\Leftrightarrow$(ii).

For every DQM score $s=a'\ell_\psi+s_\eta$, the assumed
all-path decomposition gives
$\langle\tilde\psi_c,s\rangle=b'I_\psi a=c'a$.
Thus $\tilde\psi_c\in\overline{\mathcal T}$ represents the
pathwise derivative and is the efficient influence function.
Conversely, a derivative representer
$\tilde\psi_c\in\overline{\mathcal T}$ is orthogonal to
$\overline{\mathcal T_0}$ because the derivative vanishes along
target-preserving paths. Coordinate paths then give
$c'a=\langle\tilde\psi_c,a'\ell_\psi\rangle
=\langle\tilde\psi_c,Aa\rangle$ for every $a$.
Hence $c$ annihilates $\ker(A)=\ker(I_\psi)$, implying
$c\in\operatorname{Range}(I_\psi)$ and proving
(ii)$\Leftrightarrow$(iii).
\hfill$\blacksquare$

\medskip\noindent
\textbf{Proof of Proposition~\ref{prop:jacobian_verification}.}
Lemma~\ref{lem:pathwise} applied to coordinate paths, and
orthogonality to $\overline{\mathcal T_0}$, give
\[
J_{\psi,g}a=-\langle g,a'\ell_\psi\rangle
=-\langle g,a'\ell_\psi^{\rm eff}\rangle=-L_g(Aa).
\]
Since $\ker(A)=\ker(I_\psi)$ and
$\operatorname{Range}(A)=\widetilde S_\psi$,
\[
\ker(I_\psi)\subseteq\ker(J_{\psi,g}),
\qquad
\operatorname{rank}(J_{\psi,g})
=\operatorname{rank}(L_g)\le r_\psi.
\]
Surjectivity of $A$ gives $J_{\psi,g}\neq0$ iff some component
is relevant. By Definition~\ref{def:strong_relevance}, strong
relevance is equivalent to injectivity of $L_g$, and hence to
$\operatorname{rank}(L_g)=r_\psi$.
For $m\ge d_\psi$, full column rank is therefore equivalent to
$r_\psi=d_\psi$ and strong relevance.
\hfill$\blacksquare$

\medskip\noindent
\textbf{Proof of Theorem~\ref{thm:rank_sandwich}.}
For every $a\in\mathbb R^{d_\psi}$,
\[
a\in\mathcal N_\psi
\iff a'\ell_\psi^{\rm eff}=0
\iff a'I_\psi a=0
\iff a\in\ker(I_\psi).
\]
Thus $\mathcal N_\psi=\ker(I_\psi)$ and
$r_\psi=d_\psi-\dim(\mathcal N_\psi)$.
Proposition~\ref{prop:jacobian_verification} gives the lower bound,
while $\mathcal C\subseteq\mathcal N_\psi$ gives the upper bound.
Equality of the outer bounds forces
$\mathcal C=\mathcal N_\psi$ and
$\operatorname{rank}(J_{\psi,g})=r_\psi$;
the latter implies strong relevance by the same proposition. Finally, $\ker(I_\psi)\subseteq\ker(J_{\psi,g})$ implies
that these kernels coincide iff their dimensions agree,
equivalently iff
$\operatorname{rank}(J_{\psi,g})=r_\psi$, which characterizes
strong relevance by Proposition~\ref{prop:jacobian_verification}.

\hfill$\blacksquare$

\end{appendix}
\phantomsection
	\bibliographystyle{ectabib}
	\bibliography{references}

\clearpage
\setcounter{page}{1}
\renewcommand{\thepage}{SA-\arabic{page}}
\begingroup
\setlength{\abovedisplayskip}{4pt plus 2pt minus 1pt}
	\setlength{\belowdisplayskip}{4pt plus 2pt minus 1pt}
	\setlength{\abovedisplayshortskip}{2pt plus 1pt}
	\setlength{\belowdisplayshortskip}{2pt plus 1pt}
	\setlength{\jot}{2pt}
	\begin{center}
		{\Large\textsc{Supplementary Appendix}}\\[0.6em]
		{\large When Is Neyman-Orthogonal Inference Feasible?\\Existence and Relevance}\\[0.4em]
		{\normalsize Facundo Arga\~{n}araz \quad and \quad Juan Carlos Escanciano}
	\end{center}
	
\begin{appendix}
\setcounter{section}{2}
\renewcommand{\bigskip}{\vspace{\smallskipamount}}
\noindent
This Supplementary Appendix contains additional material for the applications
and extensions. Section~\ref{app:selection} contains the sample-selection
proofs, including the AR--NO expansion and uniformity analysis.
Section~\ref{app:tva} gives primitive conditions and proofs for the
latent-variable application, while Section~\ref{app:mixture} develops
the underlying tangent-set theory for mixture models.
Section~\ref{app:cone} treats one-sided orthogonality, and
Section~\ref{app:mc_additional} reports additional Monte Carlo results.
%%%%%%%%%%%%%%%%%%%%%%%%%%%%%%%%%%%%%%%%%%%%%%%%%%%%%%%%%%%%%%
\section{Sample Selection}
\label{app:selection}

\subsection{Restricted tangent space}

Throughout this subsection the full-vector target is fixed at
$\theta=\theta_0$. Let $K_\lambda(\cdot\mid x)$ denote the conditional
law of $(Y,D)$ given $X=x$, set
\[
\mathcal H_c:=\{s\in L_2^0:\mathbb E[s\mid X]=0\},
\]
and let $\overline{\mathcal T}_{0,c}$ be the closed tangent space
generated by restricted paths holding $P_X$ fixed. Define
\[
\Phi:\mathcal H_c\to L_2(P_X)^2,\qquad
\Phi(s):=\mathbb E[\varepsilon s\mid X].
\]
Its adjoint is $\Phi^*\zeta=\zeta(X)'\varepsilon$, and Assumption
\ref{ass:ss_moments}(ii) gives
\begin{equation}
c\mathbb E \left[ \|\zeta\|^2 \right]
\le\|\Phi^*\zeta\|_{L_2}^2
=\mathbb E[\zeta'\Omega\zeta]
\le C\mathbb E \left[ \|\zeta\|^2\right].
\label{eq:ss_closed_range}
\end{equation}
Thus $\operatorname{Range}(\Phi^*)$ is closed and
\begin{equation}
(\ker\Phi)^\perp
=\operatorname{Range}(\Phi^*)
=\{\zeta(X)'\varepsilon:\zeta\in L_2(P_X)^2\}.
\label{eq:ss_phi_range}
\end{equation}

We use two elementary DQM facts. First, every restricted score decomposes
orthogonally as
\[
s=s_X+s_c,\qquad s_X=\mathbb E[s\mid X],\quad s_c\in\mathcal H_c.
\]
Since $Q_X$ is unrestricted, bounded marginal tilts are restricted paths and are
dense in $L_2^0(P_X)$, so
\begin{equation}
L_2^0(P_X)+\overline{\mathcal T}_{0,c}\subseteq\overline{\mathcal T_0}.
\label{eq:ss_conditional_decomposition}
\end{equation}
Only this inclusion is used; sufficiency is established below directly along an
arbitrary restricted DQM path.

Second, if $\{\mathbb P_\tau\}$ is DQM with score $s$, $g\in L_2$, and
$\sup_{\tau\le\tau_0}\mathbb E_\tau[g^2]<\infty$, then
\begin{equation}
\frac{\mathbb E_\tau[g]-\mathbb E[g]}{\tau}
\longrightarrow\mathbb E[gs].
\label{eq:ss_score_continuity}
\end{equation}
Indeed,
$u_\tau=(\sqrt{f_\tau}-\sqrt{f_0})/\tau\to s\sqrt{f_0}/2$
strongly in $L_2$, whereas
$v_\tau=g(\sqrt{f_\tau}+\sqrt{f_0})\rightharpoonup2g\sqrt{f_0}$
weakly in $L_2$. To verify the latter convergence, first test against
functions supported on $\{|g|\le M\}$ and then use density. Equation
\eqref{eq:ss_score_continuity} follows from
$(\mathbb E_\tau g-\mathbb E g)/\tau=\langle u_\tau,v_\tau\rangle$.

Let
\[
\mathcal B_p:=\{b_p\in L_\infty(P_X):A_0b_p\in L_\infty(P_X)\},
\qquad
\mathcal B_q:=C_c^\infty(I^+),
\]
and, for $b=(b_p,b_q)$, define
\[
m_b(X):=
\begin{pmatrix}
b_p(X)\\ A_0(X)b_p(X)+b_q(T)
\end{pmatrix},
\qquad
R(m):=\varepsilon'\Omega^{-1}m.
\]

\begin{proposition}[Restricted tangent directions]
\label{prop:ss_tangent_decomp}
Under Assumption~\ref{ass:ss_moments},
\[
\ker\Phi\subseteq\overline{\mathcal T}_{0,c},
\qquad
R(m_b)\in\overline{\mathcal T}_{0,c}
\quad
\text{for every }b\in\mathcal B_p\times\mathcal B_q.
\]
All these directions are generated by locally bounded conditional paths.
\end{proposition}

\begin{proof}
Let $k(v)=1+\tanh(v)$. For
$r=u+\zeta(X)'\varepsilon$, where $u,\zeta$ are bounded and
$\mathbb E[r\mid X]=0$, consider
\[
L_{\tau,\alpha}(Z\mid X)
:=
\frac{k\{\tau r(Z)+\alpha(X)'\varepsilon(Z)\}}
{\mathbb E[k\{\tau r(Z)+\alpha(X)'\varepsilon(Z)\}\mid X]}.
\]
Given a smooth path $\beta_\tau=(\theta_0,p_\tau,q_\tau)$, choose
$\alpha_\tau$ so that
\[
\mathbb E[L_{\tau,\alpha_\tau}
\varepsilon(Z,\beta_\tau)\mid X]=0.
\]
At $(\tau,\alpha)=(0,0)$, the derivative with respect to $\alpha$ is
$\Omega(X)$. Its uniform nonsingularity and the conditional third-moment
bound yield, by the implicit-function theorem, a measurable solution
$\alpha_\tau=o(\tau)$ whenever
\begin{equation}
\Phi(r)
=
-\left.\partial_\tau\varepsilon(Z,\beta_\tau)\right|_{\tau=0}.
\label{eq:ss_tilt_compatibility}
\end{equation}
The resulting path has score $r$ and a locally bounded likelihood ratio.
Because $k(v)>0$, the tilted conditional law has a positive selected-outcome
density, while the defining equation for $\alpha_\tau$ enforces
$\mathbb E_\tau[\varepsilon(Z,\beta_\tau)\mid X]=0$. The strict interior and
moment bounds in Assumption~\ref{ass:ss_moments}(iii) ensure that, for sufficiently
small $\tau$, $p_\tau$, $q_\tau$, and the induced selected-outcome density
$f_{1,\tau}$ satisfy the remaining restrictions of the parameter space.

For $h\in\ker\Phi$, let
\[
u_j=(-j\vee h\wedge j)
-\mathbb E[(-j\vee h\wedge j)\mid X],\qquad
h_j:=u_j-\varepsilon'\Omega^{-1}\Phi(u_j).
\]
Then $h_j\to h$ in $L_2$, $\Phi(h_j)=0$, and each $h_j$ has the form
required above. Taking $\beta_\tau=\beta_0$ in
\eqref{eq:ss_tilt_compatibility} proves
$\ker\Phi\subseteq\overline{\mathcal T}_{0,c}$.

For $b\in\mathcal B_p\times\mathcal B_q$, take
$p_\tau=p_0+\tau b_p$ and $q_\tau=q_0+\tau b_q$, choosing the
representative of $b_p$ to vanish off $\{T\in[\underline p,\bar p]\}$.
These paths remain in the model and
\[
-\left.\partial_\tau
\varepsilon(Z,\theta_0,p_\tau,q_\tau)\right|_{\tau=0}=m_b(X).
\]
Since $\Phi\{R(m_b)\}=m_b$, the same construction gives a restricted
path with score $R(m_b)$.
\end{proof}

For later use, define
\[
R_0:=Y-\mu_{\lambda_0}(X),\qquad
\rho_0(u,x):=-\partial_u\log f_{1,0}(u\mid x),
\qquad
\ell_\theta(Z):=DX\rho_0(R_0,X).
\]
The location-score identities give
\begin{equation}
\ell_\theta\in L_2^0,\qquad
\mathbb E[\varepsilon\,\delta_\theta'\ell_\theta\mid X]
=
\begin{pmatrix}0\\TX'\delta_\theta\end{pmatrix},
\label{eq:ss_target_score}
\end{equation}
for every $\delta_\theta\in\mathbb R^{d_\theta}$.
Moreover, for every $\delta_\theta\in\mathbb R^{d_\theta}$, the path
$\theta_\tau=\theta_0+\tau\delta_\theta$, with
$(p,q,f_1,Q_X)=(p_0,q_0,f_{1,0},P_X)$ held fixed, is a DQM location path
with scalar score $\delta_\theta'\ell_\theta,$ and
$\mathbb E \left[\|\ell_\theta\|^2 \right]
=\mathbb E[T\|X\|^2\mathcal I_{1,0}(X)]<\infty$.

\subsection{Existence and relevance}
\label{SS_ExiRelProofs}

\noindent\textbf{Proof of Theorem~\ref{thm:SS_existence}.}
\emph{Necessity.} Let $g\in\overline{\mathcal T_0}^{\perp}$. Then
\eqref{eq:ss_conditional_decomposition} gives $\mathbb E[g\mid X]=0$, and since
$\ker\Phi\subseteq\overline{\mathcal T}_{0,c}$, \eqref{eq:ss_phi_range} gives
$g=\zeta(X)'\varepsilon$ for some $\zeta=(\zeta_1,\zeta_2)'\in L_2(P_X)^2$.
Orthogonality to $R(m_b)$ gives
$\mathbb E[(\zeta_1+A_0\zeta_2)b_p]+\mathbb E[\zeta_2b_q(T)]=0$ for every
$(b_p,b_q)\in\mathcal B_p\times\mathcal B_q$. Taking $b_q = 0,$ bounded $b_p$ supported on
$\{|A_0|\le M\}$ and letting $M\to\infty$ yields $\zeta_1=-A_0\zeta_2$, while taking $b_p =0,$
density of $C_c^\infty(I^+)$ in $L_2(P_T)$ gives $\mathbb E[\zeta_2\mid T]=0$.
Thus $\varphi:=\zeta_2\in\mathcal V_0$ and $g=g_\varphi$. This proves necessity.

\emph{Sufficiency: preliminaries.} Let $\varphi\in\mathcal V_0$ be bounded, let
$\{\mathbb P_\tau\}$ be an arbitrary restricted DQM path with score $s$, let $Q_\tau$ denote the marginal law of $X$ under $\mathbb P_\tau$, and let
$h_\tau(x)$ denote the Hellinger distance between the conditional laws
$K_\tau(\cdot\mid x)$ and $K_0(\cdot\mid x)$ of $(Y,D)$ given $X=x$. Also set
\[
W:=Y-DX'\theta_0,\qquad
d_\tau:=p_\tau(X)-T,\qquad
r_\tau:=q_\tau-q_0.
\]
Since the terms involving $X'\theta_0$ cancel,
\begin{equation}
g_\varphi
=
\varphi(X)\{W-q_0(T)-\dot q_0(T)(D-T)\}.
\label{eq:ss_g_W}
\end{equation}

Joint DQM implies that the squared Hellinger distance between
$\mathbb P_\tau$ and $\mathbb P_0$ is $O(\tau^2)$. Decomposing the joint
square-root densities into their marginal-$X$ and conditional components, and
using $(a+b)^2\ge a^2/2-b^2$, gives the average conditional Hellinger bound
below. Hellinger contraction under the projections onto $X$ and onto $D$ gives,
respectively, the total-variation bound for $Q_\tau$ and the $L_2(Q_\tau)$
bound for $d_\tau$. Finally, if $N$ is a $\mathbb P_0$-null set, the DQM
remainder restricted to $N$ equals $\mathbb P_\tau(N)/\tau^2$. Consequently,
\begin{equation}
\mathbb E_{Q_\tau}[h_\tau^2(X)]=O(\tau^2),\;\;
\|d_\tau\|_{L_2(Q_\tau)}=O(\tau),\;\;
\|Q_\tau-P_X\|_{\mathrm{TV}}=O(\tau),\;\;
Q_\tau(N)=o(\tau^2).
\label{eq:ss_dqm_rates}
\end{equation}
Assumption~\ref{ass:ss_moments}(iii) gives, uniformly in $\tau$ and $X$,
\[
\mathbb E_\tau[W^2\mid X]
=p_\tau(X)\Bigl\{\int u^2f_{1,\tau}(u\mid X)du
+\bigl[q_\tau\{p_\tau(X)\}/p_\tau(X)\bigr]^2\Bigr\}\le C,
\]
so \eqref{eq:ss_g_W} yields $\sup_\tau\mathbb E_\tau[g_\varphi^2]<\infty$ and no
moment condition on $X$ under $Q_\tau$ is needed. Finally, $T\in I$ holds
$P_X$-a.s., so by \eqref{eq:ss_dqm_rates} the event $\{T\notin I\}$ has
$Q_\tau$-probability $o(\tau^2)$; every integrand below is bounded by
$C\|\varphi\|_\infty$ there, so this event contributes $o(\tau^2)$ throughout and
is omitted.

\emph{Sufficiency: stability of $q_\tau$.} Since $\mathbb E_\tau[W\mid X]
=q_\tau\{p_\tau(X)\}$ and $\mathbb E[W\mid X]=q_0(T)$, the conditional Hellinger
inequality and the last two displays give
$\mathbb E_{Q_\tau}|q_\tau\{p_\tau(X)\}-q_0(T)|=O(\tau)$; the common Lipschitz
bound then gives $\mathbb E_{Q_\tau} \left[|r_\tau(T)| \right]=O(\tau)$, and $|r_\tau|\le2B$ with
$\|Q_\tau-P_X\|_{\mathrm{TV}}=O(\tau)$ give $\mathbb E \left[|r_\tau(T)| \right]=O(\tau)$. Full
support of $P_T$ on $I$ and the common derivative bound imply
$\|r_\tau\|_{\infty,I}\to0$, and the common second-derivative bound with the
finite-difference inequality on $I$ then gives
\begin{equation}
\|\dot r_\tau\|_{\infty,I}\to0 .
\label{eq:ss_q_stability}
\end{equation}

\emph{Sufficiency: conclusion.} The conditional moment restrictions give the
exact identity
$\mathbb E_\tau[g_\varphi]=\mathbb E_{Q_\tau}[\varphi(X)\{q_\tau(T+d_\tau)-q_0(T)
-\dot q_0(T)d_\tau\}]$, and Taylor expansion with \eqref{eq:ss_q_stability} gives
\[
\bigl|q_\tau(T+d_\tau)-q_0(T)-\dot q_0(T)d_\tau-r_\tau(T)\bigr|
\le o(1)|d_\tau|+Cd_\tau^2 .
\]
Since $\mathbb E[\varphi\mid T]=0$ we have $\mathbb E[\varphi r_\tau(T)]=0$, so
$|\mathbb E_{Q_\tau}[\varphi r_\tau(T)]|
\le4\|\varphi\|_\infty\|r_\tau\|_{\infty,I}\|Q_\tau-P_X\|_{\mathrm{TV}}$.
With \eqref{eq:ss_dqm_rates} this gives $\mathbb E_\tau[g_\varphi]=o(\tau)$, and
score continuity \eqref{eq:ss_score_continuity} yields
$\mathbb E[g_\varphi s]=0$. The path was arbitrary and $g_\varphi \in \overline{\mathcal{T}}^{\perp}_0.$

\emph{Sufficiency: general instruments.} Put $w=1+A_0^2$,
$H_n=\{\mathbb E[w\mid T]\le n\}$ and
\[
\varphi_{nM}:=\mathbf1_{H_n} \times \bigl[(-M\vee\varphi\wedge M)
-\mathbb E[(-M\vee\varphi\wedge M)\mid T]\bigr].
\]
These are bounded with conditional mean zero, and
$\mathbb E[w(\varphi_{nM}-\varphi)^2]\to0$, first as $M\to\infty$ and then as
$n\to\infty$: conditional centering contributes at most
$n\mathbb E[|\varphi-(-M\vee\varphi\wedge M)|^2]$ on $H_n$, while the complement
vanishes by dominated convergence. By \eqref{eq:ss_closed_range},
$g_{\varphi_{nM}}\to g_\varphi$ in $L_2$, so orthogonality extends to every
$\varphi\in\mathcal V_0$ and
$\overline{\mathcal T_0}^{\perp}=\{g_\varphi:\varphi\in\mathcal V_0\}$. Since
necessity used only the constructed marginal and conditional scores, their closed
span is $\overline{\mathcal T_0}$: the locally bounded class is generating rather
than assumed to be so.

\emph{Attainability.} For $\varphi\in\mathcal V$, replace $(\theta_0,p_0,q_0)$ in
\eqref{eq:SS_moment_class} by $(\theta,p,q)$; the resulting moment has
identically zero conditional expectation, and the definition of $\mathcal V$,
local boundedness of the generating paths and \eqref{eq:ss_score_continuity} give
admissibility and local robustness. If $\sigma(X)=\sigma(T)$ then
$\mathbb E[\varphi\mid T]=0$ forces $\varphi=0$; under strict inclusion some
$\varphi=\mathbf1_A-\mathbb E[\mathbf1_A\mid T]$ is bounded and nonzero, hence
lies in $\mathcal V$, and Assumption~\ref{ass:ss_moments}(ii) gives
$\mathbb E[g_\varphi^2]\ge c\,\mathbb E[\varphi^2(1+A_0^2)]>0$.

\hfill$\blacksquare$

\medskip\noindent
\textbf{Proof of Corollary~\ref{cor:SS_theta1}.}
Theorem~\ref{thm:SS_existence} characterizes the moments orthogonal to
the full-vector restricted tangent space. For
$h\in\mathbb R^{d_\theta-1}$, \eqref{eq:ss_target_score} gives
\[
\langle g_\varphi,h'\ell_{\theta,-1}\rangle
=\mathbb E[\varphi(X)TX_{-1}']h.
\]
Thus $g_\varphi$ is also orthogonal to every regular
$\theta_{0,-1}$ direction if and only if
\eqref{eq:SS_theta1_condition} holds. For
$\varphi\in\mathcal V\setminus\{0\}$, admissibility follows from
Theorem~\ref{thm:SS_existence}, while Assumption
\ref{ass:ss_moments}(ii) ensures that $g_\varphi\neq0$.
\hfill$\blacksquare$

\medskip\noindent
\textbf{Proof of Theorem~\ref{thm:SS_relevance_compact}.}
Put $N:=\ker(\Sigma_{\tilde X})$.
The coefficient paths have scores $a'\ell_\theta$, as
established after Proposition~\ref{prop:ss_tangent_decomp}.
Since the sum of a closed subspace and a finite-dimensional
space is closed, Assumption~\ref{ass:ss_local_rep} gives
\begin{equation}
\overline{\mathcal T}
=
\overline{\mathcal T_0}
+\operatorname{span}
\{a'\ell_\theta:a\in\mathbb R^{d_\theta}\}.
\label{eq:ss_full_decomp}
\end{equation}
Moreover,
\[
a'\ell_\theta\in\overline{\mathcal T_0}
\iff
\Sigma_\varphi a=0\quad\forall\,\varphi\in\mathcal V_0
\iff
\Sigma_{\tilde X}a=0,
\]
where the last equivalence uses $\varphi=\tilde X$ and
$N\subseteq\ker(\Sigma_\varphi)$. Hence
\[
\mathcal N_{\theta_0}=N,\qquad
r_{\theta_0}=d_\theta-\dim N
=\operatorname{rank}(\Sigma_{\tilde X}).
\]
The same kernel inclusion gives
$\operatorname{rank}(\Sigma_\varphi)\le r_{\theta_0}$.
Finally, \eqref{eq:ss_full_decomp} implies that $g_\varphi$ is relevant iff
$\Sigma_\varphi\neq0$, and strongly relevant iff
$\ker(\Sigma_\varphi)=N$. Taking $\varphi=\tilde X$ gives the remaining
claims.
\hfill$\blacksquare$

\medskip\noindent
\textbf{Proof of Proposition~\ref{prop:SS_identification}.}
By \eqref{eq:ss_g_linear_theta} and
\eqref{eq:ss_jacobian_compact},
$\mathbb E[g_\varphi(Z,\theta,\eta_0)]
=-\Sigma_\varphi(\theta-\theta_0)$, where
$\Sigma_\varphi=\mathbb E[\varphi(X)T\tilde X']$.
If $v\in\ker(\Sigma_{\tilde X})$, positivity of $T$
implies $v'\tilde X=0$ a.s., and hence
$\Sigma_\varphi v=0$ for every $\varphi\in\mathcal V$.
Conversely, since $\tilde X\in\mathcal V^{d_\theta}$,
the moment based on $\tilde X$ requires
$\Sigma_{\tilde X}v=0$.
This proves the characterization of $\Theta_{\mathrm{OM}}$.
Since $\Theta$ is open and contains $\theta_0$,
$a'\theta$ is constant on
$(\theta_0+\ker(\Sigma_{\tilde X}))\cap\Theta$
if and only if
$a\in\ker(\Sigma_{\tilde X})^\perp
=\operatorname{Range}(\Sigma_{\tilde X})$.
\hfill$\blacksquare$
%%%%%%%%%%%%%%%%%%%%%%%%%%%%%%%%%%%%%%%
\subsection{AR--NO expansion and uniformity}
\label{sec:asymlin}

		We first fix \(\mathbb P\) and suppress its subscript.
		
		\begin{lemma}[Canonical representative and oracle centering]
			\label{lem:ss_canonical_representative}
			Let \(\Gamma=\mathbb E[\tilde Xm_0]\),
			\(\bar\theta=\Sigma_{\tilde X}^{\dagger}\Gamma\), and
			\(\bar\Delta=\bar\theta-\theta_0\). Under
			Assumption~\ref{ass:ss_moments},
			\[
			\Gamma=\Sigma_{\tilde X}\theta_0,
			\qquad
			\bar\theta=\Pi_{\operatorname{Range}(\Sigma_{\tilde X})}\theta_0,
			\qquad
			\tilde X'\bar\Delta=0\quad\text{a.s.}
			\]
			If \(\nu>0\), then \(\bar\theta_1=\theta_{0,1}\), and
			\(\bar q(t):=q_0(t)-t h_{0X}(t)'\bar\Delta\) satisfies
			\(TX'\bar\theta+\bar q(T)=TX'\theta_0+q_0(T)\) a.s. Moreover,
			\(\|\bar\Delta\|\le2\|\theta_0\|\), so \(\bar q\) inherits the maintained
			smoothness bounds. Moreover, \(\mathbb E[g^*]=0\). If additionally
			\(\|g^*\|_{L_4}\le C\nu\), then
			\eqref{eq:ss_corrected_variance_order} holds.
		\end{lemma}
		
		\begin{proof}
			The linear moment and \(\mathbb E[D-T\mid X]=0\) give
			\(\Gamma=\mathbb E[\tilde Xr_0']\theta_0
			=\Sigma_{\tilde X}\theta_0\). The Moore--Penrose identities yield the first
			two conclusions and \(\bar\Delta\in\ker(\Sigma_{\tilde X})\). Thus
			\(0=\bar\Delta'\Sigma_{\tilde X}\bar\Delta
			=\mathbb E[T(\tilde X'\bar\Delta)^2]\); since \(T\ge\underline p\),
			\(\tilde X'\bar\Delta=0\) a.s. If \(\nu>0\), then
			\(\Sigma^{(1)}>0\), and
			\(\Sigma_{\tilde X}(1,-b^{*\prime})'=\Sigma^{(1)}e_1\) implies
			\(e_1\in\operatorname{Range}(\Sigma_{\tilde X})\), hence
			\(\bar\Delta_1=0\). The remaining geometric claims follow by substitution and
			\(\|\bar\theta\|\le\|\theta_0\|\).
			
			Write
			\[
			v=m_0-r_{0,1}\theta_{0,1}-r_{0,-1}'\bar\theta_{-1},
			\qquad g^*=\varphi^{(1)}v.
			\]
			Equation~\eqref{eq:ss_population_profile_orthogonality} gives
			\(\mathbb E[g^*]=0\). If \(\nu=0\), then
			\(\varphi^{(1)}=0\) a.s. and the bounds are immediate. If \(\nu>0\), then
			\(\bar\Delta_1=0\) and
			\[
			v=\varepsilon_2-
			\{A_0+B_{0,-1}'\bar\Delta_{-1}\}\varepsilon_1.
			\]
			Assumption~\ref{ass:ss_moments}\textup{(ii)} therefore gives
			\[
			\Omega^*
			=\mathbb E\!\left[\{\varphi^{(1)}\}^2
			\mathbb E[v^2\mid X]\right]
			\ge c\nu^2.
			\]
			Finally,
			\[
			\Omega^*=\|g^*\|_{L_2}^2\le\|g^*\|_{L_4}^2\le C\nu^2,
			\qquad
			\mathbb E\left[|g^*|^4\right]=\|g^*\|_{L_4}^4\le C\nu^4.\qedhere
			\]
		\end{proof}
		
		We now restore the subscript \(\mathbb P\). Let \(L\) be fixed,
		\(n_\ell\asymp n\), and let \(\mathcal F_\ell\) denote the training-sample
		sigma-field. Define
		\begin{equation}
			\tilde g_{\ell,\mathbb P}
			=\hat\varphi_\ell^{(1)}
			\{\hat m_\ell-\hat r_{1,\ell}\theta_{0,1}(\mathbb P)
			-\hat r_{-1,\ell}'\bar\theta_{-1,\mathbb P}\}.
			\label{eq:app_no_profile_score}
		\end{equation}
		
		\begin{assumption}[First-step and projection rates]
			\label{ass:app_score_rates}
			Uniformly over \(\mathbb P\in\mathcal P_n^+\) and folds, in the sense of
			\eqref{eq:ss_uniform_Op}, and conditionally on \(\mathcal F_\ell\),
			\begin{align}
				&\|\hat\varphi_\ell^{(1)}-\varphi_{\mathbb P}^{(1)}\|_{\mathbb P,4}
				+\|\hat m_\ell-m_{0,\mathbb P}\|_{\mathbb P,4}
				+\|\hat r_\ell-r_{0,\mathbb P}\|_{\mathbb P,4}=O_p(\delta_n),
				\label{eq:app_component_rates}\\
				&\|\hat p_\ell-p_{0,\mathbb P}\|_{\mathbb P,4}
				+\|d_{q,\ell,\mathbb P}(T)\|_{\mathbb P,4}
				+\left\|\sup_{s\in[T,\hat p_\ell(X)]}|d'_{q,\ell,\mathbb P}(s)|
				\right\|_{\mathbb P,4}=O_p(\delta_n),
				\label{eq:app_q_rates}
			\end{align}
			where \(d_{q,\ell,\mathbb P}(s)=\hat h_{Y,\ell}(s)
			-s\hat h_{X,\ell}(s)'\bar\theta_{\mathbb P}-\bar q_{\mathbb P}(s)\).
			Also \(\bar q_{\mathbb P}''\) is uniformly bounded,
			\(\|v_{\mathbb P}\|_{\mathbb P,4}\le C\),
			\(\|r_{0,-1,\mathbb P}\|_{\mathbb P,4}\le C\),
			\(\|\varphi_{\mathbb P}^{(1)}\|_{\mathbb P,4}\le C\nu_{\mathbb P}\),
			\(\|g_{\mathbb P}^*\|_{\mathbb P,4}\le C\nu_{\mathbb P}\), and
			\(\|\varphi_{\mathbb P}^{(1)}r_{0,-1,\mathbb P}'\|_{\mathbb P,2}
			\le C\nu_{\mathbb P}\).
		\end{assumption}

\subsubsection{Primitive verification of Assumption~\ref{ass:app_score_rates}}

\paragraph{First-step error identities.}
The following lemma contains the main algebra. It is useful because it separates the
statistical problem of estimating $p_0,h_{0X},h_{0Y}$ and their derivatives from the
deterministic way in which those errors enter the debiased score.

\begin{lemma}[Exact componentwise decompositions]
  \label{lem:app_exact_decompositions}
Fix a fold and suppress its index. Write
\[
  t=\hat p(X),\qquad \Delta_p=t-T,
\]
and, for $j\in\{X,Y\}$, define
\[
  H_{j,0}=\hat h_j(t)-h_{0j}(T),
  \qquad
  H_{j,1}=\widehat{\dot h}_j(t)-\dot h_{0j}(T).
\]
Then
\begin{align}
  \hat X-\tilde X&=-H_{X,0},
  \label{eq:app_X_error_exact}\\
  \hat m-m_0
  &=-H_{Y,0}-H_{Y,1}(D-T)
    +\widehat{\dot h}_Y(t)\Delta_p,
  \label{eq:app_m_error_exact}\\
  \hat r-r_0
  &=-D H_{X,0}
    -\{tH_{X,1}+\Delta_p\dot h_{0X}(T)\}(D-T)
    +t\widehat{\dot h}_X(t)\Delta_p,
  \label{eq:app_r_error_exact}\\
  \hat\varphi^{(1)}-\varphi^{(1)}
  &=-H_{X,0,1}+b^{*\prime}H_{X,0,-1}
    -(\hat c-b^*)'\hat X_{-1}.
  \label{eq:app_phi_error_exact}
\end{align}
Moreover, for every $s$ in the common domain of the regression functions,
\begin{align}
  d_q(s)
  &=e_Y(s)-s e_X(s)'\bar\theta,
  \label{eq:app_dq_exact}\\
  d_q'(s)
  &=\dot e_Y(s)-e_X(s)'\bar\theta
    -s\dot e_X(s)'\bar\theta,
  \label{eq:app_dq_derivative_exact}
\end{align}
where
\[
  e_j(s)=\hat h_j(s)-h_{0j}(s),
  \qquad
  \dot e_j(s)=\widehat{\dot h}_j(s)-\dot h_{0j}(s).
\]
\end{lemma}

\begin{proof}
The first identity is immediate. For the next two, substitute
$D-t=(D-T)-\Delta_p$ and use
\[
 r_0=D\tilde X-T\dot h_{0X}(T)(D-T),\qquad
 \hat r=D\hat X-t\widehat{\dot h}_X(t)(D-t).
\]
After subtraction, the identity
$t\widehat{\dot h}_X(t)-T\dot h_{0X}(T)
=tH_{X,1}+\Delta_p\dot h_{0X}(T)$ gives
\eqref{eq:app_m_error_exact}--\eqref{eq:app_r_error_exact}. Adding and subtracting
$b^{*\prime}\hat X_{-1}$ gives \eqref{eq:app_phi_error_exact}. Finally,
\[
 \bar q(s)=q_0(s)-s h_{0X}(s)'(\bar\theta-\theta_0)
 =h_{0Y}(s)-s h_{0X}(s)'\bar\theta;
\]
subtracting this equality from the definition of $\hat q_{\bar\theta}$ and
differentiating prove \eqref{eq:app_dq_exact}--\eqref{eq:app_dq_derivative_exact}.
\end{proof}

\paragraph{General sufficient conditions for the estimated-function rates.}

We next state conditions directly on general first-step estimators. Note they do not impose
a particular estimation method. The random-segment formulation below is important because
the regression estimators are evaluated at the generated regressor $\hat p(X)$ rather
than at $T=p_0(X)$.

For each fold, define the random segment
\[
  \mathcal I_\ell(X)
  =\bigl[T\wedge\hat T_\ell,T\vee\hat T_\ell\bigr]
\]
and the local error envelopes
\begin{align}
  E_{j,0,\ell}(X)
  &=\sup_{s\in\mathcal I_\ell(X)}
    \|\hat h_{j,\ell}(s)-h_{0j}(s)\|,
    \label{eq:app_level_envelope}\\
  E_{j,1,\ell}(X)
  &=\sup_{s\in\mathcal I_\ell(X)}
    \|\widehat{\dot h}_{j,\ell}(s)-\dot h_{0j}(s)\|,
    \qquad j\in\{X,Y\}.
    \label{eq:app_derivative_envelope}
\end{align}
For $j=Y$, the norm inside the supremum is absolute value.

\begin{assumption}[General primitive first-step conditions]
  \label{ass:app_general_firststep}
The following conditions hold uniformly over the relevant distribution class and
folds.
\begin{compactenum}[(i)]
  \item With probability approaching one, $T$ and $\hat T_\ell$ belong to a common
  compact interval $\mathcal I^\circ$ contained in $(0,1)$. On a neighborhood of that
  interval, $h_{0X}$ and $h_{0Y}$ are twice continuously differentiable and their first
  two derivatives are uniformly bounded.

  \item For a deterministic sequence $\delta_n\downarrow0$,
  \begin{equation}
    \|\hat T_\ell-T\|_{\mathbb P,4}
    +\sum_{j\in\{X,Y\}}\sum_{k=0}^1
      \|E_{j,k,\ell}\|_{\mathbb P,4}
    =O_p(\delta_n).
    \label{eq:app_primitive_regression_rate}
  \end{equation}

  \item The population and estimated projection coefficients satisfy
  \begin{equation}
    \|b^*\|\le C, \qquad
    \|\hat c_\ell-b^*\|=O_p(\delta_n).
    \label{eq:app_primitive_c_rate}
  \end{equation}

  \item $\|X\|_{\mathbb P,4}\le C$ and $\|\bar\theta_{\mathbb P}\|\le C$.
\end{compactenum}
\end{assumption}

 Assumption
\ref{ass:app_general_firststep} does not require uniform convergence of the regression
estimators on the entire index support. It requires only local control on the random
segment joining the true and estimated indices. This is exactly the region entered by
the generated-regressor expansion.

\begin{proposition}[Verification of the estimated-function part of Assumption~\ref{ass:app_score_rates}]
  \label{prop:app_ass7_estimated}
Under Assumption~\ref{ass:app_general_firststep},
\eqref{eq:app_component_rates} and
\eqref{eq:app_q_rates} hold.
\end{proposition}

\begin{proof}
Adding and subtracting $\partial_t^r h_{0j}(\hat T_\ell)$,
the mean-value theorem gives, for $r=0,1$,
\[
\|H_{j,r,\ell}\|_{\mathbb P,4}
\le\|E_{j,r,\ell}\|_{\mathbb P,4}
+C\|\hat T_\ell-T\|_{\mathbb P,4}=O_p(\delta_n).
\]
Conditional Jensen implies
$\|\tilde X\|_{\mathbb P,4}\le2\|X\|_{\mathbb P,4}$,
so $\|\hat X_\ell\|_{\mathbb P,4}=O_p(1)$.
Because $D,T,\hat T_\ell\in[0,1]$ and the true derivatives are
bounded, the exact identities in
Lemma~\ref{lem:app_exact_decompositions} give the stated
$m$ and $r$ rates. For the instrument,
\[
\|\hat\varphi_\ell^{(1)}-\varphi^{(1)}\|_{\mathbb P,4}
\le(1+\|b^*\|)\|H_{X,0,\ell}\|_{\mathbb P,4}
+\|\hat c_\ell-b^*\|\|\hat X_{-1,\ell}\|_{\mathbb P,4}
=O_p(\delta_n).
\]
This proves \eqref{eq:app_component_rates}.
Finally, \eqref{eq:app_dq_exact}--\eqref{eq:app_dq_derivative_exact}
bound $|d_q(T)|$ by $E_{Y,0,\ell}+CE_{X,0,\ell}$ and the segment
supremum of $|d_q'|$ by
$E_{Y,1,\ell}+C(E_{X,0,\ell}+E_{X,1,\ell})$.
Taking $L_4$ norms proves \eqref{eq:app_q_rates}.
\end{proof}
\paragraph{Obtaining the projection-coefficient rate.}

Condition~\eqref{eq:app_primitive_c_rate} is a rate for a finite-dimensional
coefficient, but it is not automatic when a raw Moore--Penrose inverse is used. The
next result supplies it under a spectrally stable nuisance block. This result is
estimator-agnostic: the first steps enter only through rates for two sample matrices.

For a calibration sample $J_\ell$, define
\begin{equation}
  \hat M_\ell
  :=|J_\ell|^{-1}\sum_{i\in J_\ell}
    \hat X_{-1,i\ell}\hat r_{-1,i\ell}',
  \qquad
  \hat d_\ell
  :=|J_\ell|^{-1}\sum_{i\in J_\ell}
    \hat r_{-1,i\ell}\hat X_{1,i\ell},
  \label{eq:app_Md_hat}
\end{equation}
and $\hat c_\ell=(\hat M_\ell')_{\rho_n}^\dagger\hat d_\ell$.
Their population limits are \(M=\Sigma_{-1,-1}\) and \(d=\mathbb E[T\tilde X_{-1}\tilde X_1]\). Indeed,
using $r_0=T\tilde X+B_0(X)(D-T)$ and
$\mathbb E[D-T\mid X]=0$ gives
\begin{align*}
  \mathbb E[\tilde X_{-1}r_{0,-1}']
  &=\mathbb E[T\tilde X_{-1}\tilde X_{-1}']=M,\\
  \mathbb E[r_{0,-1}\tilde X_1]
  &=\mathbb E[T\tilde X_{-1}\tilde X_1]=d.
\end{align*}

\begin{lemma}[Stable nuisance-block coefficient]
\label{lem:app_c_rate}
Write $\lambda_{\min}^+(M)$ for the smallest positive eigenvalue,
with $\lambda_{\min}^+(0)=+\infty$. Suppose
\[
\lambda_{\min}^+(M)\ge\mu_n>0,
\qquad \|b^*\|\le C,
\]
uniformly over distributions
$\mathcal P_{n,M}^+\subseteq\mathcal P_n^+$, and
\begin{equation}
\|\hat M_\ell-M\|+\|\hat d_\ell-d\|=O_p(q_{c,n}),
\qquad q_{c,n}\ll\rho_n\ll\mu_n,
\label{eq:app_matrix_rate}
\end{equation}
uniformly over distributions and folds. Then thresholding retains
exactly $\operatorname{rank}(M)$ singular values with probability
approaching one, and
\begin{equation}
\|\hat c_\ell-b^*\|
=O_p\left(\frac{q_{c,n}}{\mu_n}\right).
\label{eq:app_c_perturbation_rate}
\end{equation}
Consequently, \eqref{eq:app_primitive_c_rate} holds whenever
$q_{c,n}/\mu_n=O(\delta_n)$.
\end{lemma}

\begin{proof}
If $M=0$, then $d=b^*=0$, and thresholding gives
$\hat c_\ell=0$ with probability approaching one.
For $M\ne0$, put $A_\ell=\hat M_\ell'$, $P=MM^\dagger$, and
$\hat P_\ell=(A_\ell)_{\rho_n}^\dagger A_\ell$.
Weyl's inequality and $q_{c,n}\ll\rho_n\ll\mu_n$
give rank recovery and
$\|(A_\ell)_{\rho_n}^\dagger\|=O_p(\mu_n^{-1})$.
Wedin's bound gives
$\|\hat P_\ell-P\|=O_p(q_{c,n}/\mu_n)$.
Since $d=M'b^*$ and $Pb^*=b^*$,
\[
\hat c_\ell-b^*
=(A_\ell)_{\rho_n}^\dagger
\{(\hat d_\ell-d)-(A_\ell-M')b^*\}
+(\hat P_\ell-P)b^*.
\]
Taking norms proves \eqref{eq:app_c_perturbation_rate}.
The final claim follows immediately.
\end{proof}
The matrix rate in Lemma~\ref{lem:app_c_rate} is itself easy to obtain with a fresh
calibration sample.

\begin{lemma}[Fresh-sample matrix rate]
  \label{lem:app_fresh_matrix}
Suppose $J_\ell$ is independent of the data used to construct the first-step
estimators, as with nested splitting, $|J_\ell|\asymp n$, and
\begin{align*}
  \|\hat X_\ell-\tilde X\|_{\mathbb P,4}
  +\|\hat r_\ell-r_0\|_{\mathbb P,4}&=O_p(\delta_n),\\
  \|\hat X_\ell\|_{\mathbb P,4}
  +\|\hat r_\ell\|_{\mathbb P,4}
  +\|\tilde X\|_{\mathbb P,4}
  +\|r_0\|_{\mathbb P,4}&=O_p(1).
\end{align*}
Then
\begin{equation}
  \|\hat M_\ell-M\|+\|\hat d_\ell-d\|
  =O_p(\delta_n+n^{-1/2}).
  \label{eq:app_fresh_matrix_rate}
\end{equation}
\end{lemma}

\begin{proof}
Condition on the first-step training sample and set
$M_\ell^\circ=\mathbb E[\hat X_{-1,\ell}\hat r_{-1,\ell}'
\mid\mathcal F_\ell]$. The product identity
\[
\hat X_{-1,\ell}\hat r_{-1,\ell}'-\tilde X_{-1}r_{0,-1}'
=(\hat X_{-1,\ell}-\tilde X_{-1})\hat r_{-1,\ell}'
+\tilde X_{-1}(\hat r_{-1,\ell}-r_{0,-1})'
\]
and Cauchy--Schwarz give $\|M_\ell^\circ-M\|=O_p(\delta_n)$.
The $L_4$ bounds give bounded conditional second moments for each
product entry. Independence, $|J_\ell|\asymp n$, and fixed dimension
therefore imply
$\|\hat M_\ell-M_\ell^\circ\|=O_p(n^{-1/2})$
by conditional Chebyshev. The same argument applies to $\hat d_\ell$.
\end{proof}

\begin{remark}[Why a stability condition is needed]
If $M$ is singular and $\hat c_\ell$ uses an unthresholded sample pseudoinverse,
\eqref{eq:app_primitive_c_rate} does not follow from
$\hat M_\ell\to M$ and $\hat d_\ell\to d$ alone. A sample matrix can be full rank even
when its population limit is singular, and inversion of the spurious small singular
values can make $\hat c_\ell$ unstable. Thresholding removes these spurious directions.
Lemma~\ref{lem:app_c_rate} requires separation only of the positive
spectrum, so the subclass $\mathcal P_{n,M}^+\subseteq\mathcal P_n^+$
permits singular nuisance blocks.
\end{remark}

\paragraph{Moment and boundedness conditions.}

We now verify the remaining clauses of Assumption~\ref{ass:app_score_rates}. Lemma~\ref{lem:ss_canonical_representative} gives
\begin{equation}
	\Gamma=\Sigma_{\tilde X}\theta_0,
	\qquad
	\bar\theta=\Pi_{\operatorname{Range}(\Sigma_{\tilde X})}\theta_0,
	\qquad
	\bar\Delta\in\ker(\Sigma_{\tilde X}),
	\qquad
	\tilde X'\bar\Delta=0\quad\text{a.s.}
	\label{eq:app_canonical_identities}
\end{equation}
If $\nu>0$, then $\bar\Delta_1=0$ and differencing $\bar q(t)=q_0(t)-t h_{0X}(t)'\bar\Delta$ gives
\begin{equation}
	\label{eq:app_qbar_second}
	  \bar q''(t)
	=q_0''(t)
	-\{2\dot h_{0X}(t)+t\ddot h_{0X}(t)\}'\bar\Delta.
\end{equation}
Also, note that, for $v$ in \eqref{eq:ss_corrected_oracle},
\begin{align}
	v
	&=\varepsilon_2-\{A_0(X)+B_0(X)'\bar\Delta\}\varepsilon_1
	=\varepsilon_2-\{A_0(X)+B_{0,-1}(X)'\bar\Delta_{-1}\}\varepsilon_1.
	\label{eq:app_v_identity}
\end{align}

\begin{assumption}[Population moment conditions]
  \label{ass:app_population_moments}
Uniformly over the population class:
\begin{compactenum}[(i)]
  \item $\|\theta_0\|\le C$, $\mathbb E[|Y|^4+\|X\|^4]\le C$, and the functions
  $q_0,h_{0X}$ have the uniformly bounded derivatives appearing in
  \eqref{eq:app_qbar_second}.

  \item The standardized target residual has bounded kurtosis:
  \begin{equation}
    \mathbb E\left[\left|\frac{\varphi^{(1)}}{\nu}\right|^4\right]\le C.
    \label{eq:app_phi_kurtosis}
  \end{equation}

  \item The oracle residual and orthogonalized nuisance regressor satisfy
  \begin{equation}
    \mathbb E[|v|^4\mid X]\le C,
    \qquad
    \mathbb E[\|r_{0,-1}\|^4\mid X]\le C
    \quad\text{a.s.}
    \label{eq:app_conditional_moments}
  \end{equation}
\end{compactenum}
\end{assumption}

\begin{proposition}[Verification of the population part of Assumption~\ref{ass:app_score_rates}]
  \label{prop:app_ass7_population}
Under Assumption~\ref{ass:app_population_moments},
\begin{equation}
  \sup_t|\bar q''(t)|\le C,
  \qquad
  \|v\|_{\mathbb P,4}+\|r_{0,-1}\|_{\mathbb P,4}\le C,
  \label{eq:app_population_unscaled}
\end{equation}
and
\begin{equation}
  \|\varphi^{(1)}\|_{\mathbb P,4}\le C\nu,
  \qquad
  \|g^*\|_{\mathbb P,4}\le C\nu,
  \qquad
  \|\varphi^{(1)}r_{0,-1}'\|_{\mathbb P,2}\le C\nu.
  \label{eq:app_population_scaled}
\end{equation}
Thus all the population moment and boundedness clauses of Assumption~\ref{ass:app_score_rates} hold.
\end{proposition}

\begin{proof}
The projection identity gives
$\|\bar\theta\|\le\|\theta_0\|$ and
$\|\bar\Delta\|\le2\|\theta_0\|\le2C$.
Equation~\eqref{eq:app_qbar_second} therefore bounds $\bar q''$,
while \eqref{eq:app_conditional_moments} bounds the unscaled
fourth moments. Equation~\eqref{eq:app_phi_kurtosis} gives
$\|\varphi^{(1)}\|_{\mathbb P,4}\le C\nu$.
Since $\varphi^{(1)}$ is $X$-measurable,
\[
\mathbb E[|g^*|^4]
\le C\mathbb E[|\varphi^{(1)}|^4]\le C\nu^4,
\qquad
\mathbb E[\{\varphi^{(1)}\}^2\|r_{0,-1}\|^2]
\le C\mathbb E[\{\varphi^{(1)}\}^2]=C\nu^2.
\]
These inequalities follow by conditioning on $X$ and applying
\eqref{eq:app_conditional_moments}; taking roots proves
\eqref{eq:app_population_scaled}.
\end{proof}
\begin{remark}[A weaker but less primitive alternative]
Assumption~\ref{ass:app_population_moments}\textup{(iii)} is convenient rather than
necessary. It can be replaced directly by the relevance-scaled conditions
\[
  \|g^*\|_{\mathbb P,4}\le C\nu,
  \qquad
  \|\varphi^{(1)}r_{0,-1}'\|_{\mathbb P,2}\le C\nu,
\]
together with bounded unscaled fourth moments of $v$ and $r_{0,-1}$. The conditional
form in \eqref{eq:app_conditional_moments} is useful because it shows why the scaling
by $\nu$ is natural: all shrinking behavior comes from the target residual
$\varphi^{(1)}$, while the conditional moments of the remaining score factors stay
bounded.
\end{remark}

The general verification is collected in
Corollary~\ref{cor:app_ass7} below.
\subsubsection{From Assumption~\ref{ass:app_score_rates} to the score conditions}

For a fixed fold, define
\begin{equation}
  \hat v_\ell
  :=\hat m_\ell-\hat r_{1,\ell}\theta_{0,1}(\mathbb P)
    -\hat r_{-1,\ell}'\bar\theta_{-1,\mathbb P},
  \qquad
  \tilde g_{\ell,\mathbb P}
  :=\hat\varphi_\ell^{(1)}\hat v_\ell.
  \label{eq:app_tilde_g}
\end{equation}
Because positive target relevance implies
$\bar\theta_{1,\mathbb P}=\theta_{0,1}(\mathbb P)$, we may equivalently write
\[
  \hat v_\ell=\hat m_\ell-\hat r_\ell'\bar\theta_{\mathbb P}.
\]

\begin{lemma}[First-step score]
  \label{lem:app_firststep_score}
  \label{lem:app_score_rates}
Suppose Assumption~\ref{ass:app_score_rates} holds, $\nu_{\mathbb P}\ge\nu_n$, and
$\delta_n=o(\nu_n)$. Uniformly over the population class and folds,
\begin{align}
  \|\tilde g_{\ell,\mathbb P}-g_{\mathbb P}^*\|_{\mathbb P,2}
  &=O_p(\delta_n),
  &
  \left|\mathbb E_{\mathbb P}
  [\tilde g_{\ell,\mathbb P}-g_{\mathbb P}^*\mid\mathcal F_\ell]\right|
  &=O_p(\delta_n^2),
  \label{eq:app_score_rates}\\
  \|\hat\varphi_\ell^{(1)}\hat r_{-1,\ell}'\|_{\mathbb P,2}
  &=O_p(\nu_{\mathbb P}),
  &
  n^{-1/2}\sum_\ell\sum_{i\in I_\ell}
  (\tilde g_{i\ell,\mathbb P}-g_{\mathbb P,i}^*)
  &=O_p(\delta_n+\sqrt n\,\delta_n^2).
  \label{eq:app_crossfit_score_rate}
\end{align}
\end{lemma}

\begin{proof}
Fix a fold, condition on $\mathcal F_\ell$, and suppress the fold and
$\mathbb P$ subscripts. All bounds below are uniform. Write
$t=\hat p(X)$,
$\hat q_{\bar\theta}(s)=\hat h_Y(s)-s\hat h_X(s)'\bar\theta$,
and $d_q=\hat q_{\bar\theta}-\bar q$.
Since $\nu>0$, Lemma~\ref{lem:ss_canonical_representative} gives
$\bar\theta_1=\theta_{0,1}$, so $\hat v=\hat m-\hat r'\bar\theta$.

\emph{Conditional mean.}
Collecting terms in the feasible residual gives
\[
\hat v
=Y-DX'\bar\theta-\hat q_{\bar\theta}(t)
-(D-t)\hat q_{\bar\theta}'(t).
\]
The canonical representation therefore implies
\begin{equation}
a_\ell(X):=\mathbb E[\hat v\mid X,\mathcal F_\ell]
=\bar q(T)-\hat q_{\bar\theta}(t)
-(T-t)\hat q_{\bar\theta}'(t).
\label{eq:app_conditional_score_identity}
\end{equation}
Write $a_\ell=R_{\bar q}-d_q(T)+R_{d_q}$, where
$R_f=f(T)-f(t)-(T-t)f'(t)$.
Taylor's theorem and the fundamental theorem of calculus give
\begin{align}
|R_{\bar q}|&\le C|T-t|^2,
\label{eq:app_qbar_remainder}\\
|R_{d_q}|&\le
2|T-t|\sup_{s\in[T,t]}|d_q'(s)|.
\label{eq:app_dq_remainder}
\end{align}
Here $[T,t]$ denotes the segment joining the two endpoints.
Thus no estimated second derivative is needed.
Because $\mathbb E[\varphi^{(1)}\mid T]=0$, the term
$\varphi^{(1)}d_q(T)$ has zero conditional expectation.
H\"older's inequality and Assumption~\ref{ass:app_score_rates} yield
\begin{align}
|\mathbb E[\varphi^{(1)}a_\ell\mid\mathcal F_\ell]|
&=O_p(\nu\delta_n^2),
\label{eq:app_phi_conditional_mean}\\
\|a_\ell\|_{\mathbb P,2}
&\le\|d_q(T)\|_{\mathbb P,2}
+C\|t-T\|_{\mathbb P,4}^2
+2\|t-T\|_{\mathbb P,4}
 \left\|\sup_{s\in[T,t]}|d_q'(s)|\right\|_{\mathbb P,4}
=O_p(\delta_n).
\label{eq:app_a_l2}
\end{align}
For $e_\varphi=\hat\varphi^{(1)}-\varphi^{(1)}$, iterated
expectations then give
\begin{equation}
|\mathbb E[e_\varphi\hat v\mid\mathcal F_\ell]|
=|\mathbb E[e_\varphi a_\ell\mid\mathcal F_\ell]|
\le\|e_\varphi\|_{\mathbb P,2}\|a_\ell\|_{\mathbb P,2}
=O_p(\delta_n^2).
\label{eq:app_ephi_mean}
\end{equation}
Since $\mathbb E[g^*]=0$ and $\nu$ is uniformly bounded,
\eqref{eq:app_phi_conditional_mean} and \eqref{eq:app_ephi_mean}
prove the conditional-mean assertion in \eqref{eq:app_score_rates}.

\emph{Second moments.}
The component rates and bounded $\|\bar\theta\|$ give
\begin{equation}
\|\hat v-v\|_{\mathbb P,4}
\le\|\hat m-m_0\|_{\mathbb P,4}
+\|\bar\theta\|\|\hat r-r_0\|_{\mathbb P,4}
=O_p(\delta_n).
\label{eq:app_vhat_rate}
\end{equation}
Using
\begin{equation}
\tilde g-g^*=e_\varphi v+\hat\varphi^{(1)}(\hat v-v),
\label{eq:app_score_product_decomp}
\end{equation}
and $\|\hat\varphi^{(1)}\|_{\mathbb P,4}
=O_p(\nu+\delta_n)=O_p(1)$, H\"older gives
$\|\tilde g-g^*\|_{\mathbb P,2}=O_p(\delta_n)$.

Similarly, with $e_r=\hat r_{-1}-r_{0,-1}$,
\[
\hat\varphi^{(1)}\hat r_{-1}'
=\varphi^{(1)}r_{0,-1}'
+e_\varphi r_{0,-1}'
+\varphi^{(1)}e_r'+e_\varphi e_r'.
\]
The population product bound and H\"older imply
$\|\hat\varphi^{(1)}\hat r_{-1}'\|_{\mathbb P,2}
=O_p(\nu+\delta_n+\nu\delta_n+\delta_n^2)=O_p(\nu)$,
because $\delta_n=o(\nu_n)$ and $\nu\ge\nu_n$.

\emph{Cross-fitted sum.}
Restore fold indices and put
$D_{i\ell}=\tilde g_{i\ell,\mathbb P}-g_{\mathbb P,i}^*$ and
$\mu_\ell=\mathbb E[D_{i\ell}\mid\mathcal F_\ell]$.
Conditional independence gives
\[
\operatorname{Var}\left(
n^{-1/2}\sum_{i\in I_\ell}(D_{i\ell}-\mu_\ell)
\biggm|\mathcal F_\ell\right)
\le\frac{n_\ell}{n}\|D_{i\ell}\|_{\mathbb P,2}^2
=O_p(\delta_n^2).
\]
Hence each centered fold sum is $O_p(\delta_n)$, while its mean
$n_\ell\mu_\ell/\sqrt n$ is $O_p(\sqrt n\,\delta_n^2)$.
Summing over the fixed number of folds proves
\eqref{eq:app_crossfit_score_rate}.
\end{proof}
Lemma~\ref{lem:app_firststep_score} establishes (U2) and the
second condition in (U4). We next verify the matrix-estimation
and projection-bias requirements.
\subsubsection{The matrix-estimation condition in (U3)}

We next verify only the matrix-rate requirement in (U3). The threshold separation
$q_n\ll\rho_n\ll\lambda_n$ is not derived here and remains a maintained assumption.
For a calibration sample $J_\ell$, let
\[
  \mathbb P_{J,\ell}f
  :=|J_\ell|^{-1}\sum_{i\in J_\ell}f(Z_i)
\]
and define
\begin{equation}
  \hat\Sigma_{-\ell}
  :=\mathbb P_{J,\ell}(\hat X_\ell\hat r_\ell'),
  \qquad
  \hat\Gamma_{-\ell}
  :=\mathbb P_{J,\ell}(\hat X_\ell\hat m_\ell).
  \label{eq:app_Sigma_Gamma_hat}
\end{equation}
The corresponding population quantities are
\begin{equation}
  \Sigma_{\tilde X,\mathbb P}
  =\mathbb E_{\mathbb P}[\tilde Xr_0']
  =\mathbb E_{\mathbb P}[T\tilde X\tilde X'],
  \qquad
  \Gamma_{\mathbb P}=\mathbb E_{\mathbb P}[\tilde Xm_0].
  \label{eq:app_Sigma_Gamma_population}
\end{equation}

The proof has the same two components as Lemma~\ref{lem:app_fresh_matrix}. It is
useful first to state the decomposition in a form that does not impose a particular
estimation method or sample-splitting scheme.

\begin{lemma}[General decomposition of the matrix rate]
  \label{lem:app_general_U3_matrix}
Suppose, uniformly over the population class and folds,
\begin{align}
  &\|\hat X_\ell-\tilde X\|_{\mathbb P,4}
   +\|\hat r_\ell-r_0\|_{\mathbb P,4}
   +\|\hat m_\ell-m_0\|_{\mathbb P,4}
   =O_p(\delta_n),
  \label{eq:app_U3_component_rates}\\
  &\|\hat X_\ell\|_{\mathbb P,4}
   +\|\tilde X\|_{\mathbb P,4}
   +\|\hat r_\ell\|_{\mathbb P,4}
   +\|r_0\|_{\mathbb P,4}
   +\|\hat m_\ell\|_{\mathbb P,4}
   +\|m_0\|_{\mathbb P,4}
   =O_p(1).
  \label{eq:app_U3_moment_bounds}
\end{align}
If, for a deterministic sequence $\omega_n\to0$,
\begin{align}
  &\|\mathbb P_{J,\ell}(\hat X_\ell\hat r_\ell')
       -\mathbb E_{\mathbb P}[\hat X_\ell\hat r_\ell'
         \mid\mathcal F_\ell]\|\nonumber\\
  &\quad+
   \|\mathbb P_{J,\ell}(\hat X_\ell\hat m_\ell)
       -\mathbb E_{\mathbb P}[\hat X_\ell\hat m_\ell
         \mid\mathcal F_\ell]\|
   =O_p(\omega_n),
  \label{eq:app_U3_empirical_rate}
\end{align}
then
\begin{equation}
  \|\hat\Sigma_{-\ell}-\Sigma_{\tilde X,\mathbb P}\|
  +\|\hat\Gamma_{-\ell}-\Gamma_{\mathbb P}\|
  =O_p(\delta_n+\omega_n).
  \label{eq:app_U3_general_rate}
\end{equation}
\end{lemma}

\begin{proof}
Let
$\Sigma_\ell^\circ
=\mathbb E_{\mathbb P}[\hat X_\ell\hat r_\ell'\mid\mathcal F_\ell]$
and
$\Gamma_\ell^\circ
=\mathbb E_{\mathbb P}[\hat X_\ell\hat m_\ell\mid\mathcal F_\ell]$.
Then
\begin{align}
\hat\Sigma_{-\ell}-\Sigma_{\tilde X,\mathbb P}
&=(\hat\Sigma_{-\ell}-\Sigma_\ell^\circ)
 +(\Sigma_\ell^\circ-\Sigma_{\tilde X,\mathbb P}),
\label{eq:app_U3_Sigma_decomp}\\
\hat\Gamma_{-\ell}-\Gamma_{\mathbb P}
&=(\hat\Gamma_{-\ell}-\Gamma_\ell^\circ)
 +(\Gamma_\ell^\circ-\Gamma_{\mathbb P}).
\label{eq:app_U3_Gamma_decomp}
\end{align}
The first terms are $O_p(\omega_n)$ by
\eqref{eq:app_U3_empirical_rate}.
For either product, use
$\hat X\hat u'-\tilde Xu_0'
=(\hat X-\tilde X)\hat u'+\tilde X(\hat u-u_0)'$,
with $u=r$ or $u=m$. Conditional Cauchy--Schwarz gives
\begin{align}
\|\Sigma_\ell^\circ-\Sigma_{\tilde X,\mathbb P}\|_F
&\le
\|\hat X_\ell-\tilde X\|_{\mathbb P,2}
\|\hat r_\ell\|_{\mathbb P,2}
+\|\tilde X\|_{\mathbb P,2}
\|\hat r_\ell-r_0\|_{\mathbb P,2}
=O_p(\delta_n),
\label{eq:app_U3_Sigma_bias}\\
\|\Gamma_\ell^\circ-\Gamma_{\mathbb P}\|
&\le
\|\hat X_\ell-\tilde X\|_{\mathbb P,2}
\|\hat m_\ell\|_{\mathbb P,2}
+\|\tilde X\|_{\mathbb P,2}
\|\hat m_\ell-m_0\|_{\mathbb P,2}
=O_p(\delta_n).
\label{eq:app_U3_Gamma_bias}
\end{align}
Here $\|\cdot\|_F$ is the Frobenius norm; the bounds follow from
$\|ab'\|_F=\|a\|\|b\|$ and $\|\cdot\|_{\mathbb P,2}
\le\|\cdot\|_{\mathbb P,4}$.
Combining the bounds proves \eqref{eq:app_U3_general_rate}.
Fixed dimension makes the Frobenius and operator rates equivalent.
\end{proof}
Condition~\eqref{eq:app_U3_empirical_rate} isolates the only part that depends on
how the calibration observations are related to the training sample. Under nested
splitting it follows directly from conditional Chebyshev's inequality.

\begin{lemma}[Fresh-calibration matrix rate]
  \label{lem:app_fresh_U3_matrix}
Suppose $J_\ell$ is independent of the observations used to construct the first-step
estimators as with nested splitting, $|J_\ell|\asymp n$, and
\eqref{eq:app_U3_moment_bounds} holds. Then
\eqref{eq:app_U3_empirical_rate} holds with $\omega_n=n^{-1/2}$. Consequently,
under \eqref{eq:app_U3_component_rates},
\begin{equation}
  \|\hat\Sigma_{-\ell}-\Sigma_{\tilde X,\mathbb P}\|
  +\|\hat\Gamma_{-\ell}-\Gamma_{\mathbb P}\|
  =O_p(\delta_n+n^{-1/2}).
  \label{eq:app_U3_fresh_rate}
\end{equation}
\end{lemma}

\begin{proof}
Let $\mathcal H_\ell$ be the sigma-field generated by the
observations used to construct the first-step estimators,
excluding the calibration sample $J_\ell$.
Conditional on $\mathcal H_\ell$, the fitted functions
are fixed and the calibration observations are iid.
For any scalar entry $W(Z)$ of
$\hat X_\ell(Z)\hat r_\ell(Z)'$ or
$\hat X_\ell(Z)\hat m_\ell(Z)$, evaluated at an independent
draw $Z$ from $\mathbb P$, H\"older and
\eqref{eq:app_U3_moment_bounds} give
$\mathbb E_{\mathbb P}[W(Z)^2\mid\mathcal H_\ell]=O_p(1)$.
Hence
\[
\operatorname{Var}_{\mathbb P}
(\mathbb P_{J,\ell}W\mid\mathcal H_\ell)
=
|J_\ell|^{-1}\operatorname{Var}_{\mathbb P}
(W(Z)\mid\mathcal H_\ell)
=O_p(n^{-1}).
\]
Conditional Chebyshev and fixed dimension imply
\begin{align}
\|\hat\Sigma_{-\ell}-\Sigma_\ell^\circ\|
&=O_p(n^{-1/2}),
\label{eq:app_U3_Sigma_sampling}\\
\|\hat\Gamma_{-\ell}-\Gamma_\ell^\circ\|
&=O_p(n^{-1/2}),
\label{eq:app_U3_Gamma_sampling}
\end{align}
where $\Sigma_\ell^\circ$ and $\Gamma_\ell^\circ$ are the
conditional expectations defined in the preceding proof.
For an independent draw, these expectations are unchanged
when conditioning on $\mathcal H_\ell$ rather than
$\mathcal F_\ell$, since the fitted functions are
$\mathcal H_\ell$-measurable. This verifies \eqref{eq:app_U3_empirical_rate};
Lemma~\ref{lem:app_general_U3_matrix} gives
\eqref{eq:app_U3_fresh_rate}.
\end{proof}
The preceding results verify the matrix rate; the spectral
separation $q_n\ll\rho_n\ll\lambda_n$ remains an additional
requirement.

\subsubsection{Projection bias in (U4)}

Lemma~\ref{lem:app_firststep_score} gives the second condition
in (U4). It remains to verify
\begin{equation}
\nu_{\mathbb P}^{-1}
\left\|\mathbb E_{\mathbb P}
[\hat\varphi_\ell^{(1)}\hat r_{-1,\ell}'\mid\mathcal F_\ell]\right\|
=O_p(\kappa_n).
\label{eq:app_U4_bias_target}
\end{equation}
Under nested splitting, calibration is fresh relative to the
nuisance-function estimates. Under simple cross-fitting,
we instead control the first-step errors in population norm.
\begin{lemma}[Verification of (U4)]
  \label{lem:app_U4_verification}
Suppose $|B_\ell|\asymp n$. Under nested splitting, condition on the nuisance-training
sample $H_\ell$ and define the vector-valued function
\[
  f_c(Z)
  :=\hat r_{-1,\ell}(Z)
  \{\hat X_{1,\ell}(Z)-\hat X_{-1,\ell}(Z)'c\}.
\]
Suppose that, with probability approaching one,
$\hat c_\ell\in\mathcal B_\ell$ for a conditionally fixed set
$\mathcal B_\ell$, and let
\[
  \mathcal C_\ell:=\{f_c:c\in\mathcal B_\ell\}.
\]
Suppose $\mathcal C_\ell$ has an envelope $F_\ell$ satisfying
\begin{equation}
  \|F_\ell\|_{\mathbb P,2}\le C\nu_{\mathbb P},
  \qquad
  \log N\{\epsilon\|F_\ell\|_{\mathbb P,2},
  \mathcal C_\ell,L_2(\mathbb P)\}
  \le C\log(C/\epsilon),
  \quad 0<\epsilon\le1,
  \label{eq:app_U4_entropy}
\end{equation}
and that the calibration residual satisfies
\begin{equation}
  \|\hat e_\ell\|=O_p(n^{-1/2}\nu_{\mathbb P}),
  \qquad
  \hat e_\ell
  :=\mathbb P_{B,\ell}
  [\hat r_{-1,\ell}\hat\varphi_\ell^{(1)}].
  \label{eq:app_U4_calibration_residual}
\end{equation}
Then (U4) holds with $\kappa_n=n^{-1/2}$.

Under simple cross-fitting, suppose the componentwise rates in
\eqref{eq:app_component_rates}, the population moment bounds in
\eqref{eq:app_population_unscaled}--\eqref{eq:app_population_scaled},
$\nu_{\mathbb P}\ge\nu_n$, and $\delta_n=o(\nu_n)$ hold. Then (U4) holds with
$\kappa_n=\delta_n/\nu_n$.
\end{lemma}

\begin{proof}
\emph{Nested splitting.}
Condition on the sigma-field $\mathcal H_\ell$ generated by the
nuisance-training sample $H_\ell$. The fitted functions are fixed
and $B_\ell$ consists of independent draws.
The entropy integral in \eqref{eq:app_U4_entropy} is bounded by
$C\int_0^1\sqrt{1+\log(C/\epsilon)}\,d\epsilon<\infty$.
The conditional maximal bound for the fixed-dimensional affine class
therefore gives
\begin{equation}
\sup_{c\in\mathcal B_\ell}
\|(\mathbb P_{B,\ell}-\mathbb P)f_c\|
=O_p\{\|F_\ell\|_{\mathbb P,2}/\sqrt{|B_\ell|}\}
=O_p(n^{-1/2}\nu_{\mathbb P}).
\label{eq:app_U4_maximal_rate}
\end{equation}
Here $\mathbb Pf_c$ is the fresh-draw expectation conditional on
$\mathcal H_\ell$. On the event $\hat c_\ell\in\mathcal B_\ell$,
$f_{\hat c_\ell}=\hat r_{-1,\ell}\hat\varphi_\ell^{(1)}$, so
\begin{equation}
\|\mathbb Pf_{\hat c_\ell}\|
\le\|\hat e_\ell\|
+\sup_{c\in\mathcal B_\ell}
 \|(\mathbb P_{B,\ell}-\mathbb P)f_c\|
=O_p(n^{-1/2}\nu_{\mathbb P}).
\label{eq:app_U4_nested_bias}
\end{equation}
After calibration, this fresh-draw expectation equals
$\mathbb E_{\mathbb P}
[\hat r_{-1,\ell}\hat\varphi_\ell^{(1)}\mid\mathcal F_\ell]$.
Dividing by $\nu_{\mathbb P}$ proves the bias condition.
The envelope also gives
$\|f_{\hat c_\ell}\|_{\mathbb P,2}=O_p(\nu_{\mathbb P})$,
proving the second condition in (U4).

With thresholded calibration, the residual is
\[
\hat e_\ell
=\{I-\hat M_\ell'(\hat M_\ell')_{\rho_n}^\dagger\}\hat d_\ell.
\]
It need not vanish; \eqref{eq:app_U4_calibration_residual}
controls the component discarded by thresholding.
\emph{Simple cross-fitting.}
Condition on $\mathcal F_\ell$ and write
$e_{\varphi,\ell}=\hat\varphi_\ell^{(1)}
-\varphi_{\mathbb P}^{(1)}$ and
$e_{r,\ell}=\hat r_{-1,\ell}-r_{0,-1,\mathbb P}$.
The population normal equations give
$\mathbb E_{\mathbb P}
[\varphi_{\mathbb P}^{(1)}r_{0,-1,\mathbb P}']=0$; hence
\begin{equation}
\begin{split}
&\mathbb E_{\mathbb P}
[\hat\varphi_\ell^{(1)}\hat r_{-1,\ell}'\mid\mathcal F_\ell]\\
&\quad=
\mathbb E_{\mathbb P}
[e_{\varphi,\ell}r_{0,-1,\mathbb P}'
+\varphi_{\mathbb P}^{(1)}e_{r,\ell}'
+e_{\varphi,\ell}e_{r,\ell}'\mid\mathcal F_\ell].
\end{split}
\label{eq:app_U4_simple_expansion}
\end{equation}
Cauchy--Schwarz bounds the three terms by
$O_p(\delta_n)$, $O_p(\nu_{\mathbb P}\delta_n)$, and
$O_p(\delta_n^2)$, respectively. Thus
\begin{equation}
\left\|\mathbb E_{\mathbb P}
[\hat\varphi_\ell^{(1)}\hat r_{-1,\ell}'\mid\mathcal F_\ell]\right\|
=O_p(\delta_n+\nu_{\mathbb P}\delta_n+\delta_n^2).
\label{eq:app_U4_simple_unscaled}
\end{equation}
Since $\nu_{\mathbb P}\ge\nu_n$, $\delta_n=o(\nu_n)$, and
$\nu_{\mathbb P}$ is uniformly bounded, division by
$\nu_{\mathbb P}$ gives $O_p(\delta_n/\nu_n)$.
For the second condition, the same product expansion, retaining
the population product, and H\"older give
\[
\|\hat\varphi_\ell^{(1)}\hat r_{-1,\ell}'\|_{\mathbb P,2}
=O_p(\nu_{\mathbb P}+\delta_n
+\nu_{\mathbb P}\delta_n+\delta_n^2)
=O_p(\nu_{\mathbb P}).
\]
This proves both parts of (U4).
\end{proof}

\begin{corollary}[General verification and rate map]
\label{cor:app_ass7}
\label{cor:app_U2_U4b}
\label{cor:app_U3_matrix}
\label{cor:app_complete_U4}
\label{rem:app_rate_map}
Suppose Assumptions~\ref{ass:app_general_firststep}
and~\ref{ass:app_population_moments} hold uniformly over
the population class and folds. Then
Assumption~\ref{ass:app_score_rates} holds with rate $\delta_n$.
Moreover:
\begin{compactenum}[(i)]
\item
With fresh calibration, the coefficient-rate requirement
in Assumption~\ref{ass:app_general_firststep}(iii) follows
from $\|b^*\|\le C$ and
\[
\lambda_{\min}^+(\Sigma_{-1,-1})\ge\mu_n>0,\qquad
\delta_n+n^{-1/2}\ll\rho_n\ll\mu_n,\qquad
\frac{\delta_n+n^{-1/2}}{\mu_n}=O(\delta_n).
\]
If $\inf_n\mu_n>0$, sufficient rate restrictions are
$n^{-1/2}=O(\delta_n)$ and $\delta_n\ll\rho_n\ll1$.

\item
If $\nu_{\mathbb P}\ge\nu_n$ and $\delta_n=o(\nu_n)$,
(U2) and the second condition in (U4) hold, and
\[
n^{-1/2}\sum_\ell\sum_{i\in I_\ell}
(\tilde g_{i\ell,\mathbb P}-g_{\mathbb P,i}^*)
=O_p(\delta_n+\sqrt n\,\delta_n^2).
\]

\item
The matrix part of (U3) holds with
$q_n=\delta_n+\omega_n$ under
\eqref{eq:app_U3_empirical_rate}.
Fresh calibration gives $\omega_n=n^{-1/2}$;
if also $n^{-1/2}=O(\delta_n)$, one may take $q_n=\delta_n$.
More generally, any deterministic $q_n\to0$ satisfying
$\delta_n+\omega_n=O(q_n)$ is valid.
The separation $q_n\ll\rho_n\ll\lambda_n$ remains an
additional requirement.

\item
Under the relevance and rate conditions in (ii),
both parts of (U4) hold with
$\kappa_n=\delta_n/\nu_n$ for simple cross-fitting,
and with $\kappa_n=n^{-1/2}$ under the nested-splitting
conditions of Lemma~\ref{lem:app_U4_verification}.
\end{compactenum}
The resulting uniform AR--NO restrictions are
\begin{equation}
\frac{\delta_n}{\nu_n}\to0,\qquad
\frac{\sqrt n\,\delta_n^2}{\nu_n}\to0,\qquad
a_n:=\frac{q_n}{\lambda_n}\to0,\qquad
(\sqrt n\,\kappa_n+1)a_n\to0.
\label{eq:app_uniform_rate_map}
\end{equation}
\end{corollary}

\begin{proof}
Propositions~\ref{prop:app_ass7_estimated}
and~\ref{prop:app_ass7_population} verify
Assumption~\ref{ass:app_score_rates}.
Lemmas~\ref{lem:app_c_rate}--\ref{lem:app_fresh_matrix}
give (i), and Lemma~\ref{lem:app_firststep_score} gives (ii).
For (iii), conditional Jensen bounds $\tilde X$ in $L_4$;
\eqref{eq:m0r0}, bounded derivatives, and the population
fourth moments bound $m_0$ and $r_0$.
The component rates give the corresponding estimated-function
bounds, so Lemmas~\ref{lem:app_general_U3_matrix}
and~\ref{lem:app_fresh_U3_matrix} apply.
Lemma~\ref{lem:app_U4_verification} proves (iv).
Finally, \eqref{eq:app_uniform_rate_map} restates the restrictions
of Theorem~\ref{thm:ss_AR}.
\end{proof}

\subsubsection{Logit--lasso and cubic B-spline first steps}
\label{sec:app_logit_spline}

We verify the rates in Corollary~\ref{rem:app_rate_map} using
logistic lasso for the propensity and cubic B-splines for the
conditional means and their derivatives.
Two additional requirements are explicit: a moment comparison
converting the logistic prediction rate to an $L_4$ rate, and
stability of the conditional regressions and their derivatives
under the generated index. Propensity consistency alone does
not imply the latter.
For the propensity dictionary $w_n(x)\in\mathbb R^{p_n}$, write
\begin{equation}
  r_{p,n}:=\left\{\frac{s_p\log(p_n\vee n)}{n}\right\}^{1/2}.
  \label{eq:app_propensity_rate}
\end{equation}
For the spline step, let $J_n$ be the number of basis functions. Let
$\lambda_{h,n}$ be a deterministic sequence such that the penalties in the
conditional-mean regressions satisfy
$\max_{\ell,W}\lambda_{W,\ell}=O_p(\lambda_{h,n})$, and define
\begin{align}
  r_{h,0,n}
  &:=J_n^{-4}+\left(\frac{J_n\log n}{n}\right)^{1/2}
    +J_n\lambda_{h,n}+r_{p,n},
  \label{eq:app_spline_level_rate}\\
  r_{h,1,n}
  &:=J_n^{-3}+\left(\frac{J_n^3\log n}{n}\right)^{1/2}
    +J_n^2\lambda_{h,n}+r_{p,n}.
  \label{eq:app_spline_derivative_rate}
\end{align}
The terms in these displays are, respectively, spline approximation error, sampling
error, penalization error, and the effect of replacing $T$ by the generated index
$\hat T_\ell$.

\begin{assumption}[Sparse-logit and local spline conditions]
\label{ass:app_logit_spline}
The following conditions hold uniformly over the population class
$\mathcal P_{n,M}^+$ of Lemma~\ref{lem:app_c_rate} and folds.

\begin{compactenum}[(i)]
\item
The propensity is exactly sparse logistic:
\[
p_0(x)=\Lambda\{w_n(x)'\gamma_0\},\qquad
\Lambda(u)=\frac{e^u}{1+e^u},\qquad
\|\gamma_0\|_0\le s_p.
\]
Its range lies in a compact interval $\mathcal I\subset(0,1)$.
The fitted propensities are clipped to a fixed compact interval
$\mathcal I^\circ\subset(0,1)$ whose interior contains $\mathcal I$.

Write $\mathcal H_\ell^{\rm tr}=I_\ell^c$ under simple
cross-fitting and $\mathcal H_\ell^{\rm tr}=H_\ell$ under nested
splitting. Partition this sample into a fixed number of inner
folds $I_{\ell j}$, indexed by $j\in\mathcal J_\ell$.
The inner propensity estimator $\hat p_{\ell j}$ is trained on
$\mathcal H_\ell^{\rm tr}\setminus I_{\ell j}$.
The logistic-lasso conditions give
\begin{equation}
\|w_n'(\hat\gamma_\ell-\gamma_0)\|_{\mathbb P,2}
+\max_{j\in\mathcal J_\ell}
 \|w_n'(\hat\gamma_{\ell j}-\gamma_0)\|_{\mathbb P,2}
=O_p(r_{p,n}).
\label{eq:app_logit_oracle}
\end{equation}
With probability approaching one, these coefficient errors
belong to cones on which
\begin{equation}
\|w_n'\Delta\|_{\mathbb P,4}
\le C\|w_n'\Delta\|_{\mathbb P,2}.
\label{eq:app_sparse_L4_comparison}
\end{equation}

\item
For $W=Y$ or a component of $X$, let
$\mathcal F_{\ell j}^p$ be the inner propensity-training
sigma-field. For Borel sets $A\subseteq\mathcal I^\circ$, define
\[
\mu_{\ell j}(A)
=\mathbb P\{\hat p_{\ell j}(X)\in A\mid\mathcal F_{\ell j}^p\},
\qquad
\nu_{W,\ell j}(A)
=\mathbb E[W\mathbf1\{\hat p_{\ell j}(X)\in A\}
          \mid\mathcal F_{\ell j}^p].
\]
Put $\pi_{\ell j}=|I_{\ell j}|/|\mathcal H_\ell^{\rm tr}|$.
The pooled regression is defined, almost everywhere under the
pooled index law, by
\begin{equation}
h_{W,\ell}^{\circ}
=\frac{d\{\sum_{j\in\mathcal J_\ell}
                 \pi_{\ell j}\nu_{W,\ell j}\}}
       {d\{\sum_{j\in\mathcal J_\ell}
                 \pi_{\ell j}\mu_{\ell j}\}}.
\label{eq:app_pooled_generated_regression}
\end{equation}
It admits an extension to a neighborhood of $\mathcal I^\circ$
with four uniformly bounded derivatives, with probability
approaching one. The true regressions $h_{0W}$ admit twice
continuously differentiable extensions there with uniformly
bounded first two derivatives.

\item
Let
$\mathcal I_\ell(X)=[T\wedge\hat T_\ell,T\vee\hat T_\ell]$,
where $\hat T_\ell=\hat p_\ell(X)$. For $r=0,1$,
\begin{equation}
\max_W
\left\|
 \sup_{t\in\mathcal I_\ell(X)}
 |\partial_t^r h_{W,\ell}^{\circ}(t)
  -\partial_t^r h_{0W}(t)|
\right\|_{\mathbb P,4}
=O_p(r_{p,n}).
\label{eq:app_generated_index_stability}
\end{equation}
The norms use a fresh draw of $X$, conditional on the
nuisance-training data.

\item
Define
\[
\mathbb P_{H,\ell}^{\rm cf}\{a_{\ell j}:j\in\mathcal J_\ell\}
=\sum_{j\in\mathcal J_\ell}
 \pi_{\ell j}\mathbb P_{I_{\ell j}}a_{\ell j}.
\]
For a quasi-uniform cubic B-spline basis $B_{J_n}$ on
$\mathcal I^\circ$, use the existing penalized estimator
\begin{equation}
\begin{split}
\hat\beta_{W,\ell}
&\in\arg\min_{\beta\in\mathbb R^{J_n}}
\left[
\mathbb P_{H,\ell}^{\rm cf}
 \{W-B_{J_n}(\hat p_{\ell j}(X))'\beta\}^2
+\lambda_{W,\ell}\|\beta\|_1
\right],\\
\hat h_{W,\ell}(t)
&=B_{J_n}(t)'\hat\beta_{W,\ell}.
\end{split}
\label{eq:app_penalized_spline_criterion}
\end{equation}
Let
\[
\hat Q_\ell
=\mathbb P_{H,\ell}^{\rm cf}
 [B_{J_n}(\hat p_{\ell j}(X))
  B_{J_n}(\hat p_{\ell j}(X))'],
\qquad
\mathcal K_\ell=\{k:\hat Q_{\ell,kk}>0\}.
\]
The principal matrix $\hat Q_{\ell,\mathcal K_\ell\mathcal K_\ell}$
is nonsingular with probability approaching one, and
\[
\|\hat Q_{\ell,\mathcal K_\ell\mathcal K_\ell}^{-1}
 \|_{\infty\to\infty}=O_p(J_n),
\]
where $\|\cdot\|_{\infty\to\infty}$ is the maximum absolute
row-sum norm. Moreover, with probability approaching one,
for a fresh draw of $X$, almost surely,
\[
\partial_t^r B_{k,J_n}(t)=0
\quad\text{for all }k\notin\mathcal K_\ell,\quad
t\in\mathcal I_\ell(X),\quad r=0,1.
\]
Thus columns unsupported by the training observations do not
contribute on the evaluation segments.

For cubic-spline approximants
$h_{W,J_n}^{\circ}=B_{J_n}'\beta_{W,J_n}^{\circ}$
to the extensions in (ii), assume
\begin{equation}
\max_{1\le k\le J_n}
\left|
\mathbb P_{H,\ell}^{\rm cf}
\left[
B_{k,J_n}(\hat p_{\ell j}(X))
\{W-B_{J_n}(\hat p_{\ell j}(X))'
       \beta_{W,J_n}^{\circ}\}
\right]\right|
=O_p\left\{
J_n^{-5}+\sqrt{\frac{\log n}{nJ_n}}
\right\}.
\label{eq:app_spline_score_bound}
\end{equation}
Finally, $\max_{\ell,W}\lambda_{W,\ell}=O_p(\lambda_{h,n})$
and $d_\theta$ is fixed.
\end{compactenum}
\end{assumption}

The pooled-law definition permits atoms caused by clipping.
No density lower bound is imposed outside the true index support.
Parts (iii)--(iv) are maintained generated-index, local design,
and score conditions; propensity consistency alone does not
establish them. The design condition concerns only spline columns
that contribute on the evaluation segments.

\begin{lemma}[Logistic-lasso propensity rate]
\label{lem:app_logit_rate}
Under Assumption~\ref{ass:app_logit_spline}(i),
\begin{equation}
\|\hat p_\ell(X)-T\|_{\mathbb P,4}
+\max_{j\in\mathcal J_\ell}
 \|\hat p_{\ell j}(X)-T\|_{\mathbb P,4}
=O_p(r_{p,n}).
\label{eq:app_logit_L4_rate}
\end{equation}
\end{lemma}

\begin{proof}
The logistic link is $1/4$-Lipschitz, and clipping to an
interval containing $T$ cannot increase prediction error.
Apply \eqref{eq:app_sparse_L4_comparison} and
\eqref{eq:app_logit_oracle} to each estimator.
\end{proof}

\begin{lemma}[Local cubic-spline level and derivative rates]
\label{lem:app_spline_rates}
Suppose Assumption~\ref{ass:app_logit_spline} holds and
$J_n\log n/n\to0$. Then, for $r=0,1$,
\begin{equation}
\max_W
\left\|
 \sup_{t\in\mathcal I_\ell(X)}
 |\partial_t^r\hat h_{W,\ell}(t)
  -\partial_t^r h_{0W}(t)|
\right\|_{\mathbb P,4}
=O_p(r_{h,r,n}).
\label{eq:app_spline_uniform_rates}
\end{equation}
Consequently,
\begin{equation}
\sum_{j\in\{X,Y\}}\sum_{r=0}^1
\|E_{j,r,\ell}\|_{\mathbb P,4}
=O_p(r_{h,0,n}+r_{h,1,n}).
\label{eq:app_spline_segment_rate}
\end{equation}
\end{lemma}

\begin{proof}
Fix $W$ and a fold. The smooth extension in
Assumption~\ref{ass:app_logit_spline}(ii) admits an approximant
$h_{W,J_n}^{\circ}=B_{J_n}'\beta_{W,J_n}^{\circ}$ satisfying
\begin{equation}
\sup_{t\in\mathcal I^\circ}
|\partial_t^r h_{W,J_n}^{\circ}(t)
 -\partial_t^r h_{W,\ell}^{\circ}(t)|
\le CJ_n^{-(4-r)},\qquad r=0,1.
\label{eq:app_spline_approximation}
\end{equation}
Local support and quasi-uniform knots give
\begin{equation}
\sup_{t\in\mathcal I^\circ}
\sum_{k=1}^{J_n}|\partial_t^rB_{k,J_n}(t)|
\le CJ_n^r,\qquad r=0,1.
\label{eq:app_spline_locality}
\end{equation}

Let $\hat S_{W,\ell}$ be the score vector in
\eqref{eq:app_spline_score_bound}, and let
$\hat z_{W,\ell}$ be a subgradient of
$\|\hat\beta_{W,\ell}\|_1$, with
$\|\hat z_{W,\ell}\|_\infty\le1$.
Columns outside $\mathcal K_\ell$ vanish at every training
observation. The KKT equations restricted to $\mathcal K_\ell$
therefore give
\[
\hat Q_{\ell,\mathcal K_\ell\mathcal K_\ell}
(\hat\beta_{W,\ell}-\beta_{W,J_n}^{\circ})_{\mathcal K_\ell}
=(\hat S_{W,\ell})_{\mathcal K_\ell}
-\frac{\lambda_{W,\ell}}2
 (\hat z_{W,\ell})_{\mathcal K_\ell}.
\]
The inverse and score bounds imply
\begin{equation}
\|(\hat\beta_{W,\ell}-\beta_{W,J_n}^{\circ})
       _{\mathcal K_\ell}\|_\infty
=O_p\left\{
J_n^{-4}+\sqrt{\frac{J_n\log n}{n}}
+J_n\lambda_{h,n}
\right\}.
\label{eq:app_spline_coefficient_rate}
\end{equation}
Unsupported columns also vanish, together with their first
derivatives, on the evaluation segments. Thus
\eqref{eq:app_spline_locality} yields
\begin{equation}
\left\|
 \sup_{t\in\mathcal I_\ell(X)}
 |\partial_t^r\hat h_{W,\ell}(t)
  -\partial_t^r h_{W,J_n}^{\circ}(t)|
\right\|_{\mathbb P,4}
=O_p\left\{
J_n^{-(4-r)}
+\sqrt{\frac{J_n^{2r+1}\log n}{n}}
+J_n^{r+1}\lambda_{h,n}
\right\}.
\label{eq:app_spline_sampling_rate}
\end{equation}
Adding the approximation error
\eqref{eq:app_spline_approximation} and the generated-index
error \eqref{eq:app_generated_index_stability} proves
\eqref{eq:app_spline_uniform_rates}.
The envelope conclusion follows because $d_\theta$ is fixed.
\end{proof}

The derivative determines the common rate: differentiation changes
the approximation error from $J_n^{-4}$ to $J_n^{-3}$ and
multiplies the sampling and penalization terms by $J_n$.
\begin{proposition}[Primitive rate for Assumption~\ref{ass:app_score_rates}]
  \label{prop:app_logit_spline_ass7}
Suppose Assumptions~\ref{ass:app_population_moments}
and~\ref{ass:app_logit_spline} hold,
$J_n\log n/n\to0$, and the sequence $\delta_n$ defined in
\eqref{eq:app_logit_spline_delta_general} tends to zero.
Suppose also that the positive nuisance-block spectrum is uniformly stable,
\[
\inf_{\mathbb P,n}\lambda_{\min}^+(M_{\mathbb P})>0,
\qquad \sup_{\mathbb P,n}\|b_{\mathbb P}^*\|<\infty,
\]
and that the calibration matrix and vector obey the rate in
Lemma~\ref{lem:app_c_rate} with
\[
q_{c,n}=O\{r_{p,n}+r_{h,0,n}+r_{h,1,n}+n^{-1/2}\},
\qquad q_{c,n}\ll\rho_n\ll1.
\]
Then Assumption~\ref{ass:app_score_rates} holds with
\begin{equation}
  \delta_n
  =r_{p,n}+J_n^{-3}
   +\left(\frac{J_n^3\log n}{n}\right)^{1/2}
   +J_n^2\lambda_{h,n}.
  \label{eq:app_logit_spline_delta_general}
\end{equation}
If
\begin{equation}
  s_p\log(p_n\vee n)\lesssim n^{1/3},
  \qquad J_n\asymp n^{1/9},
  \qquad
  \lambda_{h,n}\lesssim
  \left(\frac{\log n}{nJ_n}\right)^{1/2},
  \label{eq:app_logit_spline_tuning}
\end{equation}
then
\begin{equation}
  \delta_n=n^{-1/3+o(1)}.
  \label{eq:app_logit_spline_delta}
\end{equation}
Here and below $n^{-1/3+o(1)}$ allows powers of $\log n$.
\end{proposition}

\begin{proof}
Lemma~\ref{lem:app_logit_rate} and
Lemma~\ref{lem:app_spline_rates} verify
\eqref{eq:app_primitive_regression_rate}. Since $r_{h,0,n}\le C r_{h,1,n}$ for
$J_n\ge1$, its rate is bounded by the right-hand side of
\eqref{eq:app_logit_spline_delta_general}. Uniform stability of $M_{\mathbb P}$ and
Lemma~\ref{lem:app_c_rate} give the same rate for
$\hat c_\ell-b_{\mathbb P}^*$, verifying
\eqref{eq:app_primitive_c_rate}. The remaining clauses of
Assumption~\ref{ass:app_general_firststep} and its population counterpart follow
from the maintained compactness and moment conditions. Corollary~\ref{cor:app_ass7}
then proves Assumption~\ref{ass:app_score_rates} with the stated $\delta_n$.

Under \eqref{eq:app_logit_spline_tuning},
\[
  r_{p,n}=O(n^{-1/3}),\qquad
  J_n^{-3}\asymp n^{-1/3},\qquad
  \left(\frac{J_n^3\log n}{n}\right)^{1/2}
  =n^{-1/3}(\log n)^{1/2},
\]
and $J_n^2\lambda_{h,n}=n^{-1/3+o(1)}$.
Substitution into \eqref{eq:app_logit_spline_delta_general} gives
\eqref{eq:app_logit_spline_delta}.
\end{proof}

For calibration without fresh observations, let $\mathcal A_n$
be a deterministic, pointwise measurable class containing, with
probability approaching one, the scalar entries of
$\hat X_\ell\hat r_\ell'$ and $\hat X_\ell\hat m_\ell$ for every
fold. Suppose it has an envelope $F_n$ with
$\sup_{\mathbb P}\|F_n\|_{\mathbb P,2}\le C$ and
\begin{equation}
\sup_Q
\log N\{\epsilon\|F_n\|_{Q,2},\mathcal A_n,L_2(Q)\}
\le Cv_n\log(Cn/\epsilon),
\qquad
v_n\lesssim s_p\log(ep_n/s_p)+J_n,
\label{eq:app_logit_spline_matrix_entropy}
\end{equation}
for $0<\epsilon\le1$, where the supremum is over finitely
supported probability measures $Q$ with
$0<\|F_n\|_{Q,2}<\infty$.
The uniform-entropy maximal inequality then verifies
\eqref{eq:app_U3_empirical_rate} with
\begin{equation}
\omega_n=\left(\frac{v_n\log n}{n}\right)^{1/2}
=O(n^{-1/3+o(1)})
\label{eq:app_logit_spline_omega}
\end{equation}
under \eqref{eq:app_logit_spline_tuning}.
This complexity bound is an additional condition on the fitted
product classes. Under nested splitting,
Lemma~\ref{lem:app_fresh_U3_matrix} instead gives
$\omega_n=n^{-1/2}$.
\begin{corollary}[Verification of (U2)--(U4)]
  \label{cor:app_logit_spline_U_conditions}
  \label{cor:ss_primitive_firststep}
Suppose the conditions of Proposition~\ref{prop:app_logit_spline_ass7} hold with
\eqref{eq:app_logit_spline_tuning}, and suppose
$\nu_{\mathbb P}\ge\nu_n$ and $\delta_n=o(\nu_n)$. Under nested splitting, use a
fresh calibration sample of order $n$. Under simple cross-fitting, suppose
\eqref{eq:app_logit_spline_matrix_entropy} holds. Then:
\begin{compactenum}[(i)]
  \item (U2) holds with $\delta_n=n^{-1/3+o(1)}$;
  \item the matrix part of (U3) holds with
  \[
    q_n=n^{-1/3+o(1)};
  \]
  the additional requirement $q_n\ll\rho_n\ll\lambda_n$ remains the spectral
  threshold condition;
  \item under simple cross-fitting, (U4) holds with
  $\kappa_n=\delta_n/\nu_n$;
 \item under nested splitting, (U4) holds with $\kappa_n=n^{-1/2}$,
provided $\sqrt n\,\delta_n^2=O(\nu_n)$.
\end{compactenum}
\end{corollary}

\begin{proof}
Part (i) follows from Proposition~\ref{prop:app_logit_spline_ass7} and
Corollary~\ref{cor:app_U2_U4b}. For part (ii), under nested splitting,
Corollary~\ref{cor:app_U3_matrix} gives
$q_n=\delta_n+n^{-1/2}=n^{-1/3+o(1)}$. Under simple cross-fitting,
\eqref{eq:app_logit_spline_omega} and the general part of that corollary give
$q_n=\delta_n+\omega_n=n^{-1/3+o(1)}$.

Part (iii) is Corollary~\ref{cor:app_complete_U4}. For part (iv), it remains only to
verify the additional nested-splitting hypotheses. For every $\epsilon>0$, the uniform coefficient rate permits
a fixed $C_\epsilon<\infty$ such that
\[
\limsup_{n\to\infty}\sup_{\mathbb P}
\mathbb P(\hat c_\ell\notin\mathcal B_{\ell,\epsilon})
\le\epsilon,\qquad
\mathcal B_{\ell,\epsilon}
=\{c:\|c-b_{\mathbb P}^*\|\le C_\epsilon\delta_n\}.
\]
Fix $\epsilon$ and write
$\mathcal B_\ell=\mathcal B_{\ell,\epsilon}$ below.
For $c\in\mathcal B_\ell$, write
\[
  \hat X_{1,\ell}-\hat X_{-1,\ell}'c
  =\varphi_{\mathbb P}^{(1)}
   +(\hat X_{1,\ell}-\tilde X_1)
   -(\hat X_{-1,\ell}-\tilde X_{-1})'b_{\mathbb P}^*
   -\hat X_{-1,\ell}'(c-b_{\mathbb P}^*).
\]
The $L_4$ norm of the right-hand side is
$O_p(\nu_{\mathbb P}+\delta_n)=O_p(\nu_{\mathbb P})$. Multiplication by
$\hat r_{-1,\ell}$ and H\"older's inequality therefore produce an $L_2$ envelope
of order $\nu_{\mathbb P}$. Because $c$ has fixed dimension and the class is affine
in $c$, its covering entropy is bounded by $C\log(C/\epsilon)$. For the calibration residual, suppress the fold and distribution
indices and put
\[
A=\hat M',\qquad Q=I-MM^\dagger,\qquad
\hat Q=I-A A_{\rho_n}^\dagger,\qquad
\hat P=A_{\rho_n}^\dagger A.
\]
The matrix rate and the singular-subspace bound used in
Lemma~\ref{lem:app_c_rate} imply
\[
\|\hat Q-Q\|+\|(I-\hat P)b^*\|=O_p(\delta_n),
\qquad \|\hat Q A\|=O_p(\delta_n).
\]
Since $\hat Q A=\hat Q A(I-\hat P)$, it follows that
$\|\hat Q A b^*\|=O_p(\delta_n^2)$.

Let $f=f_{b^*}$ as defined in Lemma~\ref{lem:app_U4_verification},
and let $\mathbb P f$ denote expectation conditional on
$\mathcal H_\ell$. The component rates and population normal
equations give
$\|f\|_{\mathbb P,2}=O_p(\nu_{\mathbb P})$
and $\|\mathbb P f\|=O_p(\delta_n)$.
To control its null-space component, write
$e_X(s)=\hat h_{X,-1}(s)-h_{0X,-1}(s)$ and
$\Delta_p=\hat p(X)-T$.
Using $Q\tilde X_{-1}=0$, Taylor's formula and the segment
derivative error bound give
\[
Q\mathbb E[\hat r_{-1}\mid X,\mathcal H_\ell]
=-TQe_X(T)+R(X),\qquad
\|R(X)\|\le C\{|\Delta_p|^2+
|\Delta_p|E_{X,1,\ell}(X)\}.
\]
Thus $\|R\|_{\mathbb P,2}=O_p(\delta_n^2)$.
Since $\mathbb E[\varphi^{(1)}\mid T]=0$ and
$\|\hat X_1-\hat X_{-1}'b^*-\varphi^{(1)}\|_{\mathbb P,4}
=O_p(\delta_n)$, this yields
$\|Q\mathbb P f\|=O_p(\delta_n^2)$.
Conditional Chebyshev and $\|\hat Q\|\le1$ now give
\begin{align*}
\hat e
&=\hat Q\hat d\\
&=\hat Q(\mathbb P_{B,\ell}-\mathbb P)f
  +Q\mathbb P f+(\hat Q-Q)\mathbb P f+\hat Q A b^*\\
&=O_p(n^{-1/2}\nu_{\mathbb P}+\delta_n^2)
 =O_p(n^{-1/2}\nu_{\mathbb P}),
\end{align*}
where the last equality uses $\sqrt n\,\delta_n^2=O(\nu_n)$
and $\nu_{\mathbb P}\ge\nu_n$.
On the localization event, the proof of
Lemma~\ref{lem:app_U4_verification} therefore gives the nested
(U4) bounds with $\kappa_n=n^{-1/2}$.
Since $\epsilon$ is arbitrary, the bounds hold uniformly in probability.
\end{proof}

\begin{corollary}[Rate frontier under the concrete first steps]
  \label{col:app_logit_spline_frontier}
Under the conditions of Corollary~\ref{cor:app_logit_spline_U_conditions}, write
\[
  \nu_n=n^{-v},\qquad \lambda_n=n^{-\xi}.
\]
Up to logarithmic factors, the uniform AR--NO rate restrictions become
\begin{align*}
  &v<\frac16,\qquad \xi<\frac13
  &&\text{under nested splitting},\\
  &v+\xi<\frac16
  &&\text{under simple cross-fitting}.
\end{align*}
Along an identified sequence $\mathbb P_n$ with sharp rates
$\nu_{\mathbb P_n}\asymp n^{-v}$ and
$\lambda_{\min}^+(\Sigma_{\tilde X,\mathbb P_n})\asymp n^{-\xi}$,
the population geometry imposes $2v\le\xi$. Thus nested splitting allows $v<1/6$, while simple
cross-fitting is feasible only if $v<1/18$. Since
$\mathrm{CP}_n\asymp n\nu_{\mathbb P_n}^2$, the corresponding frontiers are
$\mathrm{CP}_n\gg n^{2/3}$ and $\mathrm{CP}_n\gg n^{8/9}$, respectively, up to
logarithmic factors.
\end{corollary}

\begin{proof}
With $\delta_n=q_n=n^{-1/3+o(1)}$, the first two conditions in
\eqref{eq:app_uniform_rate_map} reduce to $v<1/6$, and
$q_n/\lambda_n\to0$ is $\xi<1/3$. Under nested splitting,
$\sqrt n\kappa_n=1$, so the last condition adds nothing beyond
$q_n/\lambda_n\to0$.

Under simple cross-fitting,
\[
  \sqrt n\kappa_n\frac{q_n}{\lambda_n}
  =\frac{\sqrt n\,\delta_n q_n}{\nu_n\lambda_n}
  =\frac{n^{-1/6+o(1)}}{\nu_n\lambda_n},
\]
which converges to zero exactly when $v+\xi<1/6$. Along the identified sequence
specified in the statement, the population identity
\[
  \Sigma_{\mathbb P_n}^{(1)}
  =\{e_1'\Sigma_{\tilde X,\mathbb P_n}^{\dagger}e_1\}^{-1}
  \ge\lambda_{\min}^+(\Sigma_{\tilde X,\mathbb P_n})
\]
and the overlap bound
$\Sigma_{\mathbb P_n}^{(1)}\le\nu_{\mathbb P_n}^2$ imply
$\lambda_{\min}^+(\Sigma_{\tilde X,\mathbb P_n})\lesssim
\nu_{\mathbb P_n}^2$. Thus $\xi\ge2v$. Combining this with
$v+\xi<1/6$ gives $3v<1/6$, or $v<1/18$. Finally,
$\mathrm{CP}_n\asymp n\nu_{\mathbb P_n}^2=n^{1-2v}$, which gives the two stated
concentration frontiers.
\end{proof}

		Let \(A_\ell=n_\ell^{-1}\sum_{i\in I_\ell}
		\hat\varphi_{i\ell}^{(1)}\hat r_{-1,i\ell}'\).
		
		\begin{lemma}[Profiling rates]
			\label{lem:app_profile_rates}
			Under (U3)--(U4), \(q_n\ll\rho_n\ll\lambda_n\), and
			\(a_n=q_n/\lambda_n\), uniformly over \(\mathcal P_n^+\),
			\begin{align}
				\max_\ell\|\hat\theta_{-\ell}-\bar\theta_{\mathbb P}\|&=O_p(a_n),
				\label{eq:app_plugin_rate}\\
				\max_\ell\|A_\ell\|&=O_p\{\nu_{\mathbb P}(\kappa_n+n^{-1/2})\},
				\label{eq:app_A_rate}\\
				R_{3n}^*&:=-\sqrt n\sum_\ell\frac{n_\ell}{n}A_\ell
				(\hat\theta_{-1,-\ell}-\bar\theta_{-1,\mathbb P})
				=O_p\{\nu_{\mathbb P}(\sqrt n\kappa_n+1)a_n\}.
				\label{eq:app_profile_remainder}
			\end{align}
		\end{lemma}
		
		\begin{proof}
			Weyl's inequality and $q_n\ll\rho_n\ll\lambda_n$ imply rank recovery and
			$\|\hat\Sigma_{-\ell,\rho_n}^{\dagger}\|=O_p(\lambda_n^{-1})$.
			Since $\Gamma_{\mathbb P}=\Sigma_{\tilde X,\mathbb P}\bar\theta_{\mathbb P}$,
			\[
			\hat\theta_{-\ell}-\bar\theta_{\mathbb P}
			=\hat\Sigma_{-\ell,\rho_n}^{\dagger}
			(\hat\Gamma_{-\ell}-\hat\Sigma_{-\ell}\bar\theta_{\mathbb P})
			+(\hat\Pi_{-\ell,\rho_n}-\Pi_{\mathbb P})\bar\theta_{\mathbb P}.
			\]
			The first term is $O_p(q_n/\lambda_n)$ and Wedin's bound gives the same order
			for the second, proving \eqref{eq:app_plugin_rate}. Conditionally on
			$\mathcal F_\ell$, (U4) and Chebyshev imply
			$\|A_\ell\|=O_p\{\nu_{\mathbb P}(\kappa_n+n^{-1/2})\}$, which is
			\eqref{eq:app_A_rate}; multiplying this bound by
			\eqref{eq:app_plugin_rate} yields \eqref{eq:app_profile_remainder}.
		\end{proof}

		\medskip\noindent
		\textbf{Proof of Theorem~\ref{thm:ss_AR}.}
		Adding and subtracting \(\bar\theta_{-1,\mathbb P}\) and the oracle score $g^{*}_{\mathbb P}(Z_i)$ gives
		\begin{equation}
			\sqrt n\,\hat g_n^{(1)}\{\theta_{0,1}(\mathbb P)\}
			=n^{-1/2}\sum_{i=1}^n g_{\mathbb P}^*(Z_i)
			+n^{-1/2}\sum_\ell\sum_{i\in I_\ell}
			(\tilde g_{i\ell,\mathbb P}-g_{\mathbb P,i}^*)+R_{3n}^*,
			\label{eq:app_exact_ar_decomposition}
		\end{equation}
		 By (U2), conditional Chebyshev's inequality and fixed \(L\),
		\[
		n^{-1/2}\sum_\ell\sum_{i\in I_\ell}
		(\tilde g_{i\ell,\mathbb P}-g_{\mathbb P,i}^*)
		=O_p(\delta_n+\sqrt n\,\delta_n^2).
		\]
		Together with Lemma~\ref{lem:app_profile_rates} and
		\eqref{eq:ss_uniform_rates_corrected}, this yields, uniformly over
		\(\mathbb P\in\mathcal P_n^+\),
		\begin{equation}
			\sqrt n\,\hat g_n^{(1)}\{\theta_{0,1}(\mathbb P)\}
			=n^{-1/2}\sum_{i=1}^n g_{\mathbb P}^*(Z_i)+o_p(\nu_{\mathbb P}).
			\label{eq:app_corrected_linear_expansion}
		\end{equation}
		Let \(d_{i\ell}=\tilde g_{i\ell,\mathbb P}-g_{\mathbb P,i}^*\) and
		\(h_{i\ell}=\hat\varphi_{i\ell}^{(1)}\hat r_{-1,i\ell}'
		(\hat\theta_{-1,-\ell}-\bar\theta_{-1,\mathbb P})\).
		Conditional Markov's inequality, (U2), (U4), and
		\eqref{eq:app_plugin_rate} give
		\begin{equation}
			n^{-1}\sum_{\ell,i\in I_\ell}(d_{i\ell}^2+h_{i\ell}^2)
			=O_p(\delta_n^2+\nu_{\mathbb P}^2a_n^2)
			=o_p(\nu_{\mathbb P}^2).
			\label{eq:app_relative_l2}
		\end{equation}
		Thus Cauchy--Schwarz permits replacement of the feasible score by
		\(g_{\mathbb P}^*\) in the sample second moment. Lemma~\ref{lem:ss_canonical_representative}
		gives
		\[
		\operatorname{Var}_{\mathbb P}\!\left(
		\frac{n^{-1}\sum_i(g_{\mathbb P}^*(Z_i))^2}{\Omega_{\mathbb P}^*}\right)
		\le\frac{\mathbb E_{\mathbb P}\left[|g_{\mathbb P}^*|^4\right]}
		{n(\Omega_{\mathbb P}^*)^2}=O(n^{-1}),
		\]
		while \eqref{eq:app_corrected_linear_expansion} makes centering negligible. Hence
		\(\hat\Omega_n^{(1)}\{\theta_{0,1}(\mathbb P)\}/\Omega_{\mathbb P}^*
		\to_p1\) uniformly.
		Berry–Esseen bounds the normal-approximation error by a constant times
		\[
		\frac{\mathbb E_{\mathbb P}\left[|g_{\mathbb P}^*|^3\right]}
		{\sqrt n(\Omega_{\mathbb P}^*)^{3/2}}
		\le
		\frac{\{\mathbb E_{\mathbb P}\left[|g_{\mathbb P}^*|^4\right]\}^{3/4}}
		{\sqrt n(\Omega_{\mathbb P}^*)^{3/2}}=O(n^{-1/2}),
		\]
        which follows by H\"older's inequality.
Together with \eqref{eq:app_corrected_linear_expansion} and Slutsky's theorem, this
		proves the uniform null and coverage conclusions. \hfill\(\blacksquare\)
		
		\medskip\noindent
		\textbf{Proof of Proposition~\ref{prop:ss_AR_pointwise}.}
		Apply Theorem~\ref{thm:ss_AR} to the singleton class containing the fixed
		distribution. Here \(\nu_n\) and \(\lambda_n\) may be fixed below by positive
		constants and \(q_n=O(\delta_n+n^{-1/2})\). Under nested splitting the last rate in
		\eqref{eq:ss_uniform_rates_corrected} follows from \(q_n=o(1)\); under simple
		cross-fitting it follows from
		\(\sqrt n\delta_nq_n=O(\sqrt n\delta_n^2+\delta_n)=o(1)\).
		Thus the null and coverage conclusions follow from the theorem. For local power, along the regular location path with score
\(s=\bar\delta'\ell_\theta\), \eqref{eq:ss_target_score} and
\eqref{eq:ss_population_profile_orthogonality} give
\[
\mathbb E_{\mathbb P}[g_{\mathbb P}^*s]
=\bar\delta_1\mathbb E_{\mathbb P}
[T\{\varphi_{\mathbb P}^{(1)}\}^2]
=\bar\delta_1\Sigma_{\mathbb P}^{(1)}.
\]
Le Cam's third lemma and contiguity transfer the expansion and studentizer
consistency to the local path, yielding the stated noncentral
\(\chi_1^2\) limit. \hfill\(\blacksquare\)
		
	\bigskip
		\section{Latent-Variable Models}
		\label{app:tva}
		
This appendix verifies Assumption~\ref{ass:mixture} in the
Gaussian repeated-measurement model \eqref{eq:tva_model}
and proves the results of Section~\ref{TVAcexample}.
The underlying mixture tangent-set result,
Theorem~\ref{thm:mixture_tangent}, is proved in
Section~\ref{app:mixture}.

\subsection{Verification of the mixture conditions}
\label{sec:tva_conditions}

We use the notation of Section~\ref{subsec:tva_setup}.		\begin{lemma}[Verification in the Gaussian model]
			\label{lem:tva_gauss}
			Let $Z=Y$ follow \eqref{eq:tva_model} with $J=2$ and $\theta_0>0$, and let $\eta_0$
have a density bounded away from zero on the compact interval $\mathcal A$. Then
Assumption~\ref{ass:mixture}\textup{(i)--(iv) and (vi)} holds.
		\end{lemma}
		
		\medskip\noindent\textbf{Proof of Lemma~\ref{lem:tva_gauss}:}
		Put $\bar y=(y_1+y_2)/2$, $v(y)=\sum_j(y_j-\bar y)^2$,
		$d(y)=\inf_{\alpha\in\mathcal A}|\bar y-\alpha|$, $Q_c(y)=2\{d(y)+c\}^2+v(y)$, and
		$\ell=\min\{|\mathcal A|,1\}$.  Restricting the mixture integral to the length-$\ell$
		subinterval nearest $\bar y$, and using the positive lower bound on the density of
		$\eta_0$, gives
		\[
		 p_0(y)\ge C e^{-Q_\ell(y)/(2\theta_0^2)},\qquad
		 \sup_{\alpha}p(y\mid\alpha,\theta)\le C_\theta e^{-Q_0(y)/(2\theta^2)}.
		\]
		Consequently, on a sufficiently small neighborhood of $\theta_0$, both
		$p_{\theta,\eta}/p_0$ and $\sup_\alpha\|\partial_\theta
		p(\cdot\mid\alpha,\theta)\|/p_0$ are bounded by a polynomial in $\|y\|$ times
		$C_\varepsilon\exp\{a_\varepsilon\|y\|^2+C\|y\|\}$, where
		$a_\varepsilon\downarrow0$.  Gaussian tails therefore supply the $L_p$ envelope in
		(iii) for every finite $p$.  The same bound at $\theta_0$, followed by dominated
		convergence, proves the $L_2$ continuity of $\alpha\mapsto k_\alpha$ in (ii).
		Positivity and smoothness are immediate, and the envelope gives DQM by
		\citet[Lemma 7.6]{vandervaart98}, proving (i).  Finally, $Y_1-Y_2\sim
		N(0,2\theta^2)$ independently of $\eta$ and
		\[
		 H(\mathbb P_{\theta,\eta},\mathbb P_0)\ge
		 \{1-[2\theta\theta_0/(\theta^2+\theta_0^2)]^{1/2}\}^{1/2}
		 \ge C^{-1}\|\theta-\theta_0\|.
		\]
			This marginal identifies $\theta$ and verifies the added stability
condition in part~\textup{(iv)}: if
$H(\mathbb P_{\theta_n,\eta_n},\mathbb P_0)\to0$, the displayed
marginal Hellinger bound implies
$2\theta_n\theta_0/(\theta_n^2+\theta_0^2)\to1$, hence
$\theta_n\to\theta_0$. Since
			$\varphi_{Y_1}(s)=\varphi_\eta(s)e^{-\theta^2s^2/2}$, the characteristic function
			of $Y_1$ identifies $\eta$. Thus (iv)
		holds. For~(vi), let $\bar Y=(Y_1+Y_2)/2$ and $\Delta=Y_1-Y_2$. At $\theta_0$,
$\Delta$ is independent of $(\alpha,\bar Y)$. Hence, for $g\in L_2^0$, writing
$m_g(\bar Y):=\mathbb E[g(Y)\mid\bar Y]$,
\[
S^*g(\alpha)=\mathbb E[m_g(\bar Y)\mid\alpha].
\]
Since $\bar Y\mid\alpha$ is Gaussian with mean $\alpha$ and variance
$\theta_0^2/2$, the map
$\alpha\mapsto\mathbb E[m_g(\bar Y)\mid\alpha]$ is real analytic.
Because $S^*g=0$ $\eta_0$-a.s., continuity and
$\operatorname{supp}(\eta_0)=\mathcal A$ make it zero on $\mathcal A$; analyticity
and injectivity of Gaussian convolution then imply $m_g(\bar Y)=0$ a.s. Hence
\[
\ker(S^*)
=\{g\in L_2^0:\mathbb E[g(Y)\mid\bar Y]=0\}.
\]
Thus
\[
\Pi_{\ker(S^*)}g=g-\mathbb E[g\mid\bar Y],
\]
which maps bounded mean-zero functions into $L_\infty\subset L_{2q}$ for every
$q<\infty$. Lemma~\ref{lem:density}\textup{(c)} therefore proves~(vi).
		\hfill$\blacksquare$

		\subsection{Tangent set, existence, and RKHS characterization}
		
		\medskip\noindent\textbf{Proof of Proposition~\ref{prop:tva_tangent}.}
The path, DQM, and score assertions follow from
Theorem~\ref{thm:mixture_tangent}\textup{(a)}. Normalize $(t,b)$ by
$c(t,b):=1+\|t\|+\|b\|_\infty$ and use
$\bar t=t/c(t,b)$ and $\bar b=b/c(t,b)$. This rescales the score without
changing its closed linear span. Assumption~\ref{ass:mixture}(iii) supplies an
$L_p$ likelihood-ratio envelope for every finite $p$. Since every resulting score belongs to $\mathcal T$ and these scores span $V$,
taking closures gives $V\subseteq\overline{\mathcal T}$. Assumption~
\ref{ass:mixture}\textup{(vi)} and
Theorem~\ref{thm:mixture_tangent}\textup{(b)} give the reverse inclusion.
Hence $V=\overline{\mathcal T}$.
\hfill$\blacksquare$

		\medskip\noindent\textbf{Proof of Proposition~\ref{prop:tva_overid}:}
		Because $Y_1,\ldots,Y_J$ are conditionally i.i.d.\ given $\alpha$ under every
		$(\alpha,\theta)$, the law of $(Y_1,Y_2)$ under any $\mathbb P_{\theta,\eta}\in\mathcal P$
		is exchangeable, so $\mathbb E_{\theta,\eta}[g_a]=0$ at every point of the model. Let
		$\{\mathbb P_\tau\}$ be an arbitrary DQM path through $\mathbb P_0$ with score $s$.
		Since $g_a$ is bounded,
		\[
		\tau^{-1}\bigl\{\mathbb E_\tau[g_a]-\mathbb E[g_a]\bigr\}
		=\int g_a\,\frac{\sqrt{f_\tau}-\sqrt{f_0}}{\tau}\bigl(\sqrt{f_\tau}+\sqrt{f_0}\bigr)d\mu
		\ \longrightarrow\ \langle g_a,s\rangle,
		\]
		by \eqref{msd} and $\|\sqrt{f_\tau}-\sqrt{f_0}\|_{L_2(\mu)}\to0$. The left-hand side is
		identically zero, so $\langle g_a,s\rangle=0$ for every $s\in\mathcal T$, i.e.\
		$g_a\in\overline{\mathcal T}^{\perp}$. If $\mathbb P_0(Y_1\neq Y_2)>0$, then
		$a(y_1,y_2)=\arctan(y_1)$ gives $g_a\neq0$ in $L_2$, so
		$\overline{\mathcal T}\neq L_2^{0}$ and $\mathcal P$ is locally overidentified in the
		sense of \citet{chen2018overidentification}; Corollary~\ref{corchenandsantos} gives
		the final claim.
		\hfill$\blacksquare$
		
		\medskip\noindent\textbf{Proof of Proposition~\ref{prop:tva_key_equation}:}
		The necessity argument for \eqref{eq:mixture_LR} uses only the bounded
constructed paths and therefore applies to any $g\in L_2^0$: for a locally robust
$g$, it yields \eqref{eq:mixture_LR} for some $c\in\mathbb R$. If $g$ is relevant
and $c=0$, then $S^*g=0$ and $\langle g,s_\theta\rangle=0$, so
$g\in V^\perp$. Proposition~\ref{prop:tva_tangent} gives
$V=\overline{\mathcal T}$, contradicting relevance. Hence $c\neq0$, and
$g_0:=g/c$ solves \eqref{eq:tva_key_equation}; if instead $\psi_0$ is pathwise differentiable along
		every DQM path with influence function $\dot\psi$, then evaluating along the paths of
		Proposition~\ref{prop:tva_tangent} with $t=0$ gives
		$\langle S^{*}\dot\psi-r_0,b\rangle_{L_2(\eta_0)}=0$ for all $b\in\mathcal B$, hence
		$S^{*}\dot\psi=r_0$ $\eta_0$-a.s. For an arbitrary $\alpha\in\mathcal A$, the
		path $\eta_\tau=(1-\tau)\eta_0+\tau\delta_\alpha$ is DQM with score
		$k_\alpha-1$, so pathwise differentiability also gives
		$S^{*}\dot\psi(\alpha)=\langle\dot\psi,k_\alpha-1\rangle=r_0(\alpha)$.
		Thus the equality is pointwise. For part~(ii), Assumption~\ref{ass:mixture}(iv) and
		Theorem~\ref{thm:mixture_tangent}(a) give $\Omega\succ0$. Since
		$v_\theta\perp\mathcal H_\eta$ and $k_\alpha-1\in\mathcal H_\eta$,
		$S^*v_\theta=0$ pointwise. Hence $S^*g^\star=r_0$ and
		$\langle g^\star,s_\theta\rangle=\bar r_\theta$, so
		\eqref{eq:mixture_LR} yields $g^\star\in\overline{\mathcal T_0}^{\perp}$.
		Moreover, for the $b_\psi$ above,
$\langle g^\star,s_{b_\psi}\rangle
=\langle r_0,b_\psi\rangle=1$, proving relevance. Assumption
		\ref{ass:mixture}(iii) and the stated $L_{2q}$ conditions dominate
		$g^\star$ along every path. The automatic-centering argument in the proof of
		Lemma~\ref{lem:autoadm} therefore makes
		$g(z,\lambda):=g^\star(z)-\mathbb E_\lambda[g^\star(Z)]$ admissible; its true evaluation is
		$g^\star$, so the moment is locally robust.

		It remains to verify the influence-function conclusion. For an arbitrary DQM path, put
		$G_\theta(\alpha)=\mathbb E_{\theta,\alpha}[g^\star(Z)]$.
		Theorem~\ref{thm:mixture_tangent}(b) gives
		$\|\theta_\tau-\theta_0\|=O(\tau)$ and $\eta_\tau\Rightarrow\eta_0$; moreover,
		$G_{\theta_0}=r_0$ and the envelope permits uniform differentiation in $\theta$.
		Therefore
		\[
		\mathbb E_\tau[g^\star]-\{\psi(\lambda_\tau)-\psi_0\}
		=(\theta_\tau-\theta_0)'
		\int\{\dot G_{\theta_0}-\partial_\theta r(\cdot,\theta_0)\}\,d\eta_\tau
		+o(\|\theta_\tau-\theta_0\|)=o(\tau),
		\]
		because the integral converges to
		$\langle g^\star,s_\theta\rangle-\bar r_\theta=0$. DQM and the same
		moment bound give
		$\mathbb E_\tau[g^\star]=\tau\langle g^\star,s\rangle+o(\tau)$, proving that
		$g^\star$ is an influence function along every DQM path.
		\hfill$\blacksquare$		

		\bigskip \noindent \textbf{Proof of Theorem~\ref{thm:rkhs_range}:}
		Equation~\eqref{eq:kernel} gives, for any $\alpha_1,\ldots,\alpha_n$ and
		$c\in\mathbb R^n$,
		$\sum_{i,j}c_i c_jK(\alpha_i,\alpha_j)
		=\|\sum_i c_i k_{\alpha_i}\|^{2}_{L_2}\ge0$.
		Thus $K$ is positive semidefinite and $\alpha\mapsto k_\alpha\in H:=L_2$ is its feature
		map. The feature-map characterization of an RKHS
		\citep[Theorem 4.21]{steinwart2008} yields
		\[
		\mathcal H_K=\bigl\{\langle g,k_{(\cdot)}\rangle_{H}:\ g\in H\bigr\},
		\qquad
		\|f\|_{\mathcal H_K}=\min\{\|g\|_H:\ \langle g,k_{(\cdot)}\rangle_H=f\},
		\]
		and since $S^{*}g=\langle g,k_{(\cdot)}\rangle_H$, this says
		$S^{*}(L_2)=\mathcal H_K$. Since $S^*\mathbf1=\mathbf1$ and
		$\mathbb E[g]=\int S^*g\,d\eta_0$, centering $g$ gives
		$S^{*}(L_2^{0})=\{f\in\mathcal H_K:\int f\,d\eta_0=0\}$. Combining this range identity with
Proposition~\ref{prop:tva_key_equation} shows that
$r(\cdot,\theta_0)\in\mathcal H_K$ is necessary for
relevance and pathwise differentiability and, under the
integrability conditions of part~\textup{(ii)}, sufficient
for both. When $\theta_0$ is known, the minimum squared norm
of an influence function is $\|r_0\|_{\mathcal H_K}^{2}$.
		\hfill $\blacksquare$
		
\bigskip\noindent\textbf{Proof of Corollary~\ref{cor:ladder}.}
For $f\in\mathcal H_K$, write
$f(\alpha)=\langle g_f,k_\alpha\rangle$, where
$\|g_f\|_{L_2}=\|f\|_{\mathcal H_K}$. Cauchy--Schwarz proves part~\textup{(i)}.
Continuous linear functionals preserve Banach-space derivatives and holomorphic
maps, which proves parts~\textup{(ii)} and~\textup{(iii)}.
\hfill$\blacksquare$

\bigskip\noindent\textbf{Proof of Corollary~\ref{propAnal}.}
For $J=2$,
\[
	k_\alpha(y)
	=\frac{\phi_{\theta_0}(y_1-\alpha)\phi_{\theta_0}(y_2-\alpha)}
	{p_0(y)},
\]
which is entire in $\alpha$. For $\alpha=a+\mathrm ib$,
\[
	\bigl|\phi_{\theta_0}(y-a-\mathrm ib)\bigr|
	=e^{b^2/(2\theta_0^2)}\phi_{\theta_0}(y-a),
	\qquad
	\|k_{a+\mathrm ib}\|_{L_2}
	\le e^{b^2/\theta_0^2}\|k_a\|_{L_2}.
\]
Bounded support of $\eta_0$ implies that $\|k_a\|_{L_2}$ is finite and locally
bounded in $a$. Thus $\alpha\mapsto k_\alpha$ has an entire
$L_2$-valued extension, so every function in
$\mathcal H_K$ is the restriction of an entire function
and is real-analytic on $\mathcal A^\circ$.
Real analyticity is not sufficient: the function
$\alpha\mapsto(1+\alpha^2)^{-1}$ is real-analytic
on $\mathcal A$ but has no entire extension.
Indeed, such an extension $F$ would satisfy
$(1+z^2)F(z)=1$ throughout $\mathbb C$ by the
identity theorem, which is impossible at $z=\mathrm i$.
The nonexistence and regular-estimation conclusions
follow from Theorem~\ref{thm:rkhs_range}.
\hfill$\blacksquare$

		\subsection{Value-added targets}
		\label{sec:tva_targets}
		
\medskip\noindent\textbf{Proof of Proposition~\ref{prop:tva_analytic}.}
Write $\sigma_J=\theta_0/\sqrt J$. Conditional on $\alpha$,
\[
\bar Y_J=\alpha+\sigma_J Z,
\qquad Z\sim N(0,1).
\]

Suppose first that $h>\sigma_J$. For $\theta$ in a neighborhood
of $\theta_0$, $d_J(\theta)>0$, and the Gaussian convolution
identity gives
\[
\mathbb E_{\theta,\eta}
\left[
\left.
\Phi\left\{
\frac{a-\bar Y_J}{d_J(\theta)}
\right\}
\,\right|\,\alpha
\right]
=
\Phi\left(\frac{a-\alpha}{h}\right).
\]
Hence
$\mathbb E_{\theta,\eta}[q_{J,a,h}(Y,\theta)]=\psi_{a,h}$
for every $\eta$. The function $q_{J,a,h}$ is bounded, and its
$\theta$-derivative is locally bounded because
$|x\phi(x)|$ is bounded and $d_J(\theta)$ is bounded away from
zero near $\theta_0$. Together with the Gaussian moment bounds,
this verifies the admissibility requirements.

Let
$g^\star=g_{J,a,h}(Y,\psi_{a,h},\theta_0)$. Since
$\mathbb E[\varrho(Y,\theta)\mid\alpha]=0$,
\[
S^*g^\star(\alpha)
=
\Phi\left(\frac{a-\alpha}{h}\right)-\psi_{a,h}.
\]
Moreover,
\[
\frac{d}{d\theta}
\mathbb E_{\theta,\eta_0}
[g_{J,a,h}(Y,\psi_{a,h},\theta)]
\Big|_{\theta=\theta_0}
=
\langle g^\star,s_\theta\rangle
+\mathbb E[\partial_\theta q_{J,a,h}(Y,\theta_0)]
-\Gamma_{J,a,h}\mathbb E[\partial_\theta\varrho(Y,\theta_0)].
\]
The left-hand side is zero and
$\mathbb E[\partial_\theta\varrho(Y,\theta_0)]=-4\theta_0$.
By the definition of $\Gamma_{J,a,h}$,
$\langle g^\star,s_\theta\rangle=0$.
Thus \eqref{eq:mixture_LR} holds with $c=1$, and
Theorem~\ref{thm:mixture_tangent}\textup{(c)} gives local
robustness and relevance.

Now suppose $0<h<\sigma_J$ and, toward a contradiction, that a
relevant locally robust moment exists. By
Theorem~\ref{thm:rkhs_range}, the centered representer
\[
f(\alpha)
=
\Phi\left(\frac{a-\alpha}{h}\right)-\psi_{a,h}
\]
belongs to $\mathcal H_K$. For complex $z=x+\mathrm ib$, the
Gaussian feature map satisfies
\[
\|k_{x+\mathrm ib}\|_{L_2}
\le
\exp\left(\frac{Jb^2}{2\theta_0^2}\right)
\|k_x\|_{L_2}.
\]
Hence the RKHS representation
$f(z)=\langle g_f,k_z\rangle$ implies, at $x=a$,
\[
|f(a+\mathrm ib)|
\le
C\exp\left(\frac{Jb^2}{2\theta_0^2}\right)
\]
for some $C<\infty$. On the other hand,
\[
\Phi\left(-\frac{\mathrm ib}{h}\right)
=
\frac12-\frac{\mathrm i}{\sqrt{2\pi}}
\int_0^{b/h}e^{t^2/2}\,dt,
\]
so, as $|b|\to\infty$,
\[
\left|
\Phi\left(-\frac{\mathrm ib}{h}\right)
\right|
\asymp
\frac{h}{|b|}
\exp\left(\frac{b^2}{2h^2}\right).
\]
Since $h<\theta_0/\sqrt J$,
\[
\frac{1}{2h^2}>
\frac{J}{2\theta_0^2},
\]
contradicting the RKHS growth bound. Thus
$f\notin\mathcal H_K$, and Theorem~\ref{thm:rkhs_range}
rules out both relevant locally robust moments and regular
estimation.
\hfill$\blacksquare$		
        		\section{Tangent Sets of Mixture Models}
		\label{app:mixture}
		
We use the notation and Assumption~\ref{ass:mixture} of
Section~\ref{subsec:tva_setup}. Set
\[
		V:=\overline{\operatorname{span}}\bigl(\{a's_\theta:a\in\mathbb R^{d_\theta}\}
		\cup\{s_b:b\in\mathcal B\}\bigr),
		\qquad
		v_\theta:=s_\theta-\Pi_{\mathcal H_\eta}s_\theta,
		\qquad
		\Omega:=\mathbb E[v_\theta v_\theta'].
		\]
		
		\subsection{The tangent set}
		
		\begin{theorem}[Tangent sets of mixture models]
			\label{thm:mixture_tangent}
			Let Assumption~\ref{ass:mixture}(i)--(iv) hold.
			\begin{enumerate}
			\item[(a)] \emph{(Constructed paths.)} For $b\in\mathcal B$ and
				$t\in\mathbb R^{d_\theta}$ and all $\tau$ small enough, $\eta_\tau=(1+\tau b)\eta_0$
				lies in $\mathcal M$ and $\lambda_\tau=(\theta_0+\tau t,\eta_\tau)$ defines a path in
				$\mathcal P$ that is DQM with score $t's_\theta+s_b$. After normalization these paths
				have a common $L_p$ envelope for every $p<\infty$ and their scores span $V$;
				hence $V\subseteq\overline{\mathcal T}$. Moreover
				$k_\alpha-1\in\mathcal H_\eta$ for every
				$\alpha\in\mathcal A$, and $\Omega\succ0$; in particular
				$a's_\theta\notin\mathcal H_\eta$ for every $a\neq0$.
				\item[(b)] \emph{(Upper bound.)} Every DQM path
				$\{\mathbb P_{\theta_\tau,\eta_\tau}\}\subseteq\mathcal P$ through $\mathbb P_0$
				satisfies $\|\theta_\tau-\theta_0\|=O(\tau)$ and $\eta_\tau\Rightarrow\eta_0$, and
				\[
				\langle g,s\rangle=0
				\qquad\text{for every }s\in\mathcal T\text{ and every }
				g\in V^{\perp}\cap L_{2q},\ q>1.
				\]
				Consequently, if Assumption~\ref{ass:mixture}\textup{(vi)} also holds, then
$V^\perp\cap L_{2q}$ is dense in $V^\perp$ for the $q$ in that assumption,  and $\overline{\mathcal T}=V.$ The same equality holds without~\textup{(vi)} if the envelope in
Assumption~\ref{ass:mixture}\textup{(iii)} can be taken bounded.
			\item[(c)] \emph{(Restricted paths and range condition.)} Let
				Assumption~\ref{ass:mixture}(v) also hold. Choose $b_\psi\in\mathcal B$ with
$\langle r_0,b_\psi\rangle=1$, which exists by density of
$\mathcal B$, and put $\ell_\psi=s_{b_\psi}$.
The path generated by $b_\psi$ has target derivative one. Every constructed score
				$t's_\theta+s_b$ decomposes as $a\ell_\psi+s_0$, where
				$a=t'\bar r_\theta+\langle r_0,b\rangle$ and $s_0$ is the score of an exact
				target-preserving DQM path. Moreover, for
				$g\in L_2^{0}\cap L_{2q}$ with $q>1$,
				\begin{equation}
					\begin{aligned}
					g\in\overline{\mathcal T_0}^{\perp}
					&\iff
					\exists\,c\in\mathbb R:\quad
					S^{*}g(\alpha)=c\,r_0(\alpha)
					\quad\text{for every }\alpha\in\mathcal A,\\[-2pt]
					&\hspace{72pt}\text{and}\quad
					\langle g,s_\theta\rangle=c\,\bar r_\theta,
					\end{aligned}
					\label{eq:mixture_LR}
				\end{equation}
				In particular $g$ is relevant iff $c\neq0$.
			\end{enumerate}
		\end{theorem}

		\subsection{Proofs}
        		\medskip\noindent\textbf{Proof of Theorem~\ref{thm:mixture_tangent}(a):}
		Since $\int b\,d\eta_0=0$ and $\tau\|b\|_\infty<1$, $\eta_\tau$ is a probability measure
		on $\mathcal A$, so $\lambda_\tau\in\Theta\times\mathcal M$. Write
		$p^{(1)}_\tau:=p_{\theta_\tau,\eta_0}$ and
		$\beta_\tau(z):=\int b(\alpha)p(z\mid\alpha,\theta_\tau)\,d\eta_0(\alpha)/p^{(1)}_\tau(z)$,
		so that $p_\tau=p^{(1)}_\tau(1+\tau\beta_\tau)$ with $|\beta_\tau|\le\|b\|_\infty$ and
		$\beta_0=s_b$. The parametric factor is DQM with score $t's_\theta$; moreover,
		$|\sqrt{1+x}-1-x/2|\le x^2$ and dominated convergence give
		$\beta_\tau\to s_b$ in $L_2(\mathbb P_0)$. Expanding
		$\sqrt{p_\tau}=\sqrt{p^{(1)}_\tau}\sqrt{1+\tau\beta_\tau}$ therefore proves DQM
		with score $t's_\theta+s_b$. Assumption~\ref{ass:mixture}(iii), after normalizing by
		$1+\|t\|+\|b\|_\infty$, supplies the common $L_p$ envelope. Since
		$b\mapsto s_b$ is an $L_2$-contraction and
		$\mathcal B$ is dense in $L_2^{0}(\eta_0)$, these scores span $V$.
		
		If $g\perp\mathcal H_\eta$ then $S^{*}g=0$ $\eta_0$-a.s.; by
		Assumption~\ref{ass:mixture}(ii) the map $\alpha\mapsto S^{*}g(\alpha)=\langle g,k_\alpha\rangle$
		is continuous and, $\eta_0$ having support $\mathcal A$, it vanishes on all of
		$\mathcal A$. Hence $\langle g,k_\alpha-1\rangle=S^{*}g(\alpha)-\mathbb E[g]=0$ for every
		$\alpha\in\mathcal A$, so $k_\alpha-1\in(\mathcal H_\eta^{\perp})^{\perp}=\mathcal H_\eta$.
		
		Finally, suppose $a'\Omega a=0$ for some $a\neq0$, i.e.\ $a's_\theta\in\mathcal H_\eta$.
		Approximating $a's_\theta$ by $s_b$ and applying (iv) to the path $(a,-b)$ gives
		$\|a\|\le C\|a's_\theta-s_b\|_{L_2}$; taking the approximation error to zero contradicts
		$a\ne0$. Thus $\Omega\succ0$.
		\hfill$\blacksquare$
		
		\medskip\noindent\textbf{Proof of Theorem~\ref{thm:mixture_tangent}(b):}
		DQM gives
$H(\mathbb P_\tau,\mathbb P_0)
=\tau\|s\|_{L_2}/(2\sqrt2)+o(\tau)$.
The stability clause in Assumption~\ref{ass:mixture}(iv)
first gives $\theta_\tau\to\theta_0$; its local Hellinger
bound then yields $\|\theta_\tau-\theta_0\|=O(\tau)$. Compactness gives a weakly convergent subsequence of
		$\eta_\tau$, say to $\eta^*$. Continuity of the kernel and compactness of
		$\mathcal A$ give pointwise convergence of the corresponding mixture densities;
		Scheff\'e's lemma upgrades it to total variation. Since the original path converges
		to $\mathbb P_0$, identification forces $\eta^*=\eta_0$. The argument applies to
		every subsequence, so $\eta_\tau\Rightarrow\eta_0$.
		
		Now fix $q>1$, let $p$ be conjugate to $q$ and let $g\in V^{\perp}\cap L_{2q}$. By
		Assumption~\ref{ass:mixture}(iii) and H\"older's inequality,
		$\limsup_{\tau}\int g^{2}p_\tau\,d\mu\le\|g\|^{2}_{L_{2q}}\|M_p\|_{L_p}<\infty$, so
		Lemma 1.7.2 of \citet{ibragimovkhasminskii1981} applies along the path and
		\[
		\left.\frac{d}{d\tau}\mathbb E_\tau[g]\right|_{\tau=0^{+}}=\langle g,s\rangle.
		\]
		Put $G_\theta(\alpha)=\int g(z)p(z\mid\alpha,\theta)d\mu(z)$. The envelope permits
		differentiation uniformly on $\mathcal A\times\overline U$. Orthogonality to $V$
		gives $G_{\theta_0}\equiv0$ and
		$\int\dot G_{\theta_0}d\eta_0=\langle g,s_\theta\rangle=0$. Hence
		\[
		\mathbb E_\tau[g]=\int\bigl\{G_{\theta_\tau}-G_{\theta_0}\bigr\}\,d\eta_\tau
		=(\theta_\tau-\theta_0)'\int\dot G_{\theta_0}\,d\eta_\tau
		+o\bigl(\|\theta_\tau-\theta_0\|\bigr),
		\]
		and weak convergence makes the integral tend to zero. Since
		$\|\theta_\tau-\theta_0\|=O(\tau)$, this gives
		$\mathbb E_\tau[g]=o(\tau)$ and hence $\langle g,s\rangle=0$. It remains to verify the density claim under
Assumption~\ref{ass:mixture}\textup{(vi)}. Since
$\mathcal H_\eta=\overline{\operatorname{Range}(S)}$,
\[
\mathcal H_\eta^\perp=\ker(S^*),
\qquad
V=\mathcal H_\eta\oplus
\operatorname{span}\{v_{\theta,1},\ldots,v_{\theta,d_\theta}\}.
\]
Let $D:=\ker(S^*)\cap L_{2q}$. By~\textup{(vi)}, $D$ is dense in
$\ker(S^*)$. Since $v_{\theta,j}\in\ker(S^*)$ and $\Omega\succ0$, we can choose
$w_j\in D$, $j=1,\ldots,d_\theta$, sufficiently close to $v_{\theta,j}$ in
$L_2$ that, for $w=(w_1,\ldots,w_{d_\theta})'$,
\[
C:=\mathbb E[w v_\theta']
\]
is nonsingular. For any $g\in V^\perp$, choose $g_n\in D$ with
$g_n\to g$ in $L_2$ and set
\[
\tilde g_n
:=g_n-\mathbb E[g_n v_\theta']C^{-1}w.
\]
Then $\tilde g_n\in D$,
$\mathbb E[\tilde g_n v_\theta']=0$, and hence
$\tilde g_n\in V^\perp\cap L_{2q}$. Since
$\mathbb E[g_n v_\theta']\to\mathbb E[g v_\theta']=0$,
$\tilde g_n\to g$ in $L_2$. Thus
$V^\perp\cap L_{2q}$ is dense in $V^\perp$, and the preceding orthogonality
result gives
\[
\overline{\mathcal T}\subseteq
(V^\perp\cap L_{2q})^\perp=V.
\]
Part~\textup{(a)} gives the reverse inclusion, so
$\overline{\mathcal T}=V$. If the envelope in
Assumption~\ref{ass:mixture}\textup{(iii)} is bounded, the same conclusion
follows directly with $q=1$.
		\hfill$\blacksquare$

        		\medskip\noindent\textbf{Proof of Theorem~\ref{thm:mixture_tangent}(c):}
		\emph{Coordinates.} By construction,
$b_\psi\in\mathcal B$ and $\langle r_0,b_\psi\rangle=1$, so the path
$(0,b_\psi)$ has score $\ell_\psi$ and target derivative one. For a generated direction $(t,b)$, put
		$a=t'\bar r_\theta+\langle r_0,b\rangle$ and $b^\dagger=b-ab_\psi$.  To construct a
		restricted path with score $t's_\theta+s_{b^\dagger}$, define
		\[
		F(\tau,c):=\int r(\alpha,\theta_0+\tau t)\bigl\{1+\tau b^{\dagger}(\alpha)+c\,b_\psi(\alpha)\bigr\}\,d\eta_0(\alpha)-\psi_0 .
		\]
		Here $F(0,0)=F_\tau(0,0)=0$ and $F_c(0,0)=1$.  The implicit-function theorem gives
		$c(\tau)=o(\tau)$ with $F(\tau,c(\tau))=0$, so the resulting restricted path has the
		required score and supplies the asserted decomposition of each constructed direction.
		
		\emph{Necessity in \eqref{eq:mixture_LR}.} Let $g\in\overline{\mathcal T_0}^{\perp}$.
		Restricted paths of the above form with $t=0$ and $\langle r_0,b\rangle=0$ give
		$0=\langle g,s_b\rangle=\langle S^{*}g,b\rangle_{L_2(\eta_0)}$ for every
		$b\in\mathcal B$ with $\langle r_0,b\rangle=0$; since $\mathcal B$ is dense in $L_2^0(\eta_0)$ and
$\langle r_0,b_\psi\rangle=1$, the bounded elements orthogonal to $r_0$
are dense in $r_0^\perp$, so $S^*g=c\,r_0$ $\eta_0$-a.s. This equality is in fact pointwise. Fix
		$\alpha\in\mathcal A$, set
		$b_\alpha=-r_0(\alpha)b_\psi$ and
		$\eta_\tau=\{1+\tau(b_\alpha-1)\}\eta_0+\tau\delta_\alpha$.
		For all sufficiently small $\tau$, this is a probability measure and
		$\int r(\beta,\theta_0)d\eta_\tau(\beta)=\psi_0$. Its mixture density is DQM with
		score $k_\alpha-1+s_{b_\alpha}$. Orthogonality gives
		$0=S^{*}g(\alpha)+\langle S^{*}g,b_\alpha\rangle
		=S^{*}g(\alpha)-c r_0(\alpha)$, proving the equality for every $\alpha$.
		Taking a restricted path with $t\neq0$ and
		$t'\bar r_\theta+\langle r_0,b\rangle=0$ then gives
		$0=t'\langle g,s_\theta\rangle+c\langle r_0,b\rangle=t'\{\langle g,s_\theta\rangle-c\bar r_\theta\}$,
		whence $\langle g,s_\theta\rangle=c\,\bar r_\theta$.
		
		\emph{Sufficiency in \eqref{eq:mixture_LR}.} Let $g\in L_2^{0}\cap L_{2q}$ satisfy the
		right-hand side and take any restricted DQM path.  The argument in (b) gives
		$\|\theta_\tau-\theta_0\|=O(\tau)$, $\eta_\tau\Rightarrow\eta_0$, and differentiation
		of $\mathbb E_\tau g$. Expanding $G_{\theta_\tau}$ and the target restriction, using
		$G_{\theta_0}=cr_0$, gives
		\[
		\mathbb E_\tau[g]
		=(\theta_\tau-\theta_0)'
		\left\{\int\dot G_{\theta_0}\,d\eta_\tau
		-c\int\partial_\theta r(\alpha,\theta_0)\,d\eta_\tau\right\}
		+o(\|\theta_\tau-\theta_0\|).
		\]
		(The expression is identically zero when $\theta_\tau=\theta_0$.) The bracket tends
		to $\langle g,s_\theta\rangle-c\bar r_\theta=0$, so
		$\mathbb E_\tau[g]=o(\tau)$ and its derivative $\langle g,s_0\rangle$ is zero. Thus
		$g\in\overline{\mathcal T_0}^{\perp}$. Finally, if $c\neq0$, then
$\langle g,\ell_\psi\rangle
=\langle S^*g,b_\psi\rangle=c\neq0$, proving relevance.
If $c=0$, the range conditions imply $g\in V^\perp$;
part~(b), applied to this $g\in L_{2q}$, gives
$g\perp\overline{\mathcal T}$, so $g$ is irrelevant.
\hfill$\blacksquare$

\subsection{Nonlinear targets}

\begin{lemma}[Necessary range condition for nonlinear targets]
\label{lem:nonlinear_range}
Let Assumptions~\ref{Regularity_model} and
\ref{ass:mixture}\textup{(i)--(iv), (vi)} hold.
Consider a scalar target $\Psi(\theta,\eta)$.
Suppose that there exist
$r_{\Psi,0}\in L_2^0(\eta_0)\setminus\{0\}$,
$a_\Psi\in\mathbb R^{d_\theta}$, and a dense linear
subspace $\mathcal B_\Psi\subseteq\mathcal B$ of
$L_2^0(\eta_0)$ with the following property:
for every $t\in\mathbb R^{d_\theta}$ and
$b,h\in\mathcal B_\Psi$, the map
\[
F_{t,b,h}(u,v)
:=
\Psi\bigl(\theta_0+ut,(1+ub+vh)\eta_0\bigr)
\]
is continuously differentiable near $(0,0)$, with
\[
\partial_u F_{t,b,h}(0,0)
=a_\Psi't+\langle r_{\Psi,0},b\rangle_{L_2(\eta_0)},
\qquad
\partial_v F_{t,b,h}(0,0)
=\langle r_{\Psi,0},h\rangle_{L_2(\eta_0)}.
\]
If a relevant locally robust moment exists for
$\Psi(\theta_0,\eta_0)$, then there is $g_0\in L_2^0$ such that
\[
S^*g_0=r_{\Psi,0}
\qquad \eta_0\text{-a.s.}
\]
The same range condition is necessary for regular estimation.
Consequently, if $r_{\Psi,0}$ does not agree
$\eta_0$-a.s.\ with an element of $\mathcal H_K$,
neither a relevant locally robust moment nor a regular
estimator sequence exists on the stated local experiment.
\end{lemma}

\begin{proof}
Write $r=r_{\Psi,0}$ and
$\Psi_0=\Psi(\theta_0,\eta_0)$.
Since $r\ne0$ and $\mathcal B_\Psi$ is dense, choose
$h\in\mathcal B_\Psi$ with
$\langle r,h\rangle_{L_2(\eta_0)}=1$.
For any $(t,b)$ satisfying
\[
a_\Psi't+\langle r,b\rangle_{L_2(\eta_0)}=0,
\]
the implicit-function theorem gives a continuously
differentiable $v(u)$ with $v(0)=v'(0)=0$ such that
\[
\Psi\bigl(\theta_0+ut,(1+ub+v(u)h)\eta_0\bigr)=\Psi_0.
\]
The boundedness of $b$ and $h$ makes these probability
measures for all sufficiently small $u$.
Moreover, the correcting tilt changes the square-root
density by $O(|v(u)|)=o(|u|)$ in $L_2(\mu)$.
Proposition~\ref{prop:tva_tangent} therefore gives an
exactly target-preserving DQM path with score
$t's_\theta+Sb$.

Let $g$ be the true evaluation of an admissible locally
robust moment for $\Psi_0$.
Lemma~\ref{lem:pathwise} and the generating-class property
give $g\perp\overline{\mathcal T_0}$, where
$\mathcal T_0$ is the restricted tangent set for $\Psi_0$.
For every $b\in\mathcal B_\Psi$, apply the preceding
construction with $t=0$ and
$b-\langle r,b\rangle_{L_2(\eta_0)}h$.
Writing $c=\langle g,Sh\rangle$, orthogonality gives
\[
\langle S^*g,b\rangle_{L_2(\eta_0)}
=c\langle r,b\rangle_{L_2(\eta_0)}.
\]
Density yields $S^*g=cr$ in $L_2(\eta_0)$.
Taking instead $b=-(a_\Psi't)h$ gives
\[
\langle g,s_\theta\rangle=ca_\Psi.
\]
If $c=0$, then $g$ is orthogonal to both
$\operatorname{Range}(S)$ and the components of $s_\theta$.
Proposition~\ref{prop:tva_tangent} implies
$g\perp\overline{\mathcal T}$, contradicting relevance.
Hence $c\ne0$, and $g_0=g/c$ satisfies the asserted
range equation.

For regular estimation, the necessary pathwise
differentiability condition
\citep{van1991differentiable}, applied to the latent
tilts, gives an influence function $\dot\Psi\in L_2^0$
satisfying
\[
\langle S^*\dot\Psi,b\rangle_{L_2(\eta_0)}
=\langle r,b\rangle_{L_2(\eta_0)}
\qquad(b\in\mathcal B_\Psi).
\]
Density again gives the range equation.
The RKHS conclusion follows from
Theorem~\ref{thm:rkhs_range}.
\end{proof}

\paragraph{Verification for the nonlinear policy examples.}
Suppose the latent law has a density continuous and
strictly positive on a neighborhood of its unique
interior $\varphi$-quantile $q_\varphi$.
Take
$\mathcal B_\Psi=C(\mathcal A)\cap L_2^0(\eta_0)$,
which is a dense linear subspace of $L_2^0(\eta_0)$
and consists of bounded functions.
For $b,h\in\mathcal B_\Psi$, put
\[
\eta_{u,v}=(1+ub+vh)\eta_0,
\qquad
H_b(x)=\int\mathbf 1\{\alpha\le x\}b(\alpha)\,d\eta_0(\alpha).
\]
Its CDF is
$F_{u,v}(x)=F_{\alpha,0}(x)+uH_b(x)+vH_h(x)$.
This map is jointly continuously differentiable near
$(0,0,q_\varphi)$, with
$\partial_xF_{0,0}(q_\varphi)=f_{\alpha,0}(q_\varphi)>0$.
The implicit-function theorem makes the quantile
$q_\varphi(\eta_{u,v})$ continuously differentiable
in $(u,v)$.

The endogenous-threshold gain underlying the displayed
representer in the main text is
\[
G_\varphi(\eta)
=
\int
\{q_\varphi(\eta)-\alpha\}
\mathbf 1\{\alpha\le q_\varphi(\eta)\}\,d\eta(\alpha).
\]
It is also continuously differentiable on these
submodels. In direction $b$,
\[
\dot q_b
=-\frac{H_b(q_\varphi)}{f_{\alpha,0}(q_\varphi)},
\qquad
\dot G_b
=\int(q_\varphi-\alpha)\mathbf 1\{\alpha\le q_\varphi\}
b(\alpha)\,d\eta_0(\alpha)+\varphi\dot q_b.
\]
These derivatives give the quantile and bottom-share
representers displayed in the main text, after centering.
Both are nonzero, and neither target depends on $\theta$,
so $a_\Psi=0$.
Thus both satisfy Lemma~\ref{lem:nonlinear_range}.
		
\section{One-Sided Orthogonality}
\label{app:cone}

At boundaries the tangent sets need not be linear. Let
\[
\mathfrak C=\overline{\operatorname{cone}(\mathcal T)},\qquad
\mathfrak C_0=\overline{\operatorname{cone}(\mathcal T_0)},
\]
and, for a closed convex cone $C\subset L_2^0$, let
$C^\circ=\{h:\langle h,c\rangle\le0\ \forall c\in C\}$ and $\Pi_C$ denote its
polar and metric projection. Assume the generating classes are conically generating,
so their closed conical score hulls equal $\mathfrak C$ and $\mathfrak C_0$.

An admissible moment is \emph{one-sidedly locally robust} if its derivative along
every target-preserving path is nonnegative; let
$\mathcal G_{\mathrm{LR}}^+$ denote its true evaluations. It is
\emph{one-sidedly relevant} if $\langle g_0,s\rangle>0$ for some
$s\in\mathcal T$.

\begin{theorem}[One-sided existence and relevance]
\label{thm:cone}
Let Assumption~\ref{Regularity_model}(i) hold. Suppose
Assumption~\ref{ass:regclass} holds for a conically generating class and
$\mathfrak C_0^\circ\cap L_{2q}$ is dense in $\mathfrak C_0^\circ$. Then
\[
\mathcal G_{\mathrm{LR}}^+
 =\mathfrak C_0^\circ\cap\mathcal G_{\mathrm{adm}}^0,
\qquad
\mathcal G_{\mathrm{LR}}
 =\mathfrak C_0^\circ\cap(-\mathfrak C_0^\circ)\cap\mathcal G_{\mathrm{adm}}^0.
\]
Hence a nonzero one-sidedly robust moment exists iff $\mathfrak C_0\neq L_2^0$;
$g_0\in\mathcal G_{\mathrm{LR}}^+$ is relevant iff $\Pi_\mathfrak Cg_0\neq0$; and a
relevant one-sidedly robust moment exists iff $\mathfrak C_0\subsetneq \mathfrak C$.
If $\mathcal T$ and $\mathcal T_0$ are linear, this reduces to the two-sided
results; otherwise one-sided relevance may exist when two-sided relevance does not.
\end{theorem}

\begin{proposition}[One-sided local-power envelope]
\label{prop:cone_power}
Under Theorem~\ref{thm:cone}, if $\mathfrak C_0\neq L_2^0$, then for every
$s\in\mathcal T$,
\[
\sup_{g_0\in\mathcal G_{\mathrm{LR}}^+\setminus\{0\}}
\frac{[\langle g_0,s\rangle]_+}{\|g_0\|}
=\operatorname{dist}(s,\mathfrak C_0).
\]
Upper-tail oracle tests have asymptotic size at most $\zeta$ along restricted local
sequences and, for $s\notin \mathfrak C_0$, supremal local power
\[
1-\Phi\{z_{1-\zeta}-\operatorname{dist}(s,\mathfrak C_0)\}.
\]
For $s\notin\mathfrak C_0$, the envelope is attained if $\Pi_{\mathfrak C_0^\circ}s\in L_{2q}$ and approached otherwise.
\end{proposition}

\noindent\textit{Proofs.}
Lemma~\ref{lem:pathwise} gives
$dE[g(Z,\lambda_\tau)]/d\tau|_{0^+}=-\langle g_0,s_0\rangle$.
Conical generation therefore yields
$\mathcal G_{\mathrm{LR}}^+=\mathfrak C_0^\circ\cap\mathcal G_{\mathrm{adm}}^0$,
while equality of the derivative in both directions gives the characterization of
$\mathcal G_{\mathrm{LR}}$. The bipolar theorem implies
$\mathfrak C_0^\circ=\{0\}$ iff $\mathfrak C_0=L_2^0$, and density together with
Corollary~\ref{cor:richness} gives existence.

By Moreau's decomposition,
$g_0=\Pi_\mathfrak Cg_0+\Pi_{\mathfrak C^\circ}g_0$. Thus $\Pi_\mathfrak Cg_0=0$ iff
$g_0\in \mathfrak C^\circ$, in which case $\langle g_0,s\rangle\le0$ for all
$s\in\mathcal T$; if $\Pi_\mathfrak Cg_0\neq0$, then
$\langle g_0,\Pi_\mathfrak Cg_0\rangle=\|\Pi_\mathfrak Cg_0\|^2>0$, implying relevance.
Since $\mathfrak C_0\subseteq \mathfrak C$ iff $\mathfrak C^\circ\subseteq \mathfrak C_0^\circ$, with strict inclusion
preserved under polarity, the final claim of Theorem~\ref{thm:cone} follows.

For Proposition~\ref{prop:cone_power}, Le Cam's third lemma gives local mean shift
$\langle g_0,s\rangle/\|g_0\|$. Moreau's decomposition of $s$ relative to $\mathfrak C_0$
implies, for $h\in \mathfrak C_0^\circ$,
\[
\langle h,s\rangle
\le \|h\|\,\|\Pi_{\mathfrak C_0^\circ}s\|
=\|h\|\,\operatorname{dist}(s,\mathfrak C_0),
\]
with equality at $h=\Pi_{\mathfrak C_0^\circ}s$. Corollary~\ref{cor:richness} gives
attainment or approximation according as this projection belongs to $L_{2q}$.
The size and power statements follow from the Gaussian limit.
\hfill$\blacksquare$
\section{Additional Monte Carlo Results}
\label{app:mc_additional}

Table~\ref{tab:mc_dml_ar} compares AR--NO with ordinary-inverse DML in the
three theory-driven experiments of Section~\ref{sec:montecarlo}. Both procedures
have approximately correct pointwise coverage, but their behavior near singularity
differs. DML always reports a bounded Wald interval, whose upper tail can be
extremely long; AR--NO instead records severe information loss through long or
unbounded confidence sets. The table is restricted to identified coordinates.
For the unrotated \(c=0\) target and the kernel coordinate \(\beta_2\) in
Design~B, inclusion of a normalization-dependent reference is not coverage and
is therefore not tabulated. These results provide a pointwise comparison, not a
claim of uniform dominance.

\begin{table}[H]
\centering
\begin{threeparttable}
\footnotesize
\setlength{\tabcolsep}{1.5pt}
\renewcommand{\arraystretch}{1.00}
\caption{DML and AR--NO inference in the theory-driven designs}
\label{tab:mc_dml_ar}
\begin{tabular*}{\linewidth}
{@{\extracolsep{\fill}}lrrrrrrrr@{}}
\toprule
& \multicolumn{4}{c}{$n=5{,}000$}
& \multicolumn{4}{c}{$n=10{,}000$} \\
\cmidrule(lr){2-5}\cmidrule(lr){6-9}
Design
& \shortstack{Cov.\\D/A}
& \shortstack{$L_{50}$\\D/A}
& \shortstack{$L_{99}$\\D/A}
& \shortstack{AR\\Unbd.}
& \shortstack{Cov.\\D/A}
& \shortstack{$L_{50}$\\D/A}
& \shortstack{$L_{99}$\\D/A}
& \shortstack{AR\\Unbd.} \\
\midrule
A: Triangular
& 96.0/95.1 & .19/.19 & .41/.41 & 0.1
& 94.6/94.4 & .13/.13 & .21/.20 & 0.0 \\

B: Rank one
& 97.9/96.7 & .17/.20 & 39.6/$\infty$ & 1.1
& 96.1/96.3 & .11/.14 & 22.7/5.18 & 0.6 \\

C: $\mathrm{CP}=10$
& 96.7/95.4 & .69/.61 & 405.8/$\infty$ & 10.6
& 96.8/94.9 & .48/.45 & 641.9/$\infty$ & 7.2 \\

C: $\mathrm{CP}=50$
& 96.7/95.7 & .54/.50 & 756.2/$\infty$ & 5.3
& 96.9/96.0 & .41/.40 & 74.0/$\infty$ & 4.3 \\

C: $\mathrm{CP}=200$
& 97.4/97.2 & .31/.32 & 13.6/$\infty$ & 1.9
& 95.4/96.8 & .28/.29 & 10.7/$\infty$ & 1.2 \\

C: $\mathrm{CP}=500$
& 96.1/95.9 & .19/.20 & 7.61/20.9 & 0.8
& 95.6/97.4 & .19/.20 & 6.01/$\infty$ & 1.0 \\
\bottomrule
\end{tabular*}

\begin{tablenotes}[flushleft]
\footnotesize
\item \textit{Notes:} D/A denotes DML/AR--NO. Coverage and AR
unboundedness are percentages. \(L_{50}\) is median 95\% length, conditional
on boundedness for AR--NO; \(L_{99}\) assigns infinite length to unbounded
AR--NO sets. Design~C labels give population target concentration parameters.
Each cell uses 1{,}000 replications and five-fold simple cross-fitting; the
Monte Carlo standard error at 95\% coverage is 0.69 percentage points.
\end{tablenotes}
\end{threeparttable}
\end{table}

\end{appendix}
\endgroup
\end{document}